\documentclass[pre, preprint, floatfix]{revtex4-2}

\usepackage{graphicx}
\usepackage{mathtools}
\usepackage{bm}
\usepackage{booktabs}
\usepackage{siunitx}

\usepackage[hidelinks, breaklinks=true]{hyperref}

\begin{document}

\title{Observing network dynamics through sentinel nodes}

\author{Neil G.\ MacLaren$^{1}$}
\altaffiliation{Current address: U.S. Army Research Institute for the Behavioral and Social Sciences, Ft.\,Belvoir, VA, 22060, USA}
\author{Baruch Barzel$^{2,3}$}
\author{Naoki Masuda$^{1,4,5}$}
\altaffiliation{Current address: Gilbert S. Omenn Department of Computational Medicine and Bioinformatics and Department of Mathematics, University of Michigan, Ann Arbor, MI 48109, USA}
\email{naokimas@gmail.com}

\affiliation{$^1$Department of Mathematics, State University of New York at Buffalo, NY 14260-2900, USA}
\affiliation{$^2$Department of Mathematics, Bar-Ilan University, Ramat-Gan, 5290002, Israel}
\affiliation{$^3$The Gonda Multidisciplinary Brain Research Center, Bar-Ilan University, Ramat-Gan, 5290002, Israel}
\affiliation{$^4$Institute for Artificial Intelligence and Data Science, State University of New York at Buffalo, USA}
\affiliation{$^5$Center for Computational Social Science, Kobe University, Kobe, 657-8501, Japan}

\date{\today}

\begin{abstract}
  A fundamental premise of statistical physics is that the particles in a physical system are interchangeable, and hence the state of each specific component is representative of the system as a whole. This assumption breaks down for complex networks, in which nodes may be extremely diverse, and no single component can truly represent the state of the entire system. It seems, therefore, that to observe the dynamics of social, biological or technological networks, one must extract the dynamic states of a large number of nodes---a task that is often practically prohibitive. Theoretical tools are also highly restrictive, given the analytically impenetrable combination of complex heterogeneous networks with nonlinear, often hidden, dynamics. To overcome this challenge, we use machine learning techniques to detect the network's sentinel nodes, a set of network components whose combined states can help approximate the average dynamics of the entire network. The method allows us to assess the equilibrium state of a large complex system by tracking just a small number of carefully selected nodes. We find that the sentinels are mainly determined by the network structure such that they can be extracted even with little knowledge of the system's specific interaction dynamics. Therefore, the network's sentinels offer a natural probe  by which to observe the system's dynamic states. Intriguingly, sentinels tend to avoid the highly central nodes such as the hubs.
\end{abstract}

\maketitle

\section{Introduction\label{sec:intro}}

The dynamic state of a complex networked system is given in terms of the microscopic states $x_i$ of all its components (nodes) \cite{porter2010, newman2018, barrat2008}. For example, the functionality of a cell can be captured by the expression levels of all individual genes \cite{karlebach2008, trapnell2015, wagner2016}. Similarly, the state of an ecosystem is often characterized by the abundance of all species \cite{wisz2013, bascompte2023}. Therefore, to observe the state of the system, one must simultaneously track the individual states of its many nodes---a level of empirical control that we seldom possess.

We, therefore, seek efficient reduction methods, that help track the state of the system in a more compact fashion. In most applications, such reduction is achieved by focusing on the average activity of all nodes, \textit{i.e.}, $\langle x \rangle = \sum_{i = 1}^N a_i x_i$, where the weights $\{a_1, \ldots, a_N \}$ characterize the importance of each node in assessing the system’s dynamic state \cite{gao2016, laurence2019, thibeault2020, tu2021, masuda2022, vegue2023, ma2023}. For example, selecting $a_i = k_i$, a node’s weighted degree, leads to the Gao-Barzel-Barab\'{a}si (GBB) reduction \cite{gao2016}, which helps predict the critical points of transition between dynamic states. Alternatively, under the dynamics approximate reduction technique (DART), the weight $a_i$ represents the $i$th entry of the leading eigenvector of the network’s adjacency matrix \cite{laurence2019, thibeault2020}.

While these reduction methods provide crucial theoretical insight, allowing us to analytically predict the system’s critical points of transition, they offer limited advances on observability. This is because measuring $\langle x \rangle$ via GBB or DART, one still needs to access the state of all nodes $x_1,\dots,x_N$, often beyond the bounds of empirical accessibility. More selective probes have been introduced in the context of early warning signals, where one follows just a small number of nodes that show signs of looming transitions early on, prior to the rest of the network \cite{chen2012, vafaee2016, aparicio2021, maclaren2023, masuda2024}. The problem is that such early warning nodes are, by design, selected thanks to their divergent behavior around the transition points. This renders them unlikely to be a representative sample of the entire networked system.

How then can we capture the system's collective dynamics, but with a node set that is comparable to the few early warning signifiers? To solve this, here we use machine learning to impose a sparsity condition on $\{ a_1, \ldots, a_N \}$, requiring that $a_i \ne 0$ only for a small number of nodes. The result is a set of sentinel nodes, whose average allows us to approximate the equilibrium state of the entire system. We find that, with only a few strategically selected nodes, we can achieve network-wide observability. The challenge is conceptually related to the functional observability problem in control theory \cite{Fernando2010IeeeTransAutoCont, Montanari2022Pnas}. The crucial distinction, however, is that our computational approach is not confined to linear dynamics. Instead, it enables observation of nonlinear systems and their critical transitions, even in cases where the underlying nonlinear mechanisms are unknown.

Our analysis shows that an ideal sentinel node set tends to sample the network heterogeneity. For instance, it typically comprises a range of low, average, and high degree nodes. Therefore, the optimal combination of sentinel nodes is mainly determined by the network topology, largely independent of the system’s specific dynamics. Consequently, one can extract the sentinels even without \textit{a priori} knowledge about the system’s hidden dynamics. This result allows practical observability even under restriction that one lacks full knowledge of the system's internal driving dynamic mechanisms.

\section{Results}
\label{sec:results}

\subsection{Seeking the sentinel nodes\label{sub:result-overview}}

\begin{figure*}
  \centering
  \includegraphics[width = 0.75\textwidth]{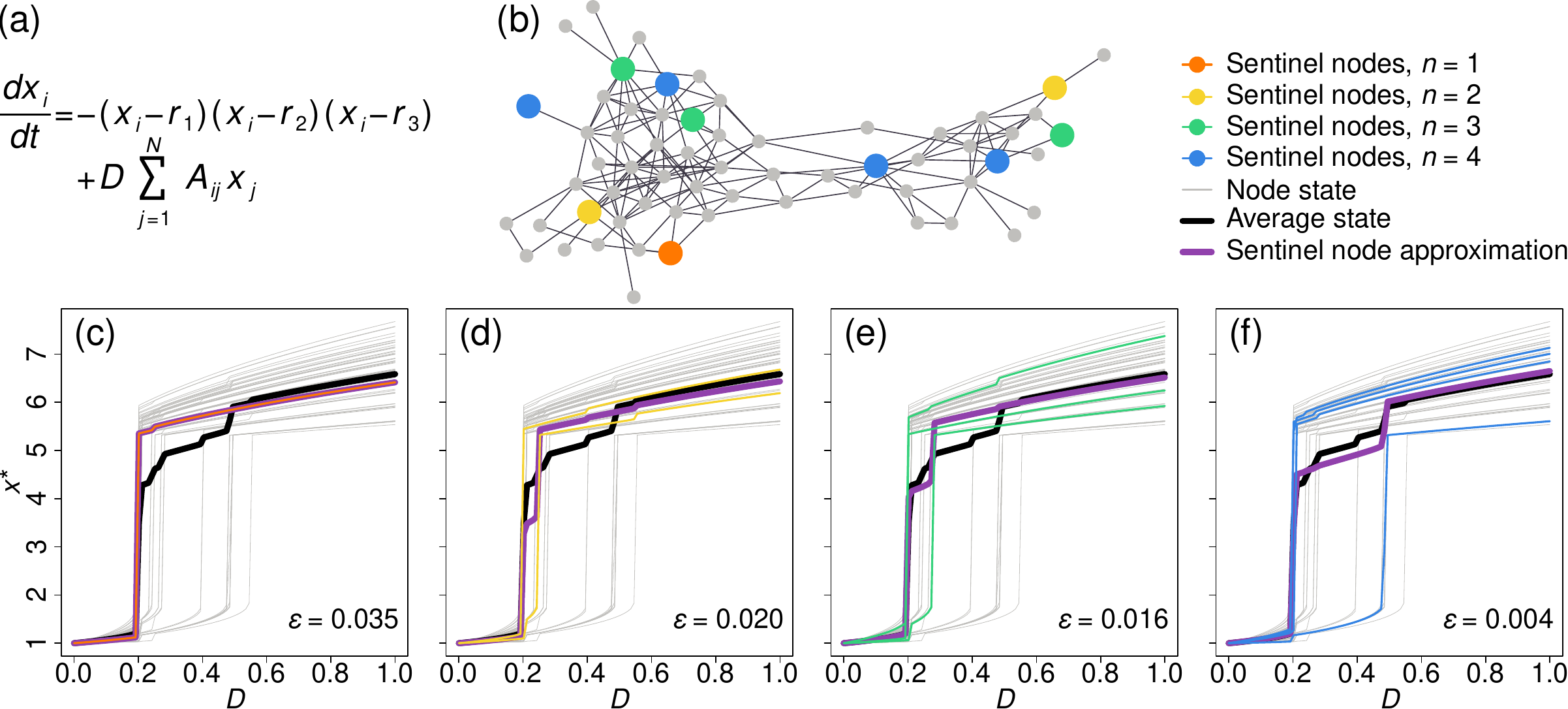}
  \caption{
    \textbf{Observing network dynamics via sentinel nodes}.\
    Approximations of the average activity, $\overline{x}$, of the coupled double-well dynamics on the dolphin network.
    (a) Coupled double-well dynamics.
    (b) Dolphin network with $N=62$ nodes. Larger circles (colored) represent nodes in four different sentinel node sets $S$ with sizes $n=1$ (orange), $n=2$ (yellow), $n=3$ (green), and $n=4$ (blue).
    (c)--(f) The equilibrium states $x_i^*$ of all nodes in the dolphin network as a function of $D$, obtained from numerical simulations (grey lines). The average state $\overline{x}$ is also shown (black solid lines). The purple solid lines represent the sentinel node approximations $\overline{x}^\prime$, obtained from the four node sets highlighted in (b). For $n = 1$, the approximation substantially deviates from the actual $\overline{x}$ (yielding $\varepsilon = 0.035$; see panel (c)) because, just a single sentinel node is used. This deviation is gradually reduced in (d), (e), and (f) as $n$ is increased, until it is practically eliminated at $n=4$ ($\varepsilon = 0.004$). In (c)--(f), we also show the individual sentinel node states (red, yellow, green and blue lines).   
  }
  \label{fig:demo}
\end{figure*}

Consider a network with $N$ components (nodes), whose activities $x_i(t)$ are driven by nonlinear dynamics. This can capture, \textit{e.g.}, mutualistic dynamics in ecological networks, epidemic dynamics on social systems, or the double-well dynamics presented in Fig.\ \ref{fig:demo}a. We denote the stable equilibrium states of the system by $x_i^*$. While the full equilibrium state of the system comprises all node activities $x_i^*$, $i = 1,\dots,N$, quite often, we only seek the average network activity 

\begin{equation}
\overline{x} = \frac{1}{N} \sum_{i=1}^N x_i^*
\label{eq:def-overlinex}
\end{equation}

\noindent
to help characterize the system's global state. For example, in contagion processes, assigning $x_i = 0$ for susceptible and $x_i = 1$ for infectious, the average $\overline{x}$ represents the fraction of infectious individuals in the population at equilibrium. In a similar fashion, $\overline{x}$ may represent the mean political opinion, the fraction of people aware of news, the average biomass in an ecosystem, or the proliferation of cancerous cells in the body, all depending on the scientific context. Quantity $\overline{x}$ is simple, intuitive, and practical, providing a natural parametrization of the network's dynamics.

Our goal is to approximate $\overline{x}$ in Eq.~\eqref{eq:def-overlinex} using a limited set $S$ of sentinel nodes, namely,

\begin{equation}
\overline{x}' = \frac{1}{n} \sum_{i \in S} x_i^*,
\label{eq:def-overlineprimex}
\end{equation}

\noindent
where $n \ll N$ is the number of sentinel nodes in $S$. To obtain $S$, we simulate the system under a range of conditions, \textit{e.g}., by varying the average edge weight $D$ in Fig.\ \ref{fig:demo}a. As $D$ is varied, $\overline{x}$ changes accordingly, and at specific points, it may also undergo a critical transition. A good sentinel node set $S$, then, satisfies two conditions:\ (i) It optimally approximates the observed $\overline{x}$ as $D$ varies, including its critical transitions; (ii) It achieves this with a small number of sentinels $n$. 

Therefore, we simulate the system under $L$ different conditions $D = D_1,\dots,D_L$, and seek the optimal node set $S$ that minimizes

\begin{equation}
\varepsilon = 
\frac{\sum_{\ell=1}^L (\overline{x}'_\ell -\overline{x}_\ell)^2}{L\sum_{\ell=1}^L \overline{x}_{\ell}},
\label{eq:def-objfun}
\end{equation}

\noindent
where $\overline{x}_\ell$ and $\overline{x}'_\ell$ are $\overline{x}$ and $\overline{x}'$ as obtained under $D_\ell$, respectively. Equation~\eqref{eq:def-objfun} captures the total difference between the approximation $\overline{x}'(D)$ and the exact $\overline{x}(D)$ across the $L$ different conditions that sample the examined range of $D$ values. It is minimal when $\overline{x}'$ successfully recovers the complete system average of Eq.~\eqref{eq:def-overlinex}, including its observed points of transition.

As an example, we consider the coupled double-well dynamics (Fig.~\ref{fig:demo}a) implemented on a dolphin social network, displayed in Fig.~\ref{fig:demo}b. We first examine the full system average $\overline{x}$ vs.\ $D$ (Fig.\ \ref{fig:demo}c-f, black solid lines). Here, the uncoupled system ($D = 0$) has two equilibrium points---one high and the other low. In our simulations we initiated the system's state around the low equilibrium, and let it relax from there. For small $D$, the system settles at that low state. However, as $D$ is increased, $\overline{x}$ undergoes a critical transition, favoring the high equilibrium point. This transition begins at around $D = 0.2$, but it is not completely abrupt. Instead, it features discrete steps, starting at $D \approx 0.2$, and only reaching a full transition at $D \approx 0.6$. This is because the individual node states $x_i^*$ (grey solid lines) do not all transition at the exact same point; some shift early on, other only at a later stage. Therefore, we observe a multistage transition \cite{lever2020, wunderling2020a, maclaren2023}. 

Next, we seek to approximate the $\overline{x}$ vs.\ $D$ curve by tracking a small number $n$ of sentinel nodes determined by our optimization algorithm. In Fig.\ \ref{fig:demo}c, we show the outcome for $n = 1$ (orange line). To be specific, we selected the best single sentinel out of the $N$ nodes, which minimizes $\varepsilon$ in Eq.~\eqref{eq:def-objfun}. We observe a gap between the sentinel approximation (purple) and the exact $\overline{x}$ (black), and hence we continue to search for larger sentinel node sets $S$. Indeed, as we increase $n$ in Fig.\ \ref{fig:demo}d-f, we find that the sentinel approximation follows the black solid line more closely. At $n = 4$, we find that $S$ successfully recovers the observed multistage transition, yielding $\varepsilon = 0.004$, a negligible discrepancy from the full system average. Notably, the smaller sets are not included as part of the larger ones, namely, the single node ($n = 1$) in Fig.\ \ref{fig:demo}c is not part of the sentinel node sets with $n = 2$ (Fig.\ \ref{fig:demo}d), $3$ (Fig.\ \ref{fig:demo}e), or $4$ (Fig.\ \ref{fig:demo}f). The sentinel property is a characteristic of the set $S$, not of its individual elements.

Figure~\ref{fig:demo} demonstrates our key result. It shows that we can approximate the average of the full state of the system by tracking just a microscopic subset of $n = 4$ nodes. This tiny sentinel node set, $S$, captures the equilibrium state of the system under the entire range of $D$ values, including the critical transition and its multistage transition pattern. This result represents a crucial advance, since it is often extremely difficult to probe the entire system, and hence being able to track $\overline{x}$ via such a small $S$ is highly desirable. Increasing $n$ results in a reduced $\varepsilon$, thus improving the approximation even further. However, for all practical purposes, we find that there is no need to go beyond $n = 4$.  

For illustration purposes, our demonstration has focused on a relatively small network ($N = 62$) coupled with a specific form of nonlinear dynamics (\textit{i.e.}, double-well). To complement this, in Supplementary Section S1, we detail our expanded testing ground, comprising a broad set of $20$ model and empirical networks, whose sizes range from $N=62$ to $N=81,171$ nodes. We test each of these networks with four nonlinear dynamics, \textit{i.e.}, coupled double-well, ecological, epidemic, and gene-regulatory. In Supplementary Section S2 (see Fig.~S1), we examine the complete array of bifurcation curves $\overline{x}$ vs.\ $D$, covering all networks and dynamics. Similar to the results obtained in Fig.\ \ref{fig:demo}, we find that our optimized sets consistently recover the full system average activity, its bifurcations, and multistage transition patterns. In Supplementary Section S3, we also include a detailed comparison of our method against the theoretically motivated GBB and DART, alongside an additional GBB variant \cite{tu2021}.

\begin{figure*}
  \centering
  \includegraphics[width = 0.75\textwidth]{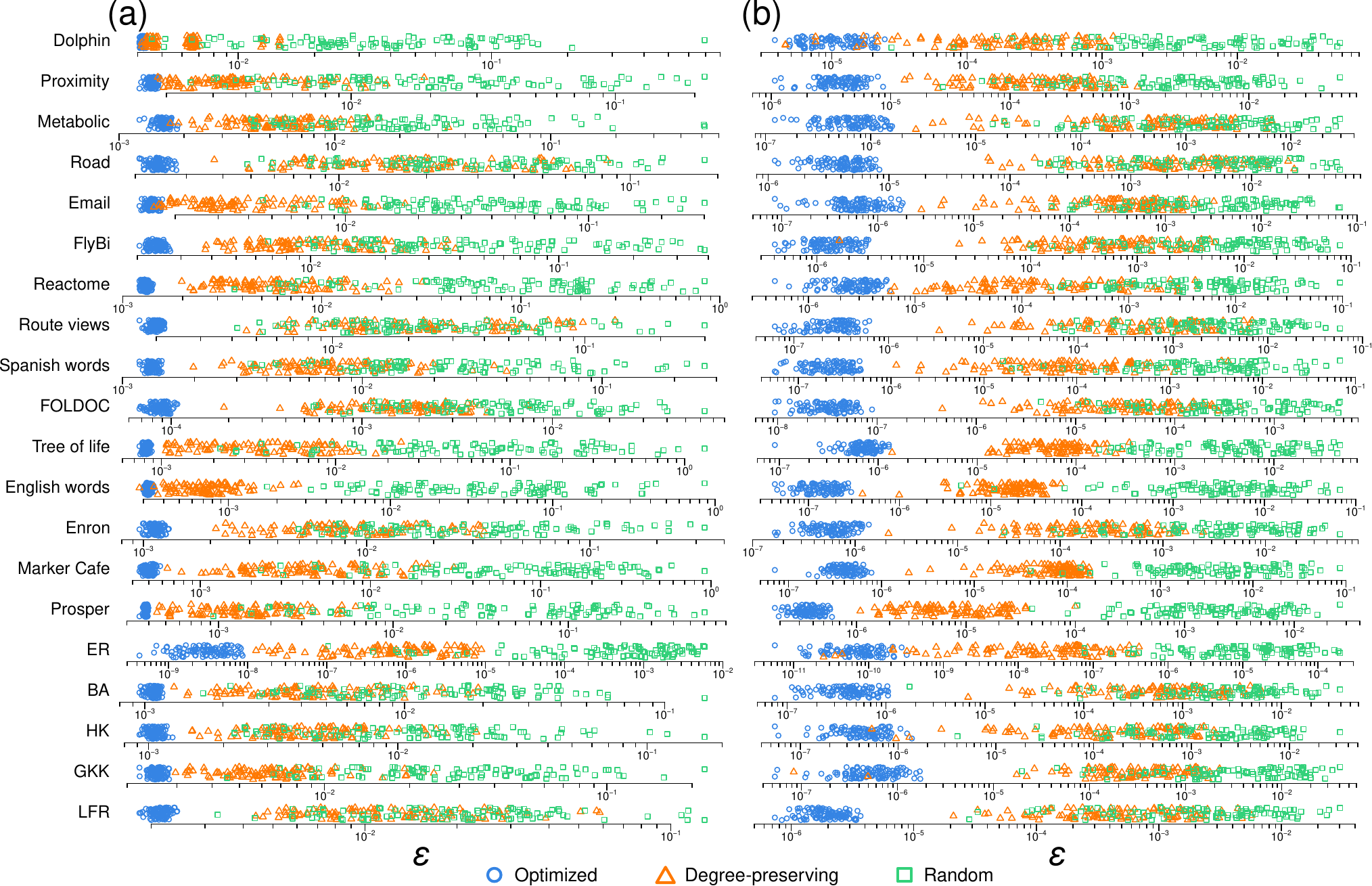}
  \caption{
    \textbf{Sentinel node approximation for different networks and dynamics}.
    (a) Approximation errors, $\varepsilon$, for 100 optimized (blue circles), degree-preserving (orange triangles), and completely random (green squares) node sets using the coupled double-well dynamics on 15 empirical and 5 model networks. The networks range in size from $N=62$ (Dolphin network) to $N=81,171$ (Prosper network); the model networks each have approximately $1,000$ nodes. ER: Erd\H{o}s-R\'{e}nyi. BA: Barab\'asi-Albert. HK: Holme-Kim. GKK: Goh-Kahng-Kim. LFR: Lancichinetti-Fortunato-Radicchi. (b) Approximation errors for 100 node sets of each type using the SIS dynamics on the same 20 networks.
  }
  \label{fig:comp-nets}
\end{figure*}

The process for detecting the optimal set $S$ is non-deterministic. Therefore, depending on the initial seed of $n$ nodes, the algorithm may settle on distinct final sets. To examine the consistency of the method's predictive power, we ran our optimization $100$ times, starting from a different selection of nodes each time. As expected, each iteration led to a different final set of nodes. The crucial point is that, despite these differences, all the detected sentinel node sets exhibited good performance. For example, on the dolphin network, our sentinel node sets all yielded an error narrowly distributed around $\varepsilon \approx 5 \times 10^{-3}$ (Fig.\ \ref{fig:comp-nets}a, top panel, blue circles)

We compare the $100$ sentinel node sets with two natural alternatives. First, we examine completely random node sets, obtained by selecting $n$ nodes with uniform probability from across the network. Such random $n$-plets show highly inconsistent results, with $\varepsilon$ ranging across two or more orders of magnitude (Fig.\ \ref{fig:comp-nets}a, green squares). These large and diverse errors were consistently observed across our $20$ networks. Hence, not surprisingly, our optimized $S$, perform substantially better than random. As a more challenging contender, we next consider degree-preserving random node sets (Fig.\ \ref{fig:comp-nets}a, orange triangles). To construct these sets, we begin with the best performing optimized node set among the $100$ runs, namely, the one with the smallest $\varepsilon$. We denote this winning node set by $\tilde{S}$. A degree-preserving set consists of randomly selected nodes that have the same degree sequence as $\tilde{S}$. The idea is to generate node sets that have similar characteristics to the best performing set detected across all optimization runs. For example, if the $n = 4$ nodes in $\tilde{S}$ have degrees $k = 1, 3, 3$, and $8$, then we generate degree-preserving sets by sampling one node with degree $k = 1$ uniformly at random from the network, two nodes with degree $k = 3$, and one final node with degree $k = 8$. Such degree-preserving node sets are superior to random sets, but are still consistently outperformed by our optimized $S$. Taken together, in Fig.\ \ref{fig:comp-nets}a, we find that the optimized sets (blue) are better than $99.3\%$ of the degree-preserving node sets, and $100\%$ of the completely random node sets.

In Fig.\ \ref{fig:comp-nets}b, we show a similar analysis for the susceptible-infectious-susceptible (SIS) epidemic dynamics, confirming similar results. This is further supported in Supplementary Section S4 (see Fig.~S5), where we repeat this examination on two other dynamics models. Quite consistently, across all network/dynamics combinations, the optimized $S$ supersedes completely random and degree-preserving random node sets $100\%$ and $96.8\%$ of the time, respectively.

In Supplementary Section S4 (see Table S1), we further verify our observations statistically through an analysis of variance (ANOVA) model of the approximation error $\varepsilon$. In this analysis, we set the dependent variable to be $\ln \varepsilon$, and considered three qualitative independent variables:\ dynamics (reference:\ coupled double-well dynamics), network (reference:\ BA network), and node set type (reference:\ completely random node sets). We exclude the Erd\H{o}s-R\'{e}nyi (ER) network because it yields small $\varepsilon$ in a majority of cases, regardless of the node set. This is presumably because of the relative similarity across all nodes characteristic in this family of networks (see Figs.~\ref{fig:comp-nets} and S5). Across all pairs of network and dynamics, we observe that $\varepsilon$ of the optimized node set is, on average, more than 900 times smaller than $\varepsilon$ of the completely random sets ($p < 10^{-7}$) and more than 40 times smaller than $\varepsilon$ of the degree-preserving sets ($p < 10^{-7}$). Finally, in Supplementary Sections S5 and S6, we validate our algorithm's performance also on weighted and directed networks. Overall, we find that our optimized sentinel node sets can uncover the global network behavior with just a handful of nodes, over a highly diverse corpus of network and dynamics, covering different scales and distinct scientific domains.

\subsection{Characteristics of good individual sentinel nodes\label{sub:characteristics}}

Our analyses in Fig.\ \ref{fig:comp-nets} and in Supplementary Section S4 (see Table S1) have shown that finding nodes with an appropriate degree sequence is an important but not sufficient condition for finding a good set $S$ of sentinel nodes. In this section, we explore further the characteristics of the nodes that our algorithm selects for inclusion in $S$. To achieve this, in Supplementary Section S7, we use a random forest model to examine an array of node characteristics that can potentially correlate with the sentinel property (see Table S10). The first, most natural, characteristic we focus on is the node degree $k_i$. We ask whether sentinel nodes tend to have a distinguishable degree centrality. We find, however, that a node's degree is a poor predictor of sentinelity. Nodes selected by our algorithm may be of low or intermediate degree, showing no specific tendency towards the average $\overline{k}$, as one may expect, given that they are selected to predict $\overline{x}$. Interestingly, we find that the largest hubs are significantly under-represented among the sentinel nodes (Supplementary Section S7; see Fig.~S14). Consistent with this observation, sentinel nodes do not particularly tend to be outliers in terms of the $x_i^*$ values, either (see Supplementary Section S8). The sentinel node sets tend to avoid the hubs, focusing primarily on small and intermediate degree nodes.

Beyond degrees, we analyze the nodes' nearest neighbor degree, the nodes' local clustering, closeness centrality, betweenness centrality, and coreness. Our findings indicate that these quantities have relatively weak predictive power and cannot be reliably used to \textit{a priori} identify potential sentinel nodes. Hence, there is little that can be said about the properties of nodes in $S$; they seem to have potentially diverse characteristics.

The reason for this, we believe, is that sentinelity is not an intrinsic property of individual nodes, but rather a characteristic of the set $S$. In other words, sentinelity emerges from the collective nature of all nodes in $S$, rather than from any specific component. Therefore, below, instead of focusing on individual sentinel nodes, we shift our attention to characterizing sentinel node sets.

\begin{figure*}
  \centering
  \includegraphics[width = 0.75\textwidth]{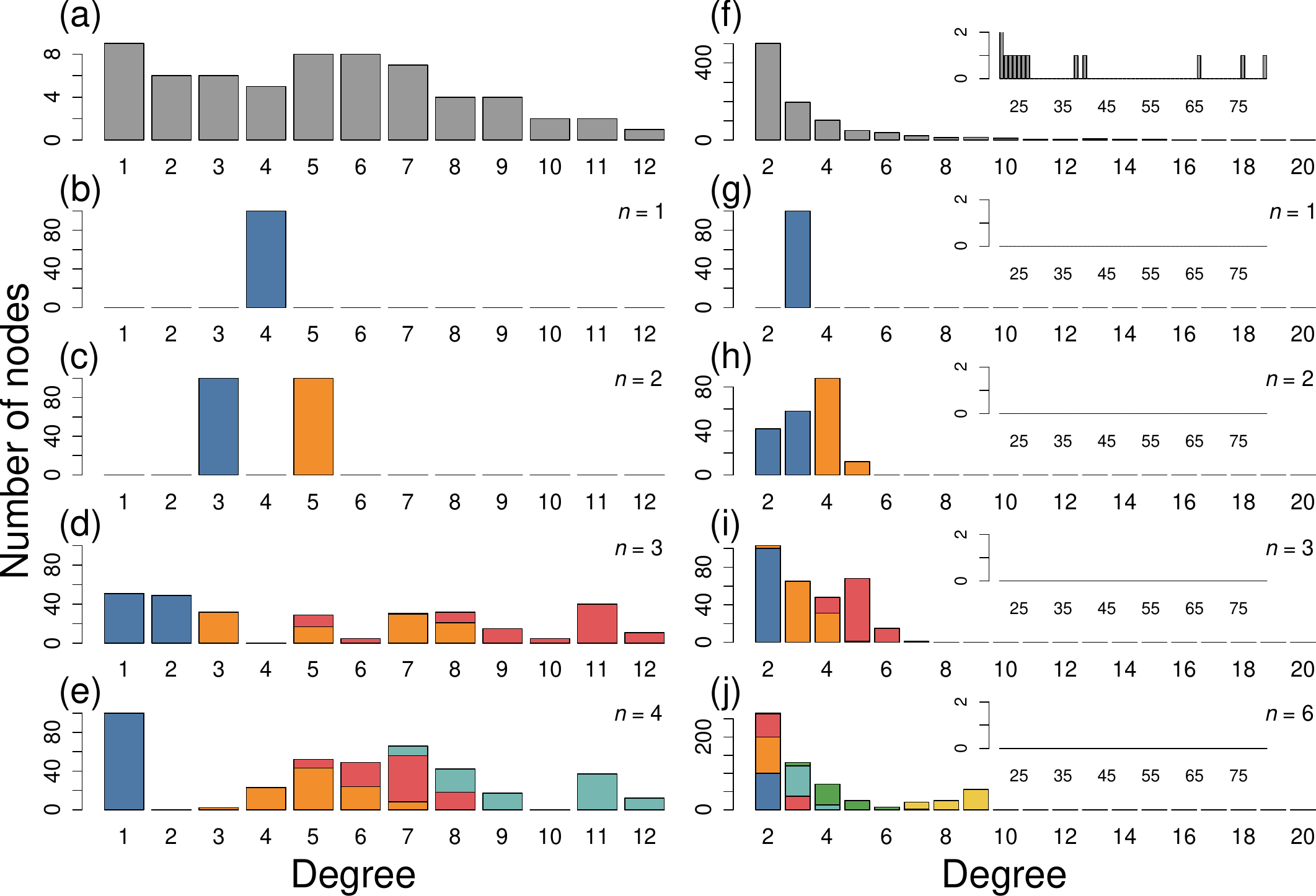}
  \caption{
    \textbf{Characteristics of sentinel nodes}.\  
    (a) Degree histogram of the dolphin network.  
    (b) Degree histogram extracted from $100$ single sentinel nodes ($n = 1$).
    (c)--(e) Similar histograms across $100$ sets with $n = 2$, $3$, and $4$ nodes. We find that sentinel node sets tend to spread across the degree sequence. Here, the bar color represents the node rank in $S$ in terms of degree:\ blue - smallest-degree node, orange - second smallest, red - third, light blue - fourth.
    (f) We repeated the analysis for a BA network of $N = 10^3$ nodes. For this network, we observe the hub nodes at the tail of the histogram, with degrees ranging from $k = 20$ to $85$ (inset).
    (g)--(j) Histograms extracted from the sentinel node sets with $n = 1,2,3$, and $6$. All histograms with $n > 2$ comprise a mixture of small and intermediate degree nodes, but lack any representation of the hub nodes (see empty insets). Bar colors are the same as in panels (b)--(e), with two additional ranks:\ green - fifth smallest node, yellow - sixth smallest node.
  }
  \label{fig:degseq}
\end{figure*}

\subsection{Characteristics of good sentinel node sets\label{sub:characteristics-set}}

We have shown that single-node features do not predict whether nodes are good sentinels or not. This is because a sentinel node set is a combinatorial construct, and not an independent collection of single nodes. Therefore, we now focus on the set features that may help characterize good sentinel node sets.

Our first observation is that nodes in $S$ tend to have a degree sequence that is balanced across the average degree in the entire network. We show this pattern, as observed in the dolphin network, in Fig.~\ref{fig:degseq}. In Fig.~\ref{fig:degseq}a, we show the degree histogram of all nodes in the network. We observe a bimodal degree distribution, with a dominant peak of nodes with degree $1$ and a secondary group of nodes scattered around the mean degree $\overline{k} = 5.13$. In Fig.~\ref{fig:degseq}b-e, we show the aggregate degree histograms realized by the nodes in $S$ obtained from $100$ independent runs of the algorithm with $n = 1$, $2$, $3$, and $4$---one panel corresponding to each $n$ value. The degrees of the chosen nodes are sorted such that the blue bars represent the smallest degree node in the optimized node set over the $100$ runs, orange the second smallest, red the third smallest, and light blue the largest degree nodes. We find that the algorithm balances the degrees of the selected nodes such that both small-degree and large-degree nodes tend to exist in the sets. This is quite clearly observed when $n \in \{3, 4 \}$, and to a lesser extent when $n = 2$.

The reason is that in order to capture the behavior of the average $\overline{x}$, we must balance nodes from different groups in the network. For example, in Fig.~\ref{fig:demo}f, the $n = 4$ nodes selected for $S$ have degrees 1, 6, 7, and 8. The $x_i^*$ values for these four nodes, shown by the four blue lines, illustrate precisely this balance:\ some appear above $\overline{x}$ and some below, depending on whether their degrees are large or small. As a result, the average over these four selected nodes offers an accurate approximation of $\overline{x}$. Qualitatively, we observe the same effect also for $n = 2$ and $3$, as shown in Fig.~\ref{fig:demo}d and e, respectively.

In Fig.~\ref{fig:degseq}f-j, we repeat the same analysis for a Barab\'{a}si-Albert (BA) network with $N = 10^3$ nodes. Here, the largest hub nodes, with $k_i = 38, 40, 66, 76$, and $81$, appear as discrete peaks in the inset of Fig.~\ref{fig:degseq}f. By examining the optimized sentinel node sets in Fig.~\ref{fig:degseq}g-j, we find that this inset is empty, implying that our optimization algorithm does not select these hub nodes under all values of $n \in \{1, 2, 3, 6\}$, in any of the $100$ runs. Instead, we observe that $S$ comprises a mixture of nodes with both small and moderately large degrees in all cases, however, avoiding the largest hubs. This result further strengthens our observation above, pertaining to individual nodes, that $S$ tends to exclude the hubs (Supplementary Section S7; see Fig.~S14).

To complete our analysis, in Supplementary Section S7 (see Fig.~S8), we repeat this process for each of our $20$ networks. We find that the effect is quite consistent:\ sentinel node sets tend to probe the network's degree sequence and select nodes with diverse degrees, both above and below the average. At the same time, however, they typically lack the largest hubs, remaining within the bulk of the degree distribution and avoiding its extreme values.

As a result, the degree histograms of the optimized node sets appear to markedly differ from that of the entire network (see Fig.~\ref{fig:degseq}). To quantify this difference, we compute the Kullback-Leibler divergence, $D_{\rm KL}$, between the optimized node sets and the original network's degree distribution (see Supplementary Section S9). In Fig.~\ref{fig:kld}a, we show $D_{\rm KL}$ for various values of $n$ between the original network degree distribution and that of the $100$ optimizes sets (blue circles). We also show the results obtained from completely random node selections (\textit{i.e.},\,$100n$ nodes; green squares).

For small $n$, the sentinel degree distribution features a discrepancy from that of the entire network, as observed through its much larger $D_{\rm KL}$ than the completely random samples; note the logarithmic scale on the vertical axis. Therefore, $S$ indeed does not simply sample nodes that are representative of the network's degree sequence. As $n$ is increased, both the optimized and the completely random node sets approach the network's degree distribution. However, throughout the entire range of examined $n$ values ($n \in \{ 1, \ldots, 12 \}$), the degree distribution of the optimized node sets remains significantly different from that of the original network while maintaining a smaller $\varepsilon$ (see Supplementary Section S10). Therefore, although our method tends to select a mixture of large and small nodes, it does not sample the degree values uniformly at random even for relatively large node sets. Similar results are also obtained for our BA network (Fig.\ \ref{fig:kld}b).

\begin{figure}
  \centering
  \includegraphics[width = 0.5\textwidth]{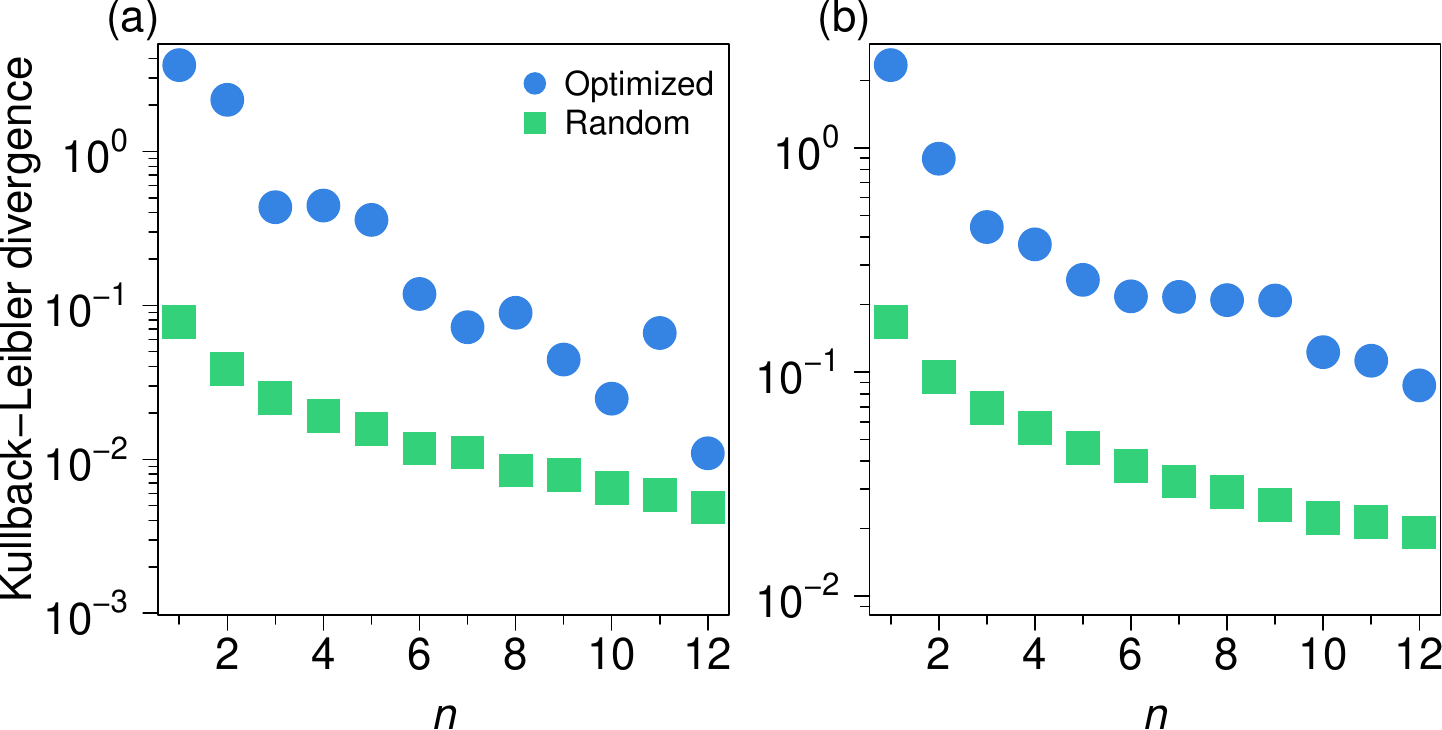}
  \caption{
\textbf{Degree distribution of optimized vs.\ random sets}.\    
We collected $100$ optimized sets of $n = 1,\dots,12$ nodes from the dolphin network and extracted their collective degree distributions. 
(a) The Kullback-Leibler divergence, $D_{\rm KL}$, between the optimized sets and the original network as $n$ varies (blue circles). The results obtained from $100$ completely random node sets are also shown (green squares); the 95\% confidence interval for each mean is smaller than the symbols and therefore not shown. For all $n$ values, and especially for the smaller sets, we find that the sentinel $D_{\rm KL}$ is significantly larger than that of the random sets.
(b) Similar results are obtained for the BA network. Here the gap between the two set types is even more pronounced, likely due to the presence of hubs in the network and their absence in $S$. 
  }
  \label{fig:kld}
\end{figure}

In Supplementary Section S11, we consider several additional node set features. The first six capture the previously considered features, \textit{i.e.}, nearest neighbor degree, clustering etc., only this time averaged over the $n$ nodes in the set. Beyond these set averages, we also consider two additional inherent node set features: (1) the maximum number of nodes in a node set which come from the same network community, and (2) the average pairwise distance between nodes in a node set (Supplementary Section S11). We found, using random forest models, that these node set features combined can, indeed, explain $\varepsilon$ to a large extent (see Table S17). In particular, the set's average degree should be neither too small nor too large for suppressing $\varepsilon$ (see Fig.~S18). We emphasize that this result does not imply that all the $n$ nodes should have intermediate degrees. As we have shown above, it is generally better for the $n$ nodes to cover a range of degree while avoiding the extremes.

With these observations, we can now construct heuristic algorithms that bypass our brute-force optimization method. Such algorithms determine sentinel node sets based on the observed sentinelity features above. For example, one can prioritize node sets that have the desired degree sequences, as motivated by Fig.\ \ref{fig:degseq}, or select nodes to ensure that they are sampled from distinct network communities. We examine these heuristic algorithms in Supplementary Section S11. We find that, while these algorithms significantly reduce $\varepsilon$, they are still substantially inferior to our optimized node sets (see Table S18). Therefore, to achieve sufficient accuracy in retrieving the global state $\overline{x}$, one must still run our full optimization algorithm.

\subsection{Transfer learning: Effectiveness of optimized node sets across different dynamics\label{sub:different-dynamics}}

Our optimization algorithm relies not only on network structure but also implicitly on the dynamics running on the network. The challenge, however, is that in practice, the actual dynamics governing the observed data are often unknown. This raises the question:\ to what extent does our optimization algorithm support transfer learning? In other words, can node sets optimized under one dynamics---referred to as the training dynamics---be effectively applied to a different dynamics, \textit{i.e}.\ the test dynamics?

We begin, once again, by examining the dolphin network, considering all potential pairings of training and test dynamics. Note that here we also use mismatched pairs of the dynamics and network (e.g.,\,the gene-regulatory dynamics on the dolphin network) to test the possibility of transfer learning. In Fig.~\ref{fig:dyn-comp}a, each panel displays the approximation error, $\varepsilon$, as obtained from all our optimized (blue circles), degree-preserving (orange triangles), and completely random (green squares) node sets. For each node set, we show the approximation error for the training dynamics on the horizontal axis and the test dynamics on the vertical axis. We also present the average result (large symbols), computed from the $100$ node sets of each type. 

As expected, node sets optimized on a specific training dynamics have consistently smaller $\varepsilon$ when evaluated on the same dynamics than on an alternative test dynamics. Indeed, they were optimized for that specific dynamics. For example, in Fig.~\ref{fig:dyn-comp}a7, we used the double-well dynamics to train the model and obtain the set $S$, and then tested $S$'s performance against the susceptible-infectious-susceptible (SIS) dynamics. 
When we use the SIS dynamics as the training dynamics, the blue circles condense around $\varepsilon = 10^{-5}$ (the horizontal axis of Fig. 5a2, a5, and a12) indicating an almost perfect prediction. However, on the vertical axis of Fig 5a7, capturing the extent of transfer learning from the double-well to SIS, we observe an average error of $\varepsilon = 5 \times 10^{-3}$. 
We also witness a broader spread across our $100$ trials, together indicating that the transfer from the training dynamics to the test dynamics is imperfect. 

The crucial point is, however, that across all training-test pairings, the optimized node sets, on average, outperform the random node sets (blue circles vs.\ green squares). This result indicates that a set optimized under one dynamics can indeed be useful for predicting the state of another dynamics. We also tested degree-preserving node sets, finding that our optimized sentinels continue to exceed their performance in ten out of the twelve tested pairings. Therefore, the optimized node sets are, to some extent, transferable between different dynamics. This result is striking in particular because different dynamics may have different bifurcation structures. For example, the coupled double-well dynamics have saddle-node bifurcations, whereas the SIS dynamics have transcritical bifurcations. Still, as Fig.~\ref{fig:dyn-comp}a7 indicates, node sets trained on one dynamics remain predictive on the other. To demonstrate this explicitly, in Fig.\ \ref{fig:dyn-comp}b, we show a specific sentinel nodes set, optimized on the dolphin network under the double-well dynamics (blue nodes). Indeed, the set is shown to successfully capture the transition pattern of its training dynamics (Fig.\ \ref{fig:dyn-comp}c). This positive result is expected. However, in Fig.\ \ref{fig:dyn-comp}d, we use the same sentinel node set to track the SIS dynamics, for which it was never trained. The figure shows that the learning against the double-well dynamics successfully transfers to the SIS dynamics.  

The root of this transferability is likely due to the fact that sentinelity is primarily a structural characteristic. Therefore, it only weakly depends on the specific dynamics, and mainly driven by the network topology. This insight is crucial when we have little knowledge of the network dynamics. In this case, we can train $S$ against a presumed dynamics and then use the obtained sentinel node set to observe the activity of the hidden dynamical system. Interestingly, the completely random node sets in each panel (green circles in Fig.\ \ref{fig:dyn-comp}a) show an approximately linear relationship. This result indicates that random node sets that happen by chance to be good sentinels for a specific training dynamics continue, on average, to be good sentinels also for the test dynamics. This observation further supports the notion that sentinelity is embedded in the topology and hence it is transferable across different dynamics.

In Fig.\ \ref{fig:dyn-comp}, we analyzed transferability on the relatively small dolphin network. This choice allowed us to track the specific node sets and visualize them, as we did in Fig.\ \ref{fig:dyn-comp}b. To complete this analysis, in Supplementary Section S12, we repeat the same experiment on a large financial network, comprising $N \approx 8 \times 10^4$ nodes, thus covering a range of scales (Fig.~S19). In Supplementary Section S12, we also show similar transferability results for a mobility-based human network (Fig.~S20) and an airport network (Fig.~S21). The positive results obtained suggest potential relevance of the proposed method to real epidemic outbreaks.

\clearpage
  \includegraphics[width = \textwidth]{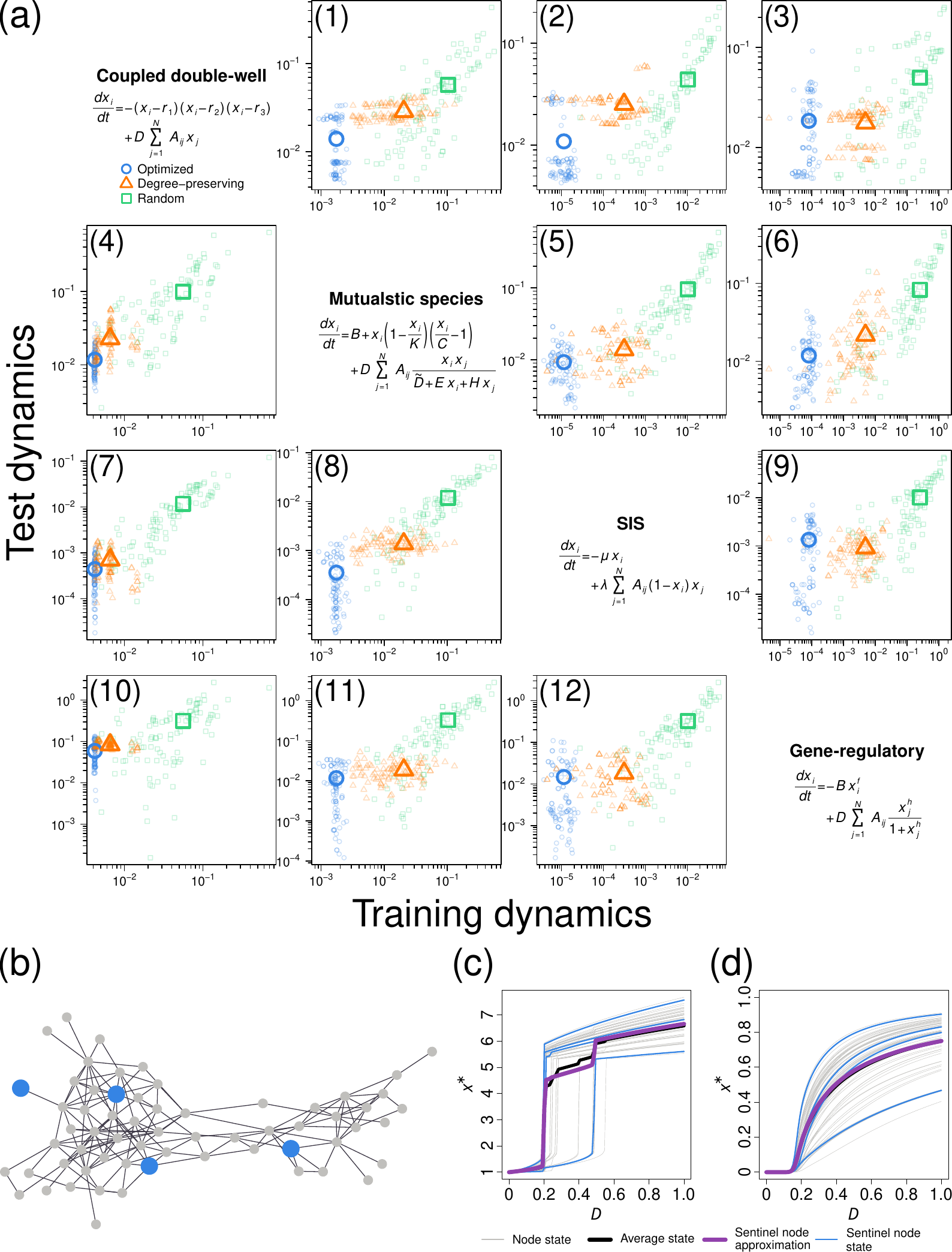} 
\clearpage

\begin{figure}
  \centering
  \caption{
    \textbf{Transfer learnability of the sentinel node approximation}.\
    We used the dolphin network and our four dynamic models to construct all potential pairs of training and test dynamics.
    (a) The approximation error $\varepsilon$ obtained from the test dynamics vs.\ that of the training dynamics for our optimized node sets (blue), degree-preserving node sets (orange), and completely random node sets (green). For example, in panel (a1), we extract the sentinels using the mutualistic species dynamics ($x$-axis, training) then examine their performance against the coupled double-well dynamics ($y$-axis, test). For each node selection method, we present results from $100$ independent sets by small symbols and the averaged $\varepsilon$ by large symbols. In all but two panels, we find that our optimized node sets perform best, both against the training dynamics and, most crucially, against the test dynamics.
    (b) Dolphin network. The large blue nodes indicate the members of the optimized node set with $n=4$ nodes that attains the lowest approximation error on the coupled double-well training dynamics.
    (c) Sentinel node approximation for the coupled double-well dynamics on the dolphin network. The states of the nodes marked in (b) are shown in blue. The states of the remaining nodes are shown in gray. The average state $\overline{x}$ (black solid line) and the sentinel node approximation (purple solid line) are also shown. As expected, the sentinel node approximation follows $\overline{x}$ quite tightly. 
    (d) Sentinel node approximation for the SIS dynamics on the same network, using the node set optimized for the coupled double-well dynamics. Despite being trained on a different dynamics, the four blue nodes still offer a reliable approximation for the SIS dynamics.
  }
  \label{fig:dyn-comp}
\end{figure}

As before, we assess the generalizability of these observations with an ANOVA. The dependent variable is $\ln \varepsilon$ evaluated on the given test dynamics. The independent variables are the training dynamics, test dynamics, network, and node set type (see Supplementary Section S12 for details). Despite the substantial increase in $\varepsilon$ due to the lack of knowledge about the test dynamics, optimized node sets still achieve $\varepsilon$ that is 7.43 times smaller than completely random node sets ($p < 10^{-7}$) and 1.52 times smaller than degree-preserving node sets ($p < 10^{-7}$) (see Table S19). We obtained similar results on the feasibility of transfer learning in the case of weighted networks (Table S3) and directed networks (Table S5). In summary, these results suggest that our optimization algorithm can select good sentinel node sets when we do not know the test dynamics, or even the type of bifurcation. 

We emphasize that these results in no way imply that the the observed network behavior is independent of the dynamics. Clearly, the network activity $\overline{x}$, which we wish to predict, and its bifurcation patterns, all arise from the interplay between structure and dynamics. Therefore, we expect that under different dynamics the system's behavior may be fundamentally distinct. This phenomenon is clearly observed in Fig.\ \ref{fig:dyn-comp}c,d, where the same network hosts qualitatively different behavior when coupled with different dynamics (double-well vs.\ SIS). The point is that the sentinels that can be used to observe these distinct $\overline{x}$ curves do transfer across dynamics, because they are mainly determined by topology and at large independent of the system's dynamics.

In the case of undirected and unweighted networks, we found that degree-preserving random node sets sometimes perform better than optimized node sets in terms of transfer learning. This phenomenon was observed specifically on large networks, or when our gene-regulatory dynamics was used either for the training or for the test dynamics (see Fig.~S22). This observation can offer guidelines for further strengthening the transfer learning. For example, given a large network, one may benefit by running our optimization algorithm on one dynamics to determine sentinel node sets. Then, by extracting each set's degree sequence, one can draw a degree-preserving node set, thus combining the strength of both our optimization method and the advantages of degree preservation.

In real systems, we often encounter uncertainties such as hidden nodes or missing edges. Therefore, we systematically examine the robustness of our transfer learning against such missing network components. We find that even if one constructs sentinel node sets with $\approx 10\%$ of the nodes/links unobserved, the obtained optimized sets are still more accurate then random ones at approximating $\overline{x}$ for the full network, \textit{i.e}., where all components are present (Supplementary Section S13). Optimized node sets are also transferable under uncertainties in the coupling strength $D$ when only some part of the network undergoes changes in $D$ (Supplementary Section S14).

\subsection{Sentinel nodes in empirical time series data}

In empirical complex systems, the dynamics generating the data at each node is usually unknown. The underlying network upon which the dynamics takes place is also subject to uncertainties and is often only partially mapped. Furthermore, the dynamics may not be in equilibrium, and hence the measured $\overline{x}$ may not properly represent the system's equilibrium. To assess the applicability of our sentinel node approximation under these challenging conditions, we collected multivariate time series data from the brain activity of healthy human individuals using functional magnetic resonance imaging (fMRI). The data, extracted from the Human Connectome Project~\cite{Van_Neuroimage2013wu}, cover $1,003$ participants, each tracked at a spatial resolution of $N = 300$ nodes. An explicit and established map of the network structure is lacking in these data. Therefore, for each participant, we used the first $4,700$ time points to estimate the network structure, and the remaining $100$ time points for recording the node activities $x_i$ (Supplementary Section S15).

The second challenge is that, as noted above, we do not have the specific dynamics for this system. Fortunately, this is precisely the scenario where transfer learning is designed to help, allowing us to obtain $S$ under one dynamics and approximate $\overline{x}$ under another. Therefore, we implemented the coupled double-well dynamics on the reconstructed brain network, and simulated the state of the system over the same range of $D$ values as previously used (Fig.\ \ref{fig:demo}). We then extracted the optimized sentinel node sets and tested their prediction against the observed brain time series data. To assess the performance of our optimized node sets, we measured the error $\varepsilon$ and compared it to that obtained from completely random node sets (Fig.\ \ref{fig:brain}; see Supplementary Section S15, including Tables S20 and S21, for statistical results). 
We find that, under the optimized sentinel node sets, $\varepsilon$ is on average $6.09\%$ smaller than under the completely random node sets. With degree-preserving node sets, $\varepsilon$ is on average $3.41\%$ smaller than that of random sets. These results were quantitatively similar when we determined optimized node sets using any of the other three dynamics or a network variant of the Wilson-Cowan dynamics for neural masses \cite{laurence2019} (see Supplementary Section S15). At first glance, these improvements may seem modest, but to truly appreciate their significance, we must consider the conditions under which they were achieved. First, fMRI data are relatively noisy. Second, our attempt is to predict $\overline{x}$ without any \textit{a priori} data on the network or the relevant dynamics. And yet, our method was still able to detect nodes with a discernible predictive advantage. In fact, we have shown that the results do not notably depend on the dynamics. Furthermore, there is no reason to assume that the brain dynamics adhere to any of our four considered model dynamics or the Wilson-Cowan dynamics. Therefore, even this seemingly small enhancement in accuracy is sufficient to demonstrate the potential practical utility of our transfer learning in empirical environments.

\begin{figure}
  \centering
  \includegraphics[width = 0.5\textwidth]{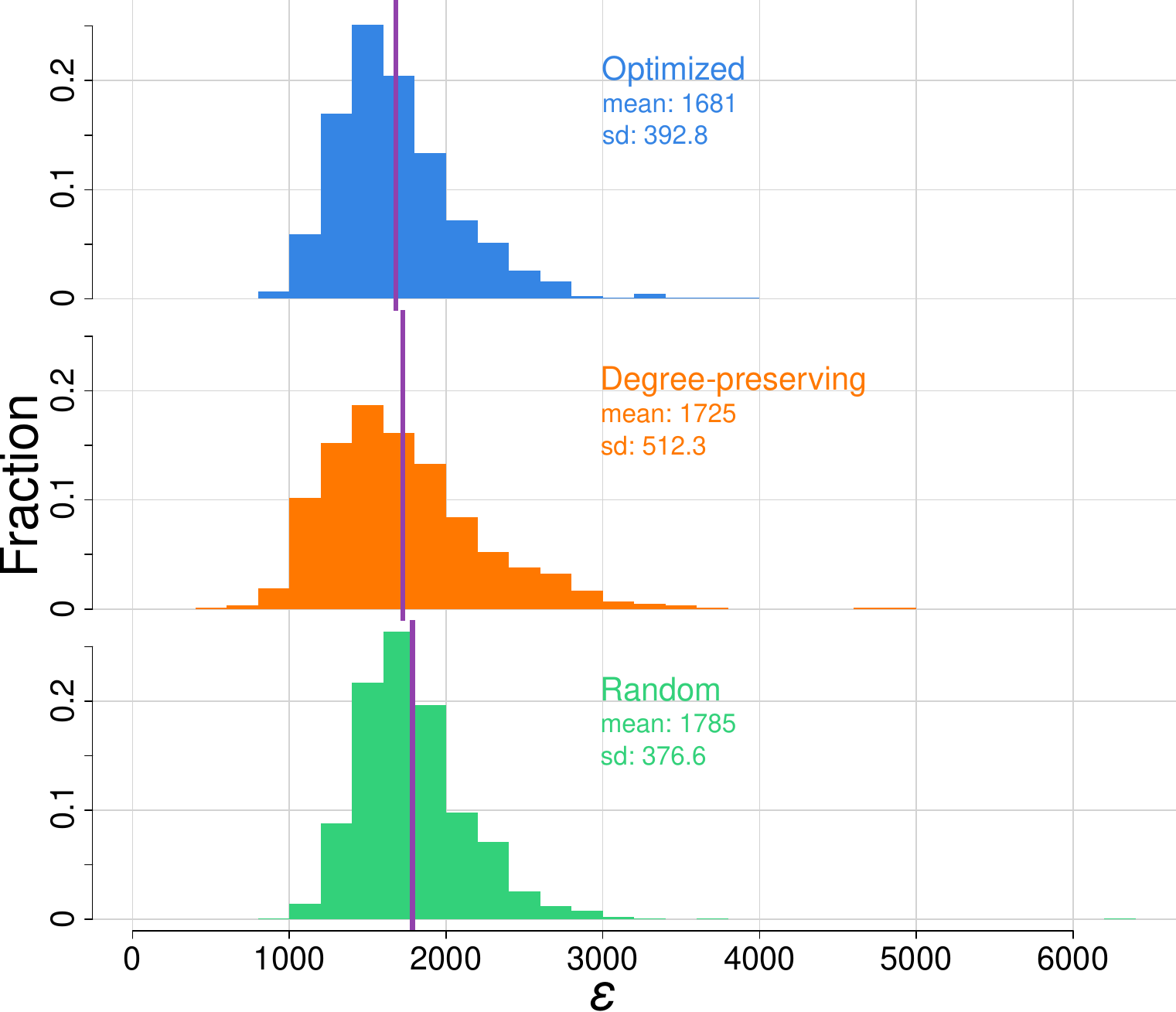}
  \caption{
    \textbf{Sentinel node sets in empirical data}.\
    We tested our optimization algorithm on multivariate time series fMRI data recorded from $1,003$ healthy human individuals. We estimated networks from a training set of each individual's time series data, simulated the coupled double-well dynamics, and then selected 100 sentinel node sets based on the simulation results. The distribution of the average approximation error $\varepsilon$ of each individual's 100 sentinel node sets, across all the individuals, is shown in blue; the corresponding distributions for degree-preserving and completely random node sets are shown in orange and green, respectively. The mean of each distribution is indicated by the vertical purple lines.
  }
  \label{fig:brain}
\end{figure}

\subsection{Weighted averaging\label{sub:weighted}}

Up to this point, we restricted ourselves to unweighted averaging over the sentinel node set, in which all node activities within $S$ add uniformly towards the approximation of $\overline{x}$. More generally, we may relax this restriction and assign non-uniform weights to all sentinels, approximating the system's state via $\overline{x}' = \sum_{i = 1; i\in S}^N a_i x_{i, \ell}^*$. Here, the weights $a_i$ are subject to $\sum_{i = 1; i\in S}^N a_i = 1$ and $a_i \geq 0$ $\forall i \in S$, \textit{i.e}.\ positive and normalized. This setting provides an additional degree of freedom to our optimization, which can potentially improve the approximation accuracy. To optimize the weights, we add a quadratic programming step to our original algorithm (see the Methods Section).

Such weighted averages are inspired by existing methods, such as GBB or DART, that have indicated that, to predict the system state, it is optimal to assign different weights to all nodes. For example, nodes are weighted by their degree ($a_i = k_i/\sum_{\ell=1}^N k_{\ell}$) in GBB \cite{gao2016, zhang2022, ma2023} and by the $i$th entry of the principal eigenvector in DART \cite{laurence2019, thibeault2020, masuda2022, thibeault2023, vegue2023}. The main difference is that in these applications one considers all nodes ($n = N$) and assignes the weights based on theoritically motivated heuristics. In contrast, in our application, we continue to demand $n \ll N$ and tune the weights using machine learning tools.

We find that the additional optimization of node weights leads to substantially smaller approximation error, $\varepsilon$, than the unweighted averages examined in the previous sections. For comparison, across all dynamics and networks, $\varepsilon$ is, on average, $46$ times smaller when we optimize both the node set and the individual node weights than when we only optimize for the node set and constrain all $a_i = 1$ (Supplementary Section S16; see Table S22).

Intriguingly, this advantage does not extend to the transferability across different dynamics. Specifically, although weight optimization improves performance on the training dynamics, it exhibits no meaningful reduction in $\varepsilon$ against the test dynamics ($p = 0.591$; see Table S23). Our supposition is that while the composition of $S$ is determined mainly by the network structure, the specific weight to assign to each node in $S$ does depend on the dynamics, and therefore weight optimization against a specific training dynamics offers no advantage when applied to the test dynamics. Consequently, although optimizing node weights drastically improves the approximation of $\overline{x}$, it is only advantageous if we know the system's dynamics, but cannot help in case of unknown dynamics.

\section{Discussion}
\label{sec:discussion}

As our understanding of complex system dynamics advances \cite{barrat2008, porter2010, barzel2013, wang2015, harush2017, hens2019, dsouza2023}, a crucial bottleneck that we continue to face is our limited empirical access to the states of their multitude of relevant parameters. This bottleneck hinders the development of new theoretical analyses and empirical validation of existing theories on dynamics. A similar challenge in biology and medicine has been that it is often difficult to track every single metabolite, protein, or cell. In that context, the identification of biomarkers \cite{oconnor2017, califf2018} has offered dramatic breakthroughs, allowing us to
gain insight into the state of the entire system  by just tracking a small set of indicative parameters. Our work here envisions an equivalent opportunity in the context of complex networks. It shows that, by tracking a mere $\lfloor \ln N \rfloor$ of sentinel nodes, we can approximate the average state of the network, making monitoring dynamics scalable even for large networks.

We have found that the detected sentinels are largely agnostic to the model of the dynamics describing the system's internal mechanisms. This result indicates that one can transfer the learned set of nodes from one dynamics to infer the state of the system under another dynamics. Such transfer learning is especially useful for systems with unknown dynamics, a rather common challenge in the context of complex networked systems \cite{barzel2015, wai2019, lai2021, gao2022, koch2023}. To take advantage of transfer learning, one first needs to select a relevant model of the dynamics, numerically simulate the system under this model, and use our optimization method to extract the sentinel node set. This set, found under the selected model, will then help approximate the system's state even if it is driven by some other, potentially unknown, dynamics.

Our method is heavily based on numerical analysis, and therefore its complexity depends on the network size $N$. In Supplementary Section S17, we analyze this dependence and show that both the dynamic simulation and the node set optimization scale as $O(N^2)$. Therefore, even for large networks, the computational bottleneck is non-restrictive, and the entire analysis pipeline is scalable. In fact, our full analysis, provided in the Supplementary Information, covers networks with as many as $N \approx 10^5$ nodes, a clear testament to the method's scalability. To further capitalize on this potential advantage, future work should aim to extract a master model, \textit{i.e}., a nonlinear system of equations whose sentinel nodes optimally transfer to as many other types of dynamics as possible. We also seek to broaden the family of dynamics that our optimization can cover, for example, by considering dynamical systems with multiple equilibria or oscillations. Should the optimized sentinel node sets depend on the specific equilibrium point? Can oscillatory dynamics also be monitored in such a parsimonious fashion? The implicit assumption of monostable dynamics captures an important limitation of the present study.

While our methodology shows good performance when tested against numerically simulated dynamics, our long-term objective is to apply it to empirical multivariate time-series data. Our preliminary results using fMRI data offer an encouraging step in this direction. However, at this stage, the optimized sentinel node sets provide only a modest improvement over the random or degree-preserving alternatives. This limitation primarily stems from the highly constrained nature of the available data. Future applications employing richer datasets, specifically data with reliable information pertaining to the underlying causal network, are expected to enhance the method's effectiveness.

Our work differs from other approaches for approximating the population activity of or efficiently observing a network by attempting to accurately estimate $\overline{x}$. First, there are methods to select a small number of sentinel nodes for constructing early warning signals for anticipating regime shifts \cite{chen2012, vafaee2016, aparicio2021, maclaren2023, masuda2024}, but there is no \textit{a priori} reason to consider that these sentinel nodes are also good at representing $\overline{x}$. Second, the theory of network observability allows us to determine the node set, observation of which enables one to reconstruct the system's complete activity \cite{liu2013}. The method has also been applied for selecting sentinel nodes with which to construct early warning signals \cite{aparicio2021}. However, this network observability concerns inference of the initial condition, i.e., the activity of each node at time $0$. In fact, $\overline{x}$ in the equilibrium or the time course of it, rather than its initial value, would be of practical interest in applications. Functional observers aim to approximate the equilibrium of any given weighted linear average of $x_i$, including $\overline{x}$ as a special case, using a linear dynamics of small order \cite{Fernando2010IeeeTransAutoCont, Montanari2022Pnas}. Functional observability methods are mathematically founded and more versatile than our method in that they can approximate a wider variety of quantities. However, they are limited to linear and fixed dynamics. In contrast, our method accommodates nonlinear dynamics, tracks state transitions as a function of the system's control parameters, and---most importantly---yields insight even when the underlying dynamical equations are unknown. Together, these features offer a substantial practical advantage.

Our approach also differs from earlier efforts to identify sentinel sites or individuals during the initial stages of an epidemic, particularly under realistic constraints such as limited observations of infected nodes and incomplete network information. While our algorithm assumes access to the network structure, it can tolerate partial knowledge (see Supplementary Section S13). Extending the method to transient dynamics---such as those observed in the early phases of an outbreak---is left for future work. One possible direction is to assume that the degree of each node is known, because this is often easier to obtain than the full network structure. A synthetic network can then be generated using the configuration model \cite{newman2018}, allowing the algorithm to be applied in this approximate setting. This procedure is conceptually related to the degree-preserving random node sets explored in this paper and may offer a viable approximation of epidemic dynamics at early stages.

Existing reduction methods often place much weight on the hub nodes. Specifically, GBB \cite{gao2016} uses a degree weighted average, and DART \cite{laurence2019, thibeault2020} weighs all nodes based on the principal eigenvector, both favoring observing $x_i^*$ of the hubs \cite{goltsev2012, pastorsatorras2016}. Therefore, in degree-heterogeneous networks, the observed global state of the system is dominated by a small set of highly central nodes. This phenomenon resembles our formalism, which also achieves observability via a limited set of sentinels. However, these methods and our sentinel node approximation select very different nodes. While the nodes dominating GBB and DART tend to be the hubs, our sentinels tend to avoid the hubs. This discrepancy is rooted in the fact that the two approaches aim to observe distinct characteristics of the system. GBB and DART seek a theoretical closure for the dynamical equations, which is indeed driven by the behavior of the hubs. Our method, in contrast, seeks to observe the average $\overline{x}$, which is best represented by majority of small- and intermediate-degree nodes, rather than by the network outliers.

This contrast touches on the key trade-off between the GBB/DART approach and our machine learning-based observability. GBB and DART are derived from theoretical principles, and hence offer full interpretability, namely, we understand their selected observable $\overline{x}^\prime$. Our approach is based on optimization, thus sacrificing the interpretability for efficiency. Indeed, without running our full algorithm, one cannot construct the sentinel node set $S$, just from theoretical principles. This discussion captures the general balance between theory and machine learning-based approaches, trading accuracy and efficiency for opaqueness. We acknowledge that an important limitation of the present study is the lack of theoretical understanding of why our sampling strategy balancing different node degrees performs well. Despite these limitations, however, we did extract several interpretable insights relating to the expected features of sentinel nodes, such as the avoidance of hubs and the set-typical degree distribution. As commonly observed with machine learning frameworks, these insights provide a retrospective understanding of $S$, but cannot truly contribute to \textit{a priori} predicting it. As the use of our proposed framework increases, we hope to obtain a richer mechanistic understanding of the sentinel nodes, which can help further reduce the search space for $S$. Analyzing weakly connected networks by theoretical means together with subtyping networks may help in achieving such goals.

Our framework accommodates directed and weighted networks, noisy dynamics, and, the common scenario in which the system's driving equations are unknown. Furthermore, without essential changes, our method allows any weighted sum of $\{ x_1^*, \ldots, x_N^* \}$ as the target quantity to be approximated. Together, these features make our method versatile enough for potential applications. For example, in ecosystem monitoring, our algorithm could help focus monitoring efforts on a small number of key species \cite{hale2018}. As we have shown, the selected sentinel species may not be the most abundant or well-connected species \cite{mills1993, cotteejones2012}. Assuming that an estimated network of species interactions is available, which is often the case \cite{delmas2019}, one can derive a small number of sentinel species by simulating a ground truth using a mutualistic species dynamics model on the measured network. Then, the populations of the chosen sentinel species are expected to provide a good approximation of population dynamics of the entire ecosystem.

From a methodological perspective, our drastic reduction from $N$ to just a small number of nodes is rooted in a combined toolbox that mixes theoretical modeling with machine learning techniques. The theoretical models, captured within the low-dimensional dynamical systems derived by GBB, DART, or other methods, represent our analytical insights into the system’s internal mechanisms. Such insights, however, are limited by the prohibitive complexity of the system. This is precisely where the machine learning component helps complement the theoretical approach, lending its computational power to help simplify the scale of the problem. This synergy between theoretical insights and the predictive power of optimization and learning methods represents, in our view, one of the main avenues towards predicting, observing, and influencing complex network behavior.

\section{Methods\label{sec:methods}}

\subsection{Node Selection\label{sub:nodeselection}}

Consider an undirected and unweighted network with $N$ nodes, $M$ edges, and an adjacency matrix $A = (A_{ij})$ with $A_{ij} \in \{0, 1\}$, $A_{ii} = 0$, and $A_{ij} = A_{ji}$. Each node is characterized by a continuous variable $x_i$ representing, for example, the abundance of a species or the fraction of a population that is infected with a pathogen.
A straightforward estimate of the network's aggregate state is the unweighted average of all node states at equilibrium, $\overline{x}$, given by Eq.~\eqref{eq:def-overlinex}.
Our goal is to approximate $\overline{x}$ by
a linear combination of a small number of equilibrium node states, $x_{i}^*$, given by Eq.~\eqref{eq:def-overlineprimex}, with a small error and across a range of a control parameter that represents, e.g., an environmental state.
Examples of the control parameter are the strength of mutualistic interactions between species and the infection rate of a contagion process.
Choice of the control parameter, its range, and its effect on $x_i^*$ varies by model and is further described in section \ref{sub:models}. 

We seek a set $S$ of sentinel nodes with $|S| = n$, $n \ll N$.
We assume that each node in $S$ has equal weight in determining $\overline{x}'$, as in Eq.~\eqref{eq:def-overlineprimex}. We relax this assumption in section~\ref{sub:weighted}.
We quantify the discrepancy between $\overline{x}$ and $\overline{x}'$ by a normalized mean squared error over a given range of the control parameter, i.e.,
$\varepsilon = \frac{\sum_{\ell=1}^L (\overline{x}'_\ell -\overline{x}_\ell)^2}{L\sum_{\ell=1}^L \overline{x}_{\ell}}$, called the approximation error.
Subscript $\ell \in \{1, \ldots L\}$ specifies a value of an evenly spaced control parameter, and $x_{i,\ell}^*$ represents the equilibrium value of $x_i$ for the $\ell$th value of the control parameter; $\ell$ is omitted in the following text when there should be no confusion.
We choose a control parameter which, when varied over the first and last (i.e., $L$th) values, causes bifurcations in $x_i^*$. We set $L = 100$ in our numerical experiments.

Given all $x_{i, \ell}^*$ values (with $i\in \{1, \ldots, N\}$ and $\ell \in \{1, \ldots, L\}$) computed, we determine $S$ by combinatorial simulated annealing. Specifically, we first initialize $S$ by selecting $n$ nodes uniformly at random from the network and calculate the approximation error for $S$, which we denote by $\varepsilon(S)$. We arbitrarily set $n = \lfloor \ln N \rfloor$, where $\lfloor \cdot \rfloor$ is the floor operation, unless we state otherwise. Then, in each $h$th iteration of the algorithm, we select an $i$th node uniformly at random from $S$ and replace it with a $j$th node that does not belong to $S$ and is chosen uniformly at random. The tentative new node set is given by $S' = S \setminus \{i\} \cup \{j\}$. We compute $\varepsilon$ for $S'$, which we denote by $\varepsilon(S')$. 
We accept $S'$ as the new $S$ according to the Metropolis criterion, i.e., with probability $1$ if $\varepsilon(S') < \varepsilon(S)$ and with probability $q = \exp\{-[\varepsilon(S') - \varepsilon(S)]/t\}$ otherwise. The normalization factor $t = t_0/\ln(h + e - 1)$ decreases as $h \in \{1, \ldots, h_{\max}\}$ increases. Therefore, an $S'$ with a larger approximation error than $S$ is less likely to be accepted as $h$ increases, encouraging convergence to a local minimum of $\varepsilon$. We set $t_0=10$.
%
%
We also set $h_{\max} = 50N$ and confirmed that this number of iterations was sufficient for reaching a local minimum of $\varepsilon$.

\subsection{Dynamical system models\label{sub:models}}

We test our method on the following four models of dynamical systems on networks.

Many systems have alternative stable states at a single value of a control parameter, the transition between which displays hysteresis. For example, a certain species may be present or extinct in an ecosystem under a given environmental condition, depending on the system's state in the past \cite{scheffer2009}. Recovery from the extinct state may be difficult without substantial improvement of the environment. Similarly, tropical ecosystems may show sudden transitions from rainforest to grassland in response to just small changes in local water availability \cite{wunderling2022}.
A coupled double-well dynamics model on networks \cite{kouvaris2012, brummitt2015, kronke2020, kundu2022a, kundu2022b} has been employed for modeling interacting climate regions \cite{wunderling2022} and biological species \cite{lever2020} and is given by
\begin{equation}
  \label{eq:dw}
  \frac{dx_i}{dt} = -(x_i - r_1)(x_i - r_2)(x_i - r_3) + D\sum_{j=1}^N A_{ij}x_j,
\end{equation}
where $r_1$, $r_2$, and $r_3$ determine the location of the equilibria and satisfy $r_1 < r_2 < r_3$; $D$ is the coupling strength parameter. We set $r_1 = 1$, $r_2 = 3$, and $r_3 = 5$. In the absence of coupling, this model has stable equilibria at $x = r_1$ and $x = r_3$, and an unstable equilibrium at $x = r_2$. In the presence of coupling, all $x_i$s tend to be near $r_1$ when $D$ is sufficiently small, and they tend to be near $x_3$  when $D$ is large. 
We use $D$ as the control parameter in the range $[0, 1]$.

A similar model proposed for mutualistic species interactions is given by
\begin{widetext}
  \begin{equation}
    \label{eq:mutualistic}
    \frac{dx_i}{dt} = B + x_i \left( 1 - \frac{x_i}{K}\right) \left(\frac{x_i}{C} - 1 \right) + D\sum_{j=1}^N A_{ij}\frac{x_ix_j}{\tilde{D} + Ex_i + Hx_j},
  \end{equation}
\end{widetext}
where $x_i$ represents the abundance of the $i$th species and $B$, $K$, $C$, $D$, $\tilde{D}$, $E$, and $H$ are constants \cite{gao2016}. The constant $B$ represents migration rate and $K$ the carrying capacity. The Allee constant, $C$, represents the ease with which a species can become established in the environment. In the absence of migration and species interactions, $x_i > 0$ at equilibrium if and only if the initial $x_i > C$; otherwise $x_i = 0$ is the only stable equilibrium. We set $B=1$, $K=5$, $\tilde{D}=5$, $E=0.9$, and $H=0.1$ following \cite{gao2016}. We use $D$ as the control parameter in the range $[0, 3]$; to observe the transition of all nodes, we use a more extended range of $D$ for this model than for the other models described in this section.

The deterministic susceptible-infectious-susceptible (SIS) model on networks, also called the individual-based approximation of the stochastic SIS model, is given by
\begin{equation}
  \label{eq:SIS}
  \frac{dx_i}{dt} = -\mu x_i + \lambda \sum_{j=1}^N A_{ij}(1 - x_i)x_j,
\end{equation}
where $x_i$ represents the probability that the $i$th node is infectious, $\lambda$ is the infection rate, and $\mu$ is the recovery rate \cite{pastorsatorras2015}. The second term on the right-hand side expresses the rate at which the $j$th node infects the $i$th node. We set $\mu = 1$ without loss of generality; changing $(\lambda, \mu)$ to $(c \lambda, c \mu)$ with a constant $c$ is equivalent to scaling the time by a factor of $c$, so it does not affect the equilibrium.  We use $\lambda$ as the control parameter in the range $[0, 1]$. One obtains $x_i = 0$ $\forall i$ when $\lambda$ is below a value called the epidemic threshold. Above the epidemic threshold, all $x_i^*$ ($<1$) values are positive given a positive initial value. The parameter $\lambda$ has the same role as $D$ in the other dynamics models.

A model of gene regulatory dynamics is given by
\begin{equation}
  \label{eq:genereg}
  \frac{dx_i}{dt} = -Bx_i^f + D\sum_{j=1}^N A_{ij}\frac{x_j^h}{1 + x_j^h},
\end{equation}
where $x_i$ represents the expression level of the $i$th gene \cite{gao2016}. We set $B = 1$, $f = 1$, and $h = 2$ following \cite{gao2016}. We use $D$ as the control parameter in the range $[0, 1]$. Given sufficiently large initial values, all $x_i^*$ will remain large when $D$ is above a threshold value, a situation which represents a living cell. When $D$ is small, all $x_i$s approach zero, representing cell death. 

To compute equilibrium values of $x_i$ at each control parameter value, we initially set each $x_i$ to a model-dependent standard value (coupled double-well: $x_i = 1$, SIS: $x_i = 0.01$, mutualistic species: $x_i = 0.001$, gene-regulatory: $x_i = 2$). Then, we solved the ODEs using the implicit Adams method provided by the deSolve package for R \cite{soetaert2010}. We used a total simulation time of $T = 15$, which we found was sufficient to allow each of the above dynamics on each network described below to relax to a point where no further change was noticeable. We used the value of each $x_i$ at $T = 15$ as $x_{i,\ell}^*$.

\subsection{Networks\label{sub:networks}}

We chose fifteen empirical and five model networks to test our method. All networks were coerced to be undirected, simple (i.e., no self- or multi-edges), and unweighted. We used the largest connected component.

In the social network of wild dolphins \cite{lusseau2003}, which we refer to as the dolphin network, each node is a dolphin individual. Two nodes are defined to be adjacent if two individuals were observed together more often than expected by chance. This network has $N = 62$ nodes, $M = 159$ edges, and an average degree $\overline{k} = 5.13$. The coefficient of variation (CV), defined as the standard deviation divided by the mean, of the degree is 0.58.

The BA model generates networks with power-law degree distributions \cite{barabasi1999}.
We initialized a BA network with a complete graph of three nodes and three edges, and added a single node with $m = 2$ edges in each time step. The final network had $N = 1,000$, $M = 1,996$, $\overline{k} = 3.99$, and CV of the degree is equal to $1.33$.

We used eighteen other networks as well as the dolphin and BA networks in the ANOVA analyses. See SI section S1
%
%
for the description of these eighteen networks.

\subsection{Optimization of node weights\label{sub:quadoptm}}

In section \ref{sub:weighted}, we approximate $\overline{x}_\ell$ with a weighted average of the node states in $S$, $\overline{x}'_\ell = \sum_{i=1; i\in S}^N a_i x_{i,\ell}^*$. The GBB and DART assign weights to all the nodes in the network. In contrast, here we determine weights only for the nodes in $S$ by a quadratic programming optimization.

We insert the quadratic programming step into our node selection method as follows. Assume that we have a set of nodes $S$, $|S| = n$. This $S$ can be random, as in the initial iteration of the algorithm, or a candidate set $S'$ from any subsequent iteration. Let $\bm a = \{a_i\}_{i\in S}$ be an $n$-dimensional vector of weights. We require that these weights are non-negative and sum to $1$. The mean squared error over the $L$ values of the control parameter is given by
\begin{widetext}
  \begin{equation}
    \epsilon = \frac{1}{L} \sum_{\ell=1}^L \left(\sum_{i=1; i\in S}^N a_i x_{i,\ell}^* - \overline{x}_{\ell}\right)^2
    = \frac{1}{L} \left( \bm a^{\top} X^{\top} X \bm a - 2 \bm{\overline{x}}^{\top} X \bm a
      +  \bm{\overline{x}}^{\top} \bm{\overline{x}} \right),
  \end{equation}
\end{widetext}
where $X$ is the $L \times n$ matrix of $x_{i,\ell}^*$ (with $\ell \in \{ 1, \ldots, L \}$, $i \in S$), 
$\bm{\overline{x}} = (\overline{x}_1, \ldots, \overline{x}_L)^{\top}$, 
and $^{\top}$ represents the transposition. Therefore, we determine $\bm a$ by
\begin{equation} \label{eq:quadoptm}
  \begin{aligned}
    \min \quad & \frac{1}{2} \bm a^{\top} X^{\top} X \bm a - \bm{\overline{x}}^{\top} X \bm a,\\
    \text{s.t.} \quad & \bm 1^{\top} \bm a = 1,\\
    & \bm a \geq \bm 0,
  \end{aligned}
\end{equation}
where $\bm{1} = (1, \ldots, 1)^{\top}$. We use the qpOASES algorithm \cite{ferreau2014} to solve for $\bm a$ via the ROI package \cite{theussl2020} in R. In an iteration of our algorithm, we first tentatively update $S$ by carrying out one step of the algorithm for the unweighted averaging, optimize the node weights, $\bm{a}$, given the updated $S$ by solving Eq.~\eqref{eq:quadoptm}, and calculate $\varepsilon$ for the updated $S$ and optimized $\bm{a}$. We then compare the obtained $\varepsilon$ with the value of $\varepsilon$ before updating $S$ and optimizing $\bm{a}$, and apply the Metropolis criterion to determine whether or not we adopt the tentatively updated $S$ and $\bm{a}$.
We iterate this process as described in Sec.~\ref{sub:nodeselection}.

Note that this additional optimization step substantially increases computation time. However, in practice, local minima are reached after fewer iterations of the simulated annealing algorithm. Therefore, we use $25N$ iterations of the algorithm when we optimize the node weights, instead of the $50N$ iterations used when we do not optimize the node weights.

\section*{Data availability}

Empirical networks used in this study are available from \href{https://networks.skewed.de}{https://networks.skewed.de} or the original sources as described in the SI. All other data was produced by simulation and is available at \href{https://zenodo.org/uploads/15848170}{https://zenodo.org/uploads/15848170} \cite{maclaren2025data}.

\section*{Code availability}

Optimization, simulation, and analysis code used in this study is available at \newline \href{https://github.com/ngmaclaren/observe}{https://github.com/ngmaclaren/observe} \cite{maclaren2025git}.

\section*{Acknowledgments}

B.B. was supported by the Israel Science Foundation (grant no.\,499/19), the Israel-China ISF-NSFC joint research program (grant no.\,3552/21), and by the VATAT grant for data science research. N. Masuda was supported by the Japan Science and Technology Agency (JST) Moonshot R\&D (under grant no.\,JPMJMS2021), the National Science Foundation (under grant no.\,2052720), and JSPS KAKENHI (under grant nos.\,JP 21H04595, 23H03414, and 24K14840). This work was performed in part at Center for Computational Research, the State University of New York at Buffalo.

\section*{Author contributions}

B.B. and N. Masuda conceived the study. N.G. MacLaren and N. Masuda developed the method and analyzed the data.
N.G. MacLaren wrote the code and ran simulations. 
All the authors wrote the manuscript.

\section*{Competing interests}

The authors declare no competing interests.

\bibliographystyle{unsrt}
\bibliography{refs.bib}

\appendix

\renewcommand\refname{Supplementary References}
\renewcommand{\thetable}{S\arabic{table}}
\renewcommand{\thefigure}{S\arabic{figure}}
\renewcommand{\thesection}{S\arabic{section}}
\renewcommand{\thesubsection}{\Alph{subsection}}
\renewcommand{\theequation}{S\arabic{equation}}

\raggedbottom

\setlength{\tabcolsep}{10pt}

\begin{center}
  \vspace*{12pt}
  {\large \bf Supplementary Materials for:\\
    \vspace{12pt} Observing network dynamics through sentinel nodes}
  \vspace{12pt} \\
  Neil G. MacLaren, Baruch Barzel, and Naoki Masuda
\end{center}

\section{Undirected and unweighted networks used in the analysis\label{sec:SInetworks}}

In the main text, we have shown results from two undirected and unweighted networks, i.e., the dolphin network and a network generated by the Barab\'{a}si-Albert (BA) model. We used eighteen other undirected and unweighted networks, of which fourteen are empirical networks and the other four are model networks, in our statistical analyses. For each network described in this section, we removed edge weights and the multiplicity of edges, coerced the network to be undirected if it was not so already, and selected the largest connected component.

We downloaded five empirical networks, including the dolphin network described in the main text, from the KONECT repository \cite{KONECT} at \url{http://konect.cc}. Besides the dolphin network, the other four empirical networks are as follows:
\begin{description}
\item [Proximity] A network of visitors at a museum \cite{isella2011}. Each visitor is a node. Two nodes are adjacent (i.e., directly connected by an edge) if any face-to-face contact of 20 seconds or more was recorded between them. There are 69 days of recording \cite{isella2011}. We use the day with the largest number of contacts, which is the network provided in the KONECT repository. The network has $N = 410$, $M = 2,765$, $\overline{k} = 13.49$, and CV of the degree equal to $0.62$.
\item [Metabolic] A metabolic network of the nematode {\em Caenorhabditis elegans} \cite{jeong2000}. In this network, nodes are metabolic compounds. Two nodes are adjacent if one metabolite is a product of the other. Note that the original study treated this network as directed, but we use the undirected version here. This network has $N = 453$, $M = 2,025$, $\overline{k} = 8.94$, and CV of the degree equal to $1.87$.
\item [Road] A network of major European roads \cite{subelj2011}. Each node represents a city. Two cities are adjacent if there is a road connection between them. Our version of this network has $N = 1,039$, $M = 1,305$, $\overline{k} = 2.51$, and CV of the degree equal to $0.48$.
\item [Email] A network of email exchanges at the University of Rovira i Virgili \cite{guimera2003}. Nodes in this network are email accounts. Two nodes are adjacent if at least one email was sent between them. Our version of this network has $N = 1,133$, $M = 5,451$, $\overline{k} = 9.62$, and CV of the degree equal to $0.97$.
\end{description}

We downloaded an additional ten networks from the Netzschleuder repository \cite{Netzschleuder} as follows:
\begin{description} 
\item [FlyBi] A network of protein interactions for the fruit fly {\em Drosophila melanogaster} \cite{tang2023}. A node is a protein. Two nodes are adjacent if the two proteins were observed to interact in a biochemical analysis. The network has $N = 2,705$, $M =  8,458$, $\overline{k} = 6.25$, and CV of the degree equal to $1.76$.
\item [Reactome] A network of human protein interactions \cite{joshitope2005}. Nodes represent proteins. Edges represent interactions. The network has $N = 5,973$, $M = 145,778$, $\overline{k} = 48.81$, and CV of the degree equal to $1.39$. 
\item [Route views] A network of routine message passing between internet routers \cite{leskovec2005}. A node is a collection of internet protocol addresses belonging to the same entity. Two nodes are adjacent if border gateway protocol (BGP) messages passed between the two entities. This network has the largest number of nodes in the data set and was collected on January 2, 2000. The network has $N = 6,474$, $M = 12,572$, $\overline{k} = 3.88$, and CV of the degree equal to $6.44$. 
\item [Spanish words] A network of word adjacencies drawn from the novel ``Don Quixote'' by Miguel de Cervantes \cite{milo2004}. Nodes are words. Two nodes are adjacent if one word appeared next to the other in the novel. The network has $N = 11,558$, $M = 43,050$, $\overline{k} = 7.45$, and CV of the degree equal to $7.77$. 
\item [FOLDOC] A network of cross-references in an online encyclopedia, the Free On-line Dictionary of Computing \cite{batagelj2002}. A node is encyclopedia entry, which is a webpage. Two nodes are adjacent if there is a hyperlink from one page to the other. The network has $N = 13,356$, $M = 91,471$, $\overline{k} = 13.70$, and CV of the degree equal to $1.13$. 
\item [Tree of life] A protein interaction network, similar to those used above \cite{zitnik2019}. We chose the network with the largest number of nodes available in the Netzschluder repository. The network has $N = 16,415$, $M = 440,122$, $\overline{k} = 53.62$, and CV of the degree equal to $1.62$. 
\item [English words] A network of word associations \cite{kiss1973}. A node is an English word. Two nodes are adjacent if a study participant gave one word as a response to another, stimulus word. The network has $N = 23,132$, $M =  297,094$, $\overline{k} = 25.69$, and CV of the degree equal to $1.74$. 
\item [Enron] A network drawn from the Enron email corpus \cite{klimt2004}. Nodes are email addresses. Nodes are adjacent if at least one email was sent from one address to the other. The network has $N = 33,696$, $M =  180,811$, $\overline{k} = 10.73$, and CV of the degree equal to $3.50$. 
\item [Marker Cafe] An online social network \cite{fire2014}. A node is a user of the Marker Cafe social network. Two nodes are adjacent if one user is in the other's ``circle.'' The network has $N = 69,317$, $M = 1,644,794$, $\overline{k} = 47.46$, and CV of the degree equal to $3.72$. 
\item [Prosper] A network of loans in the online marketplace prosper.com \cite{redmond2013}. A node is a user. Two nodes are adjacent if one user lent the other money. The network has $N = 89,171$, $M =  3,329,970$, $\overline{k} = 74.69$, and CV of the degree equal to $1.87$. 
\end{description}

We used instances of five undirected random network models, including the BA model, with $N = 1,000$ nodes. We removed any self-loops or multi-edges and retained the largest connected component. We generated the Erd\H{o}s-R\'{e}nyi (ER), BA, Holme-Kim (HK), and Lancichinetti-Fortunato-Radicchi (LFR) models using NetworkX \cite{networkx}, and the Gao-Kahng-Kim (GKK) model using igraph \cite{igraph}. The BA network is described in the main text. The remaining four networks are as follows:
\begin{description}
\item [ER] We generated an undirected ER network with the probability of connecting two edges set to $p = 0.05$. The resulting network was connected and had $N = 1,000$,  $M = 25,132$, $\overline{k} = 50.26$, and CV of the degree equal to $0.14$.
\item [HK] The HK model is a variation of the BA model that aims to produce high clustering (i.e., large density of triangles) \cite{holme2002}. We set $m = 2$ and a target local clustering coefficient of $0.1$. The final network had $N = 1,000$, $M = 1,996$, $\overline{k} = 3.99$, CV of the degree equal to $1.40$, and an average local clustering coefficient of $0.12$.
\item [GKK] The GKK model is a node fitness model \cite{goh2001}. A node is assigned a fitness $f_i = (i + i_0 -1)^{-\alpha}$, where $i_0 = N^{1-\frac{1}{\alpha}}\left[ 10\sqrt{2}(1 - \alpha) \right]^{\frac{1}{\alpha}}$ constrains the maximum degree \cite{chung2002, cho2009}. For each $i, j \in \{1, \ldots, N\}$, edge $(i, j)$ is present with probability $\frac{f_if_j}{(\sum_{\ell=1}^N f_\ell)^2}$. We set $\alpha = 1.25$, $N = 1,000$, and $M = 2,500$. The largest connected component of the generated network had $N = 949$, $M = 2,496$, $\overline{k} = 5.26$, and CV of the degree equal to $0.84$.
\item [LFR] The LFR benchmark model produces networks that have both heterogeneous degree distributions and community structure with heterogeneous community sizes \cite{lancichinetti2008}. We set the expected power-law exponents to $-3$ and $-1.5$ for the degree distribution and the distribution of community sizes, respectively. We set the probability of connecting nodes between communities to 0.1, the expected average degree to 4, and the minimum community size to 20. The final network had $N = 998$, $M = 1,988$, $\overline{k} = 3.98$, and CV of the degree equal to $0.66$.
\end{description}

\clearpage

\section{Comparison between the sentinel node approximation and $\overline{x}$ for all dynamics and networks\label{sec:comparison-all-dynamics-networks}}

\begin{figure}
\centering
  \includegraphics[width = 0.9\textwidth]{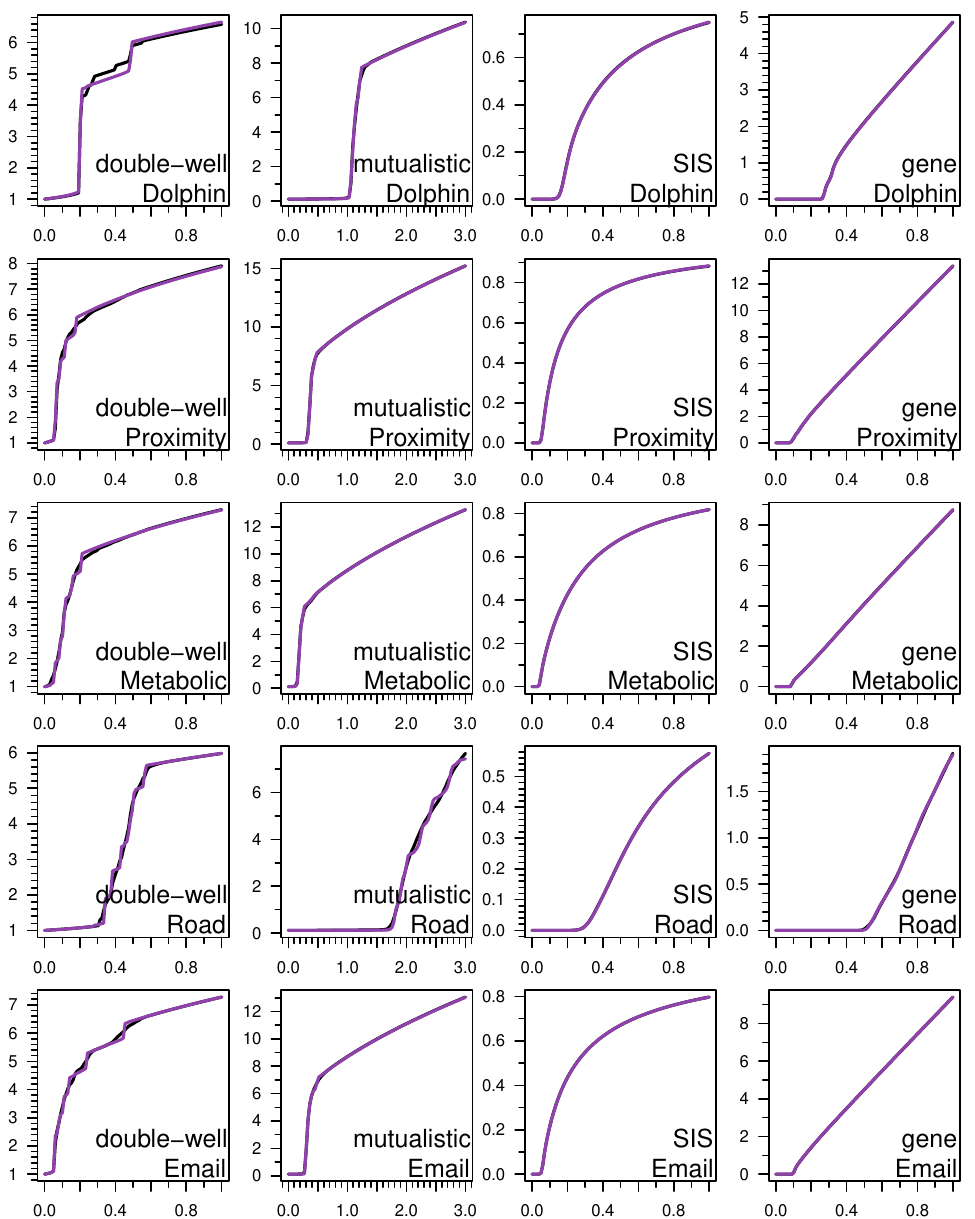}
  \caption{Comparison between the sentinel node approximation and $\overline{x}$ for each pair of dynamics and network.
    The sentinel node approximation using $n = \lfloor \ln N \rfloor$ nodes is shown in purple.
    The black lines represent the average equilibrium state of the nodes, $\overline{x}$.  
  }
  \label{fig:comp-all-dyn-net}
\end{figure}

We show in Fig.~\ref{fig:comp-all-dyn-net} the comparison between our sentinel node approximation with $n = \lfloor \ln N \rfloor$ nodes and $\overline{x}$ for each pair of the four dynamics and 20 networks. We find that the sentinel node approximation captures the main bifurcations of the original dynamical system fairly well in all cases.

\clearpage

\begin{center}
  \includegraphics[width = 0.9\textwidth]{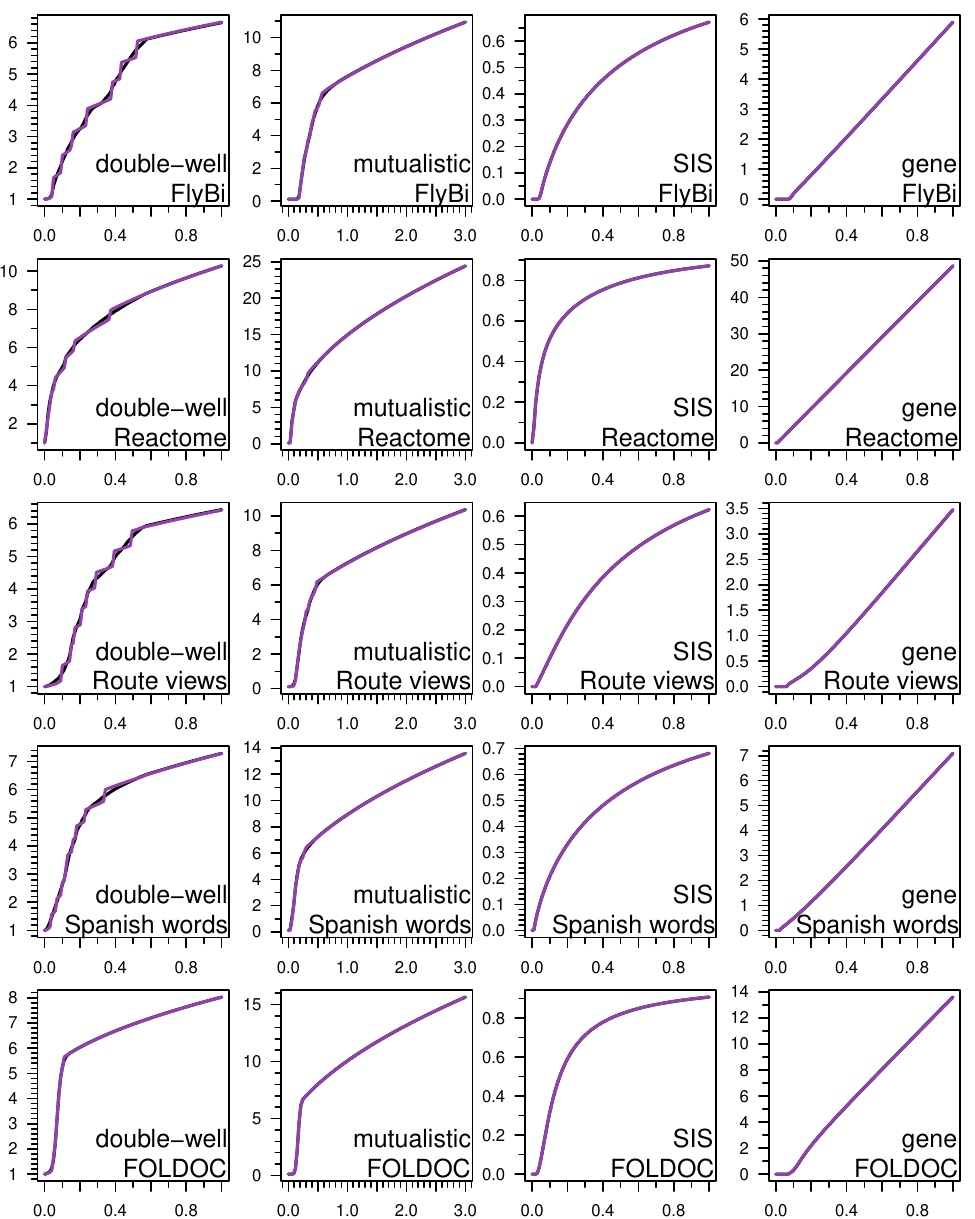}
Figure~\ref{fig:comp-all-dyn-net}: (Continued.)
\end{center}

\clearpage

\begin{center}
  \includegraphics[width = 0.9\textwidth]{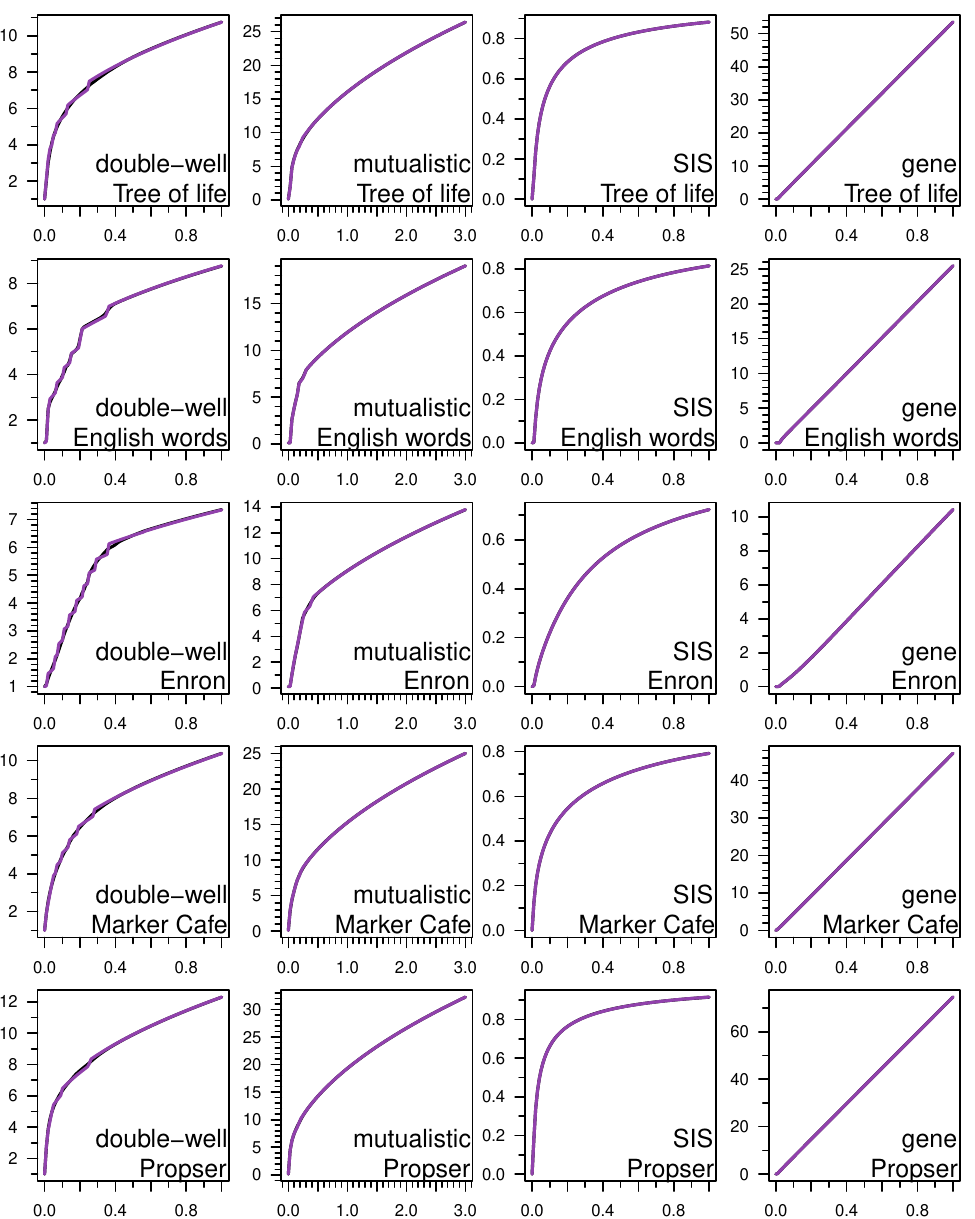}
Figure~\ref{fig:comp-all-dyn-net}: (Continued.)
\end{center}

\clearpage

\begin{center}
  \includegraphics[width = 0.9\textwidth]{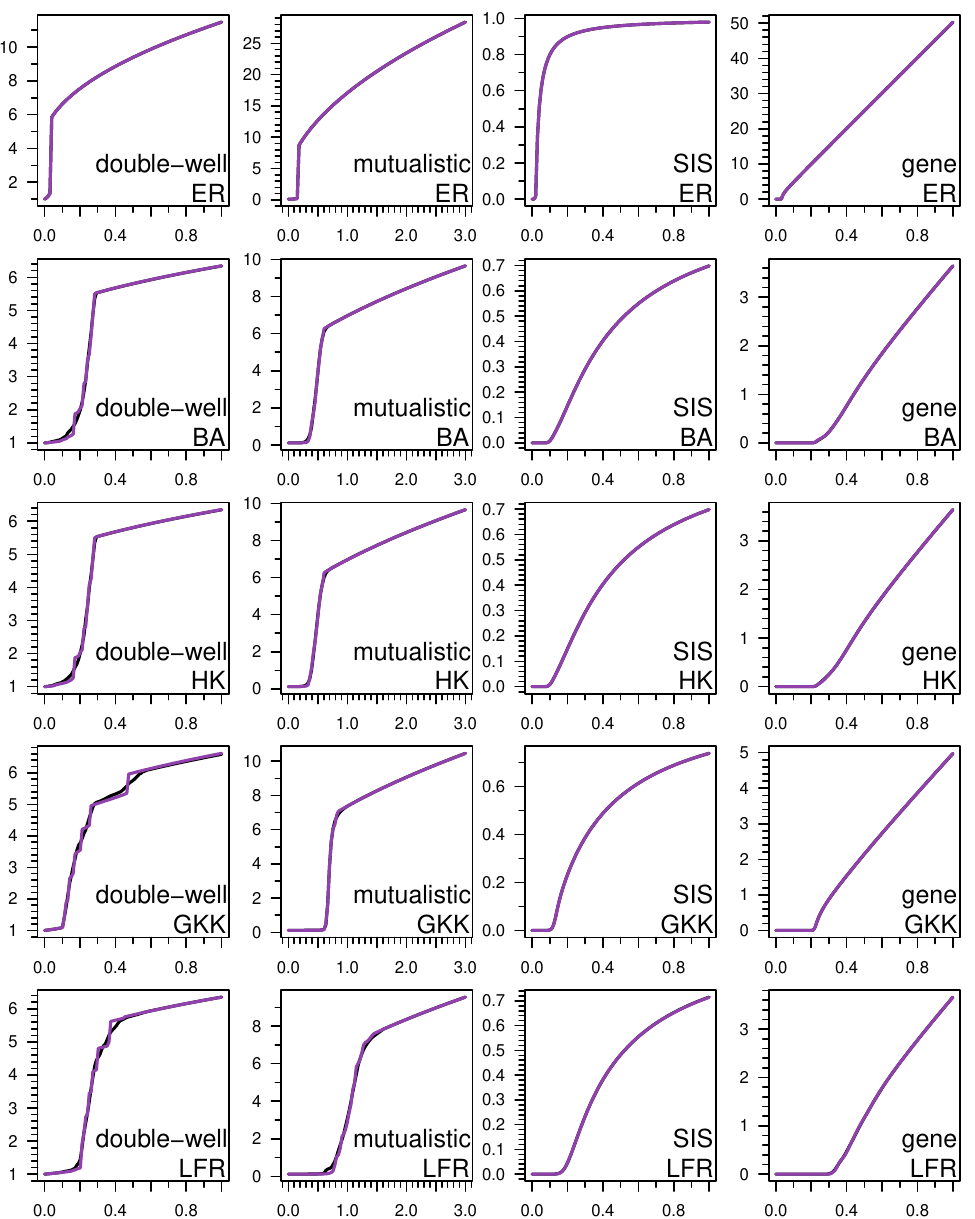}
Figure~\ref{fig:comp-all-dyn-net}: (Continued.)
\end{center}

\clearpage

\section{GBB, DART, and their variants\label{sec:variants-GBB-DART}}

Two known methods for dynamics reduction on networks are the Gao-Barzel-Barab\'asi (GBB) reduction \cite{gao2016, zhang2022, ma2023} and the dynamics approximation reduction technique (DART) \cite{laurence2019, thibeault2020, masuda2022, thibeault2023, vegue2023}. These two methods attempt to match a different weighted sum of $x_1$, $\ldots$, $x_N$ from $\overline{x}$. Specifically, the observables of the GBB and DART are in the form $\sum_{i=1}^N a_i x_i$. For undirected networks, which we assume, $a_i$ is proportional to the degree of the $i$th node in the case of the GBB and the $i$th entry of the leading eigenvector of the adjacency matrix in the case of the DART.

\begin{figure}[b]
\centering
  \includegraphics[width = 0.6\textwidth]{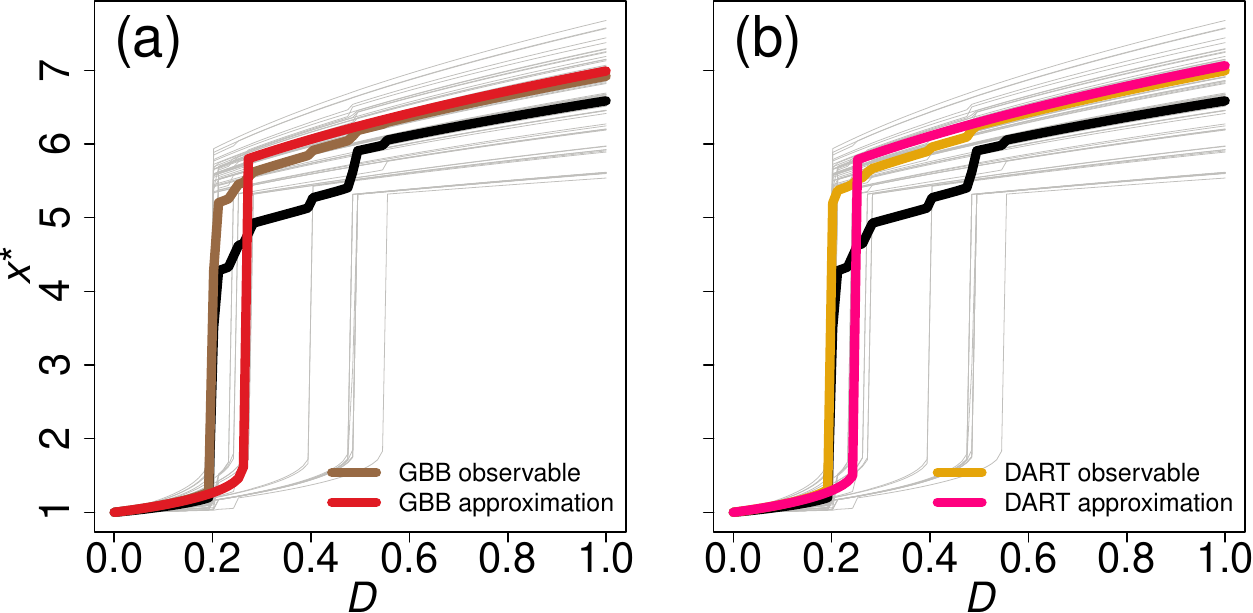}
  \caption{Alternative approximations to $\overline{x}$. The approximations for the GBB and DART methods are shown in red and pink, respectively. The brown and dark yellow lines show the observables (i.e., a particular weighted average of $\{x_1^*, \ldots, x_N^*\}$) that the GBB and DART schemes intend to approximate. The $x_i^*$ values are shown in gray. The unweighted average $\overline{x}$ is shown in black. (a) GBB. (b) One-dimensional DART.}
  \label{fig:gbb-dart}
\end{figure}

Using the dolphin network as an example (see Fig.~1 in the main text), we show the results obtained from the GBB and one-dimensional DART approximations in Fig.~\ref{fig:gbb-dart}a and \ref{fig:gbb-dart}b, respectively. The GBB approximation is faster than our algorithm to determine $S$, and the same holds true for the one-dimensional DART at least for sparse networks (see section \ref{sec:time-complexity}). However, even with just $n=1$ sentinel node, our method performs better than the GBB and DART at approximating $\overline{x} = \sum_{i=1}^N x_i / N$, reducing $\varepsilon$ by more than 75\% 
(GBB: $\varepsilon = 0.171$; DART: $\varepsilon = 0.142$). The different performance is mainly owing to the fact that, while GBB and DART track the collective state of the system (i.e., $\langle x \rangle$ in the Introduction section of the main text; brown and dark yellow lines in Fig.~\ref{fig:gbb-dart}) and thus predict a single transition point, our method can discern the multistage nature of the actual observed transition (black solid line).

Our method also outperforms a variant of the GBB \cite{tu2021} and a multi-dimensional version of DART \cite{laurence2019}. First, we consider a variant of the GBB that uses the Chebyshev polynomials~\cite{tu2021}, which we call the Chebyshev reduction. Consider a class of dynamical systems on networks given by
\begin{equation}
\frac{dx_i}{dt} = F(x_i) + \sum_{j=1}^N A_{ij} G(x_i, x_j).
\end{equation}
The four dynamical system models considered in the present study (i.e., coupled double-well, mutualistic species, SIS, and gene-regulatory) belong to this class. The original GBB~\cite{gao2016} for undirected networks is given by
\begin{equation}
\frac{d \langle x \rangle}{dt} = F(\langle x\rangle) + \beta G(\langle x \rangle, \langle x \rangle),
\label{eq:GBB}
\end{equation}
where
$\langle x\rangle = \sum_{i=1}^N a_i x_i$, $a_i = k_i x_i / \sum_{\ell=1}^N k_{\ell}$, $\forall i$, $\beta = \sum_{\ell=1}^N k_{\ell}^2 / \sum_{\ell'=1}^N k_{\ell'}$. The Chebyshev reduction approximates $F(\langle x\rangle)$ and $G(\langle x \rangle, \langle x \rangle)$ in Eq.~\eqref{eq:GBB} by Chebyshev polynomials of a given order. However, it should be noted that $G(\langle x \rangle, \langle x \rangle)$ is approximated through two-dimensional Chebyshev polynomial approximation to $G(x, y)$, not simply by fitting a one-variable Chebyshev polynomial of $x$ to $G(x, x)$.

When the original $F$ or $G$ is a polynomial of the order less than or equal to that of the Chebyshev polynomials used, the Chebyshev reduction is exact. Therefore, the Chebyshev reduction of order up to three and that of order up to two are equivalent to the original GBB for the coupled double-well and SIS dynamics, respectively, for which both $F$ and $G$ are polynomials. Given this, we only demonstrate the Chebyshev reduction for the mutualistic interaction and gene-regulatory dynamics. These dynamics have a non-polynomial $G$, which we approximate by the Chebyshev polynomials. In contrast, $F$ is polynomial in these dynamics and does not need to be approximated. We use the HK network for the mutualistic interaction dynamics and the FlyBi network for the gene-regulatory dynamics. To fit Chebyshev polynomials to $G$, we need to specify the range of $x$ that we linearly transform into $[-1, 1]$. We set this range to $[0, x_{\max}]$, where $x_{\max} = 26.87$ and $x_{\max} = 124.58$ for the HK and FlyBi networks, respectively.
%
%
With these $x_{\max}$ values, the interval $[0, x_{\max}]$ represents and covers the ranges of $x_i$ that we observe in numerical simulations as we vary the control parameter.

\begin{figure}
\centering
  \includegraphics[width = 0.95\textwidth]{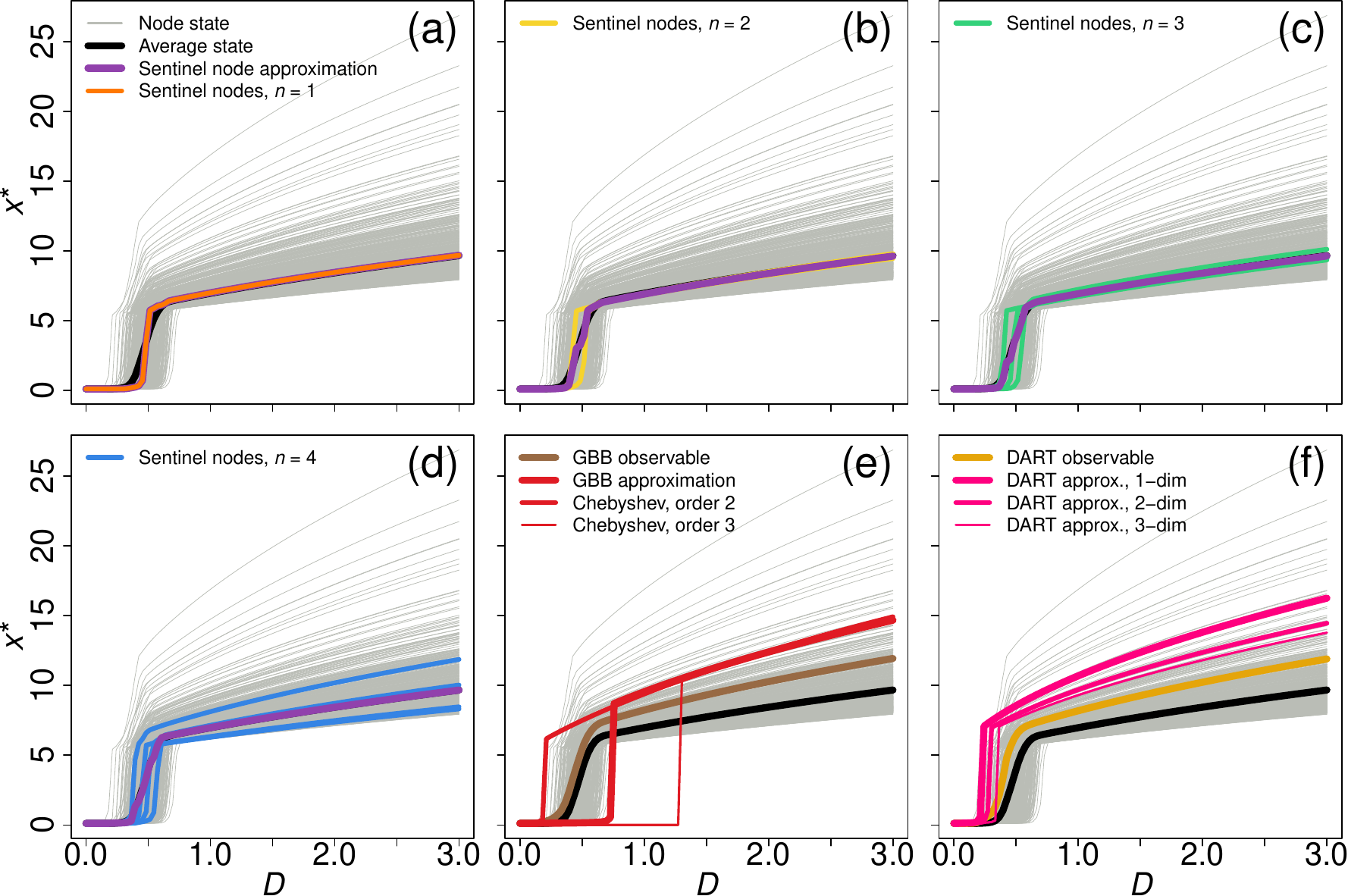}
  \caption{Approximations of the average activity, $\overline{x}$, of the mutualistic interaction dynamics on the HK network.
The gray lines indicate $x_i^*$ at each value of the control parameter. The black lines indicate $\overline{x}$. Note that the gray and black lines are the same in all the panels. The purple lines in (a)--(d) show the sentinel node approximation with the selected sentinel nodes, whose $x_i^*$ values are shown in orange (a), yellow (b), green (c), and blue (d). (e) GBB and Chebyshev reductions.
(f) DART of different dimensions. The brown line in (e) and the dark yellow line in (f) show the observables (i.e., a particular weighted average of $\{x_1^*, \ldots, x_N^*\}$) that the GBB and DART schemes, respectively, intend to approximate.
In (d), the purple line almost completely covers the black line.}
  \label{fig:demo-mutualistic-HK}
\end{figure}

\begin{figure}
\centering
  \includegraphics[width=0.95\textwidth]{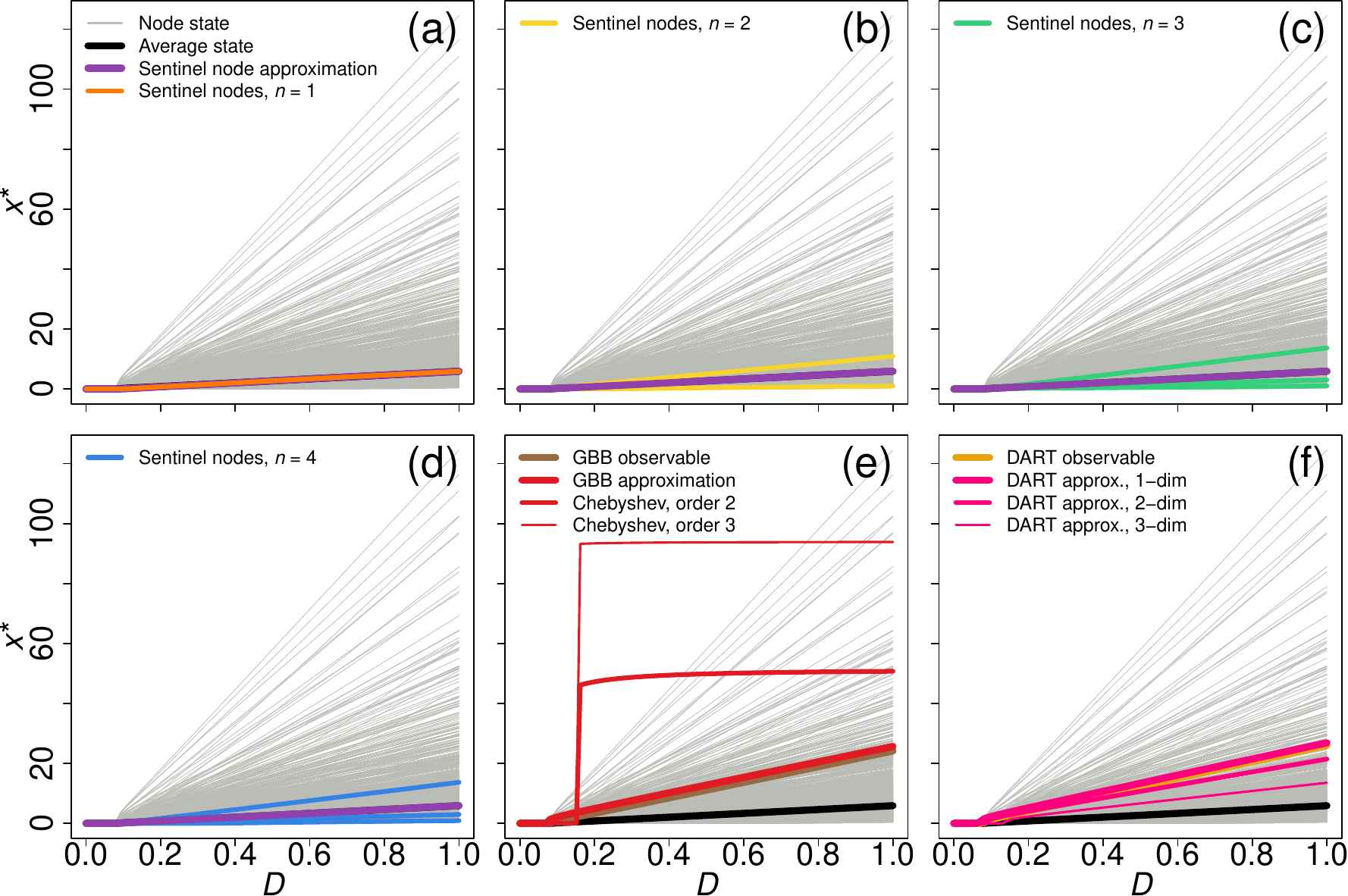}
  \caption{Approximations of the average activity, $\overline{x}$, of the gene-regulatory dynamics on the FlyBi network.
See the caption of Fig.~\ref{fig:demo-mutualistic-HK} for the legends. In (a), the orange line covers the black and purple lines. In (a)--(d), the purple line covers the black line. In (f), the thick pink line almost completely covers the dark yellow line.}
  \label{fig:demo-gene-FlyBi}
\end{figure}

We show the numerical results for the mutualistic interaction dynamics on the HK network in 
Fig.~\ref{fig:demo-mutualistic-HK}. Figure~\ref{fig:demo-mutualistic-HK}a--d shows that our sentinel node approximation works progressively better with a larger number of the sentinel nodes, $n$, as we showed in Fig.~1c--f in the main text for the coupled double-well dynamics on the dolphin network. Figure~\ref{fig:demo-mutualistic-HK}e shows the behavior of the GBB and Chebyshev reductions. We used the Chebyshev polynomials of order up to 2 and 3 for each variable (i.e., each of $x$ and $y$ in $G(x, y)$).
We find that the Chebyshev reduction does not improve over the GBB reduction in approximating either $\overline{x}$ or the target of the GBB reduction, $\langle x \rangle$. The results are similar for the gene-regulatory dynamics on the FlyBi network, as shown in Fig.~\ref{fig:demo-gene-FlyBi}e. These results are intuitive because, in the present modeling context, the Chebyshev reduction is an approximation to the GBB reduction, whose goal is not to approximate $\overline{x}$.

We now assess a multidimensional version of DART given by Eqs.~(30)--(32) in \cite{laurence2019}. The one-dimensional DART aims to approximate the dynamics of $\langle x \rangle$ and uses the leading eigenvector of the transposed adjacency matrix, $A^{\top}$, as $(a_1, \ldots, a_N)$ for a one-dimensional dynamics similar to Eq.~\eqref{eq:GBB}~\cite{laurence2019, thibeault2020}. The $d$-dimensional DART consists in separately building the one-dimensional DART for the $d$ eigenvectors of $A^{\top}$ that are associated with the $d$ largest eigenvalues in modulus and averaging the one-dimensional approximation to $\langle x \rangle$ over the $d$ one-dimensional equations. 
Among the multidimensional variants of DART, we did not use the ``cycle reduction''~\cite{laurence2019} because it assumes the presence of multiple eigenvalues with the same largest modulus for the adjacency matrix. An adjacency matrix would have such eigenvalues only if the network is highly symmetric or multipartite. In contrast, we are interested in general unipartite networks in the present study. We also avoided the multidimensional DART in \cite{thibeault2020} because it implicitly assumes community structure of the network, which informs the reduced adjacency or Laplacian matrix (i.e., Fig. 3 in \cite{thibeault2020}, and the derivation of the $3d$-dimensional reduced dynamics for the Lorenz model on p.22 with Fig. 15). Our focus is on networks that may or may not have community structure. For the same reason, we did not use the multidimensional reduction method for networks with community structure proposed in \cite{vegue2023}.

We set $d=2$ and $d=3$ in our demonstration. We show numerical results for the mutualistic interaction dynamics on the HK network and the gene-regulatory dynamics on the FlyBi network in Fig.~\ref{fig:demo-mutualistic-HK}f and \ref{fig:demo-gene-FlyBi}f, respectively. We find that the $d$-dimensional DART with a larger $d$ is more accurate at approximating $\overline{x}$. However, our sentinel node approximation provides a much more accurate approximation even with $n=1$ than the DART with $d \in \{1, 2, 3 \}$, endorsing the fact that goal of DART is to approximate a specific $\langle x\rangle$, not $\overline{x}$.

To be quantitative, we calculated the approximation error in these numerical results. The approximation error against $\overline{x}$ in the case of mutualistic interaction dynamics on the HK network, shown in Fig.~\ref{fig:demo-mutualistic-HK},
is equal to $\varepsilon = 1.35 \times 10^{-2}$, $2.48 \times 10^{-3}$, $9.16 \times 10^{-4}$, and $5.20 \times 10^{-4}$ for the sentinel node approximation with $n=1$, $2$, $3$, and $4$, respectively; $\varepsilon = 2.03$ for GBB;
$\varepsilon = 2.23$ and $3.28$ for the Chebyshev reduction of order up to 2 and 3, respectively;
$\varepsilon = 3.72$, $2.06$, and $1.46$ for the DART with $d=1$, $2$, and $3$, respectively.
The approximation error against the preferred observables for GBB and DART, i.e., $\langle x \rangle$, in the same simulations is as follows:
$\varepsilon = 0.897$ for GBB; $\varepsilon = 0.725$ and $2.43$ for the Chebyshev reduction of order up to 2 and 3, respectively;
$\varepsilon = 1.42$, $0.657$, and $0.535$ for the DART with $d=1$, $2$, and $3$, respectively.
In the case of the gene-regulatory dynamics on the FlyBi network, shown in Fig.~\ref{fig:demo-gene-FlyBi},
the approximation error against $\overline{x}$ is: 
$\varepsilon = 1.28 \times 10^{-3}$, $3.87 \times 10^{-5}$, $1.42 \times 10^{-5}$, and $2.50 \times 10^{-5}$ for the sentinel node approximation with $n=1$, $2$, $3$, and $4$, respectively; $\varepsilon = 48.9$ for GBB; $\varepsilon = 666$ and $2,530$ for the Chebyshev reduction of order up to 2 and 3, respectively; $\varepsilon = 55.0$, $30.4$, and $7.38$ for the DART with $d=1$, $2$, and $3$, respectively.
For the same simulations, the approximation error against $\langle x \rangle$ is equal to:
$\varepsilon = 0.213$ for GBB; $\varepsilon =102$ and $489$ for the Chebyshev reduction of order up to 2 and 3, respectively;
$\varepsilon = 0.0741$, $0.0562$, and $0.985$ for the DART with $d=1$, $2$, and $3$, respectively. These $\varepsilon$ values are consistent with our main conclusion that the Chebyshev reduction of order up to 2 and 3 or the DART with $d=2$ and $d=3$ does not provide an accurate approximation to $\overline{x}$ compared to the sentinel node approximation.

\clearpage

\section{Distribution of approximation errors for various dynamics and networks\label{sec:compare-networks}}

\begin{figure}[t]
  \centering
  \includegraphics[width = 0.95\textwidth]{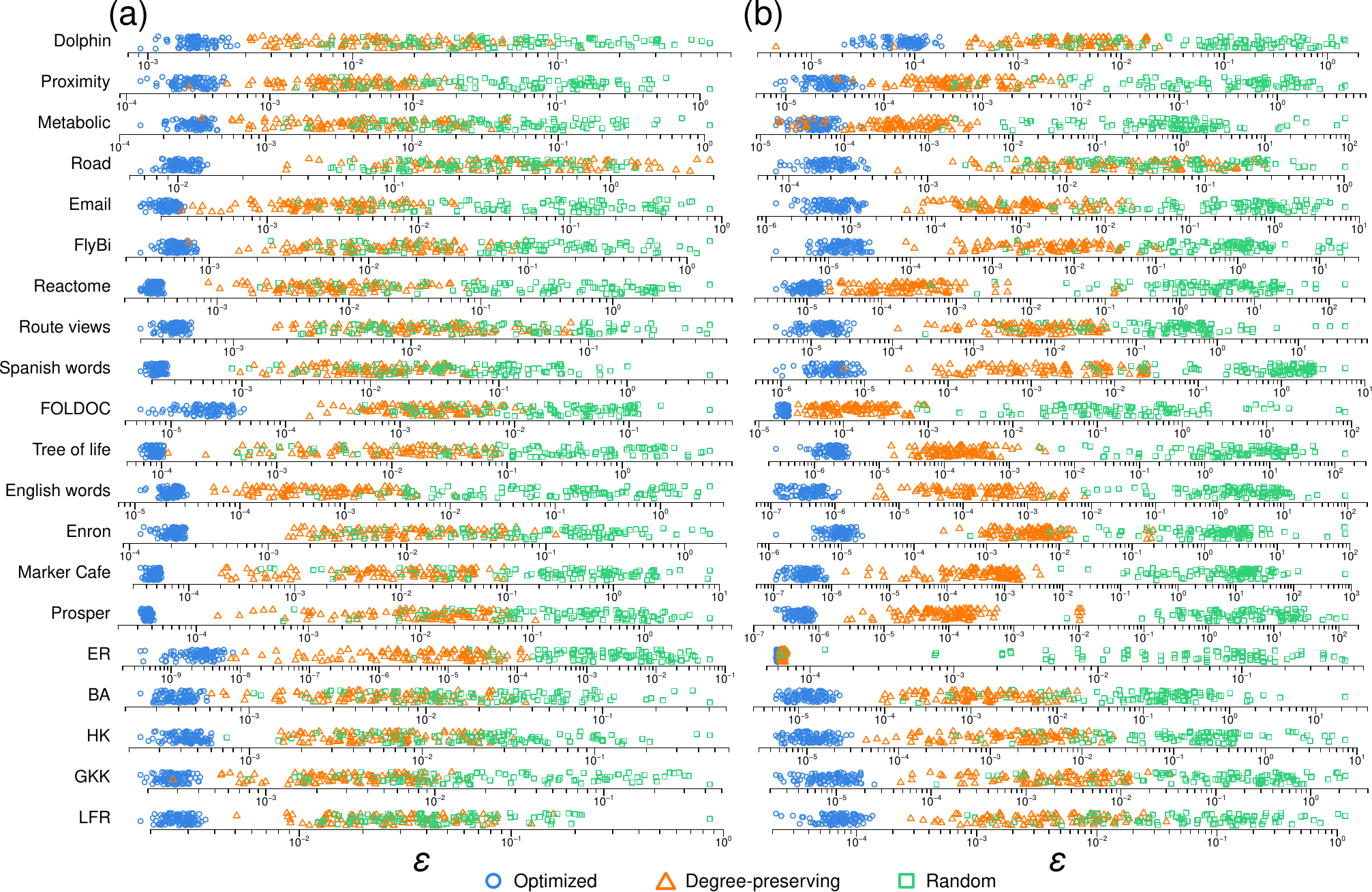}
  \caption{
    Approximation error, $\varepsilon$, for two other dynamics on 20 undirected unweighted networks. (a) Mutualistic species dynamics. (b) Gene-regulatory dynamics. For each pair of dynamics and network, we compare three node set types (optimized: blue circles, degree-preserving: orange triangles, completely random: green squares).
  }
  \label{fig:comp-nets-SI}
\end{figure}

In Figs.~2a and b in the main text, we demonstrated that the state averaged over the nodes in the optimized node set can closely approximate the state averaged over all the nodes in the coupled double-well dynamics and the SIS dynamics on all 20 undirected and unweighted networks. We show the corresponding results for the mutualistic species and gene-regulatory dynamics in Fig.~\ref{fig:comp-nets-SI}a and b, respectively.

Across the two dynamics and all networks, the optimized node sets have relatively small approximation error. This result is similar to that for the coupled double-well and SIS dynamics shown in Fig.~2 in the main text. However, the relationship between degree-preserving and completely random node sets differs. In particular, while the difference between the degree-preserving and completely random node sets is large for many networks, it is small for the road network and the LFR network. This result may be because the degree sequence may not contain sufficient information to characterize these networks: the road network has a complicated community structure imposed by geography \cite{subelj2011}, and the LFR network has a marked community structure by design \cite{lancichinetti2008}. Either of these cases may make particular nodes more suitable or less so for inclusion in $S$ for reasons other than the node's degree. In addition, the road network has a narrow degree distribution, probably making the degree a less important indicator of a node's importance in representing the dynamics of the entire network. Even in these cases, our algorithm finds node sets with relatively low approximation error.

To formally assess the differences between node set types in terms of approximation error, we generated 100 optimized, degree-preserving, and completely random node sets for each combination of dynamics and network. We computed the approximation error for each node set. We excluded the ER network from this analysis due to the tendency to obtain tiny approximation errors regardless of the node set type. We then conducted a multi-way analysis of variance (ANOVA) with three independent variables: dynamics (reference: coupled double-well dynamics), network (reference: BA network), and node set type (reference: completely random node sets). The dependent variable is the approximation error, $\varepsilon$. We use $\ln \varepsilon$ because tiny (i.e., near zero) error values are present; the error values obey skewed distributions for most dynamics, networks, and node set types; and the variance of $\varepsilon$ tends to be large when the mean of $\varepsilon$ is large.

Our ANOVA model predicts $\ln \varepsilon$ well ($R^2 = 0.76$). All independent variables are statistically significant (dynamics: $df = 3$, $F = 6,827.3$, $p < 10^{-7}$; network: $df = 18$, $F = 179.81$, $p < 10^{-7}$; node set type: $df = 2$, $F = 24,770$, $p < 10^{-7}$), partly due to a large sample size.

We show the differences between the average error for each pair of node set types, as computed by a Tukey's honestly significant difference test, in Table \ref{tab:HSD-short}. Each row of the table specifies the estimated difference in average $\ln \varepsilon$ between the two node set types, the 95\% confidence interval of the estimated mean difference, and a $p$ value for the difference adjusted for multiple comparisons. For example, the first row states that, across dynamics and networks, optimized node sets have an average $\ln \varepsilon$ value that is 6.892 smaller than that of completely random node sets. On the natural scale, the average $\varepsilon$ for optimized node sets is $1/e^{-6.892} = 984.4$ times smaller than the average $\varepsilon$ for completely random node sets.

\begin{table}[b]
\centering
  \caption{
    Differences in average $\ln \varepsilon$ between three node set types across dynamics and networks. The differences are computed by a Tukey's honestly significant difference test. The 95\% confidence intervals (CI) of the differences and associated $p$ values, adjusted for multiple comparisons, are also shown.
  }
  \label{tab:HSD-short}
  \begin{tabular}{lScc}
    \toprule
    & {Difference} & CI & $p$\\
    \midrule
    Optimized $-$ Random & -6.892 & $[-6.964, -6.819]$ & $< 10^{-7}$\\
    Degree-preserving $-$ Random & -3.149 & $[-3.222, -3.076]$ & $< 10^{-7}$\\
    Degree-preserving $-$ Optimized & 3.743 & $[3.670, 3.816]$ & $< 10^{-7}$\\
    \bottomrule
  \end{tabular}
\end{table}

\newpage
\clearpage

\section{Sentinel node set optimization on weighted networks\label{sec:SIedgeweights}}

In the main text, we focused on unweighted networks (i.e., networks in which all the edges have the same weight equal to $1$). In this section, we show that our sentinel node approximation also performs well for weighted networks.

We collected ten weighted networks from online repositories. From the KONECT project \cite{KONECT}, we used a dominance network of Japanese  macaques \cite{takahata1991} (Macaque dominance: $N=62$, $M=1,167$), a friendship network of high school students in Illinois, USA \cite{coleman1964} (High school friendship: $N=70$, $M=274$), a friendship network of Australian university students \cite{freeman1998} (University friendship: $N=217$, $M=1,839$), and the proximity network from the main text but with edge weights retained \cite{isella2011} (Weighted proximity: $N=410$, $M=2,765$). From the Netzschleuder repository \cite{Netzschleuder}, we used a network of interpersonal contacts between windsurfers \cite{freeman1988} (Windsurfer contact: $N=43$, $M=336$), a network of contacts between individuals involved in the train bombing in 2004 in Madrid, Spain \cite{hayes2006} (Terrorist contact: $N=64$, $M=243$), a network of drug interactions gathered from health records in Blumenau, Brazil \cite{correia2019} (Drug interaction: $N=75$, $M=181$), a coauthorship network \cite{newman2006} (Coauthorship: $N=379$, $M=914$), a neuronal network of {\em Caenorhabditis elegans} \cite{cook2019} (Neuronal: $N=460$, $M=1,432$), and a product export network \cite{hausmann2013} (Export: $N=774$, $M=1,779$). As we did for unweighted networks in the main text, we coerced each network to be undirected, discarded any self-edges, and analyzed only the largest connected component.

As before, we simulated each dynamics on each network, using the parameters described in the main text, obtaining $x_{i,\ell}^*$ $\forall \ell \in \{1\ldots L\}$. We then selected 100 node sets of size $n = \lfloor \ln N \rfloor$ of each type.

\begin{figure}[b]
  \centering
  \includegraphics[width = 0.55\textwidth]{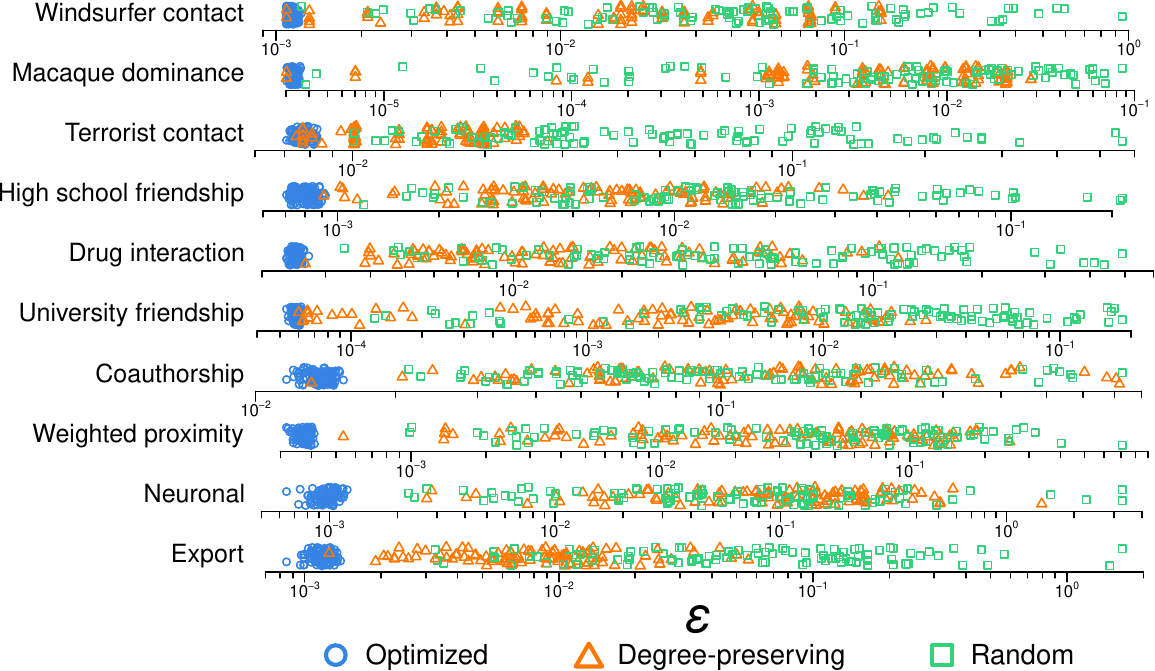}
  \caption{
    Approximation error on weighted networks for the coupled double-well dynamics for node sets with $n = \lfloor \ln N \rfloor$ nodes. We show the results for optimized (blue circles), degree-preserving (orange triangles), and completely random (green squares) node sets on all networks. Each marker represents the approximation error for one node set. Each row shows the results for one network.
  }
  \label{fig:comp-nets-weighted}
\end{figure}

We show the approximation error for the coupled double-well dynamics on the ten weighted networks in Fig.~\ref{fig:comp-nets-weighted}. As was the case for unweighted networks, our optimization algorithm consistently (i.e., over 100 independent algorithm runs and for each network) finds sentinel node sets with lower approximation error than degree-preserving and completely random node sets.

To systematically verify that the sentinel node approximation performs better than the other node set selection methods, we ran the same statistical analysis as the one we used for unweighted networks (see section \ref{sec:structural-features}\ref{sub:SI-heuristic}). The ANOVA results were similar to those for unweighted networks (model $R^2 = 0.73$; dynamics: $df = 3$, $F = 10,300$, $p < 10^{-7}$; network: $df = 9$, $F = 966.3$, $p < 10^{-7}$; node set type: $df = 6$, $F = 6,243$, $p < 10^{-7}$).  The optimized node sets had an average approximation error that was 960 times smaller than completely random node sets ($b = -6.867$, $e^b = 0.001$, $p < 10^{-7}$) and 262 times smaller than degree-preserving node sets (difference in coefficients: $-5.569$, $e^b = 0.004$, $p < 10^{-7}$); see Table \ref{tab:HSD-home-weighted}. We note that the ordering of node set types in terms of approximation error is similar to that for unweighted networks, except that there is less difference between the heuristic node sets and the completely random node sets in the case of weighted networks. For example, both the $k$-constrained and community-based node sets are not significantly different from the completely random node sets.

\begin{table}
  \centering
  \caption{Differences between node set types in terms of average $\ln \varepsilon$ when the network has edge weights and we know the test dynamics. CI: confidence interval.}
  \label{tab:HSD-home-weighted}
  \begin{tabular}{lScc}
    \toprule
    & {Difference} & CI & $p$\\
    \midrule
    Optimized $-$ Random & -6.867 & $[-6.999, -6.735]$ & $< 10^{-7}$\\
    Degree-preserving $-$ Random & -1.298 & $[-1.430, -1.166]$ & $< 10^{-7}$\\
    $k$-constrained $-$ Random & -0.093 & $[-0.225, 0.039]$ & $0.370$\\
    $k$-quantiled $-$ Random & -0.407 & $[-0.539, -0.275]$ & $< 10^{-7}$\\
    $k_{\rm nn}$-constrained $-$ Random & -0.196 & $[-0.328, -0.064]$ & $2.36 \times 10^{-4}$\\
    Community-based $-$ Random & -0.100 & $[-0.232, 0.032]$ & $0.283$\\
    Degree-preserving $-$ Optimized & 5.569 & $[5.437, 5.701]$ & $< 10^{-7}$\\
    $k$-constrained $-$ Optimized & 6.774 & $[6.642, 6.906]$ & $< 10^{-7}$\\
    $k$-quantiled $-$ Optimized & 6.460 & $[6.328, 6.592]$ & $< 10^{-7}$\\
    $k_{\rm nn}$-constrained $-$ Optimized & 6.670 & $[6.538, 6.802]$ & $< 10^{-7}$\\
    Community-based $-$ Optimized & 6.767 & $[6.635, 6.899]$ & $< 10^{-7}$\\
    $k$-constrained $-$ Degree-preserving & 1.205 & $[1.073, 1.337]$ & $< 10^{-7}$\\
    $k$-quantiled $-$ Degree-preserving & 0.891 & $[0.759, 1.023]$ & $< 10^{-7}$\\
    $k_{\rm nn}$-constrained $-$ Degree-preserving & 1.101 & $[0.969, 1.233]$ & $< 10^{-7}$\\
    Community-based $-$ Degree-preserving & 1.198 & $[1.066, 1.330]$ & $< 10^{-7}$\\
    $k$-quantiled $-$ $k$-constrained & -0.314 & $[-0.446, -0.182]$ & $< 10^{-7}$\\
    $k_{\rm nn}$-constrained $-$ $k$-constrained & -0.104 & $[-0.236, 0.028]$ & $0.238$\\
    Community-based $-$ $k$-constrained & -0.007 & $[-0.139, 0.125]$ & $> 0.999$\\
    $k_{\rm nn}$-constrained $-$ $k$-quantiled & 0.210 & $[0.078, 0.343]$ & $5.41 \times 10^{-5}$\\
    Community-based $-$ $k$-quantiled & 0.307 & $[0.175, 0.439]$ & $< 10^{-7}$\\
    Community-based $-$ $k_{\rm nn}$-constrained & 0.097 & $[-0.035, 0.229]$ & $0.318$\\
    \bottomrule
  \end{tabular}
\end{table}
    
The performance of our algorithm on weighted networks is also similar to that on unweighted networks when we do not know the test dynamics. In fact, by running the same analysis as that for the unweighted networks (see section \ref{sec:alternate-dynamics}), we find similar ANOVA results ($R^2 = 0.69$; training dynamics: $df=3$, $F=1,720$, $p < 10^{-7}$; test dynamics: $df=3$, $F=16,430$, $p < 10^{-7}$; network: $df=9$, $F= 1,817$, $p<10^{-7}$; node set type: $df= 2$, $F = 4,808$, $p < 10^{-7}$). The average approximation error for the optimized node sets is 9.67 times smaller than for completely random node sets ($b = -2.269$, $e^b = 0.096$, $p < 10^{-7}$; Table \ref{tab:HSD-other-weighted}) and 3.66 times smaller than degree-preserving node sets (difference in coefficients: $-1.297$, $e^b = 0.273$, $p < 10^{-7}$).

\begin{table}
  \centering
  \caption{Differences between node set types in terms of average $\ln \varepsilon$ when the network has edge weights and we do not know the test dynamics. CI: confidence interval.}
  \label{tab:HSD-other-weighted}
  \begin{tabular}{lScc}
    \toprule
    & {Difference} & CI & $p$\\
    \midrule
    Optimized $-$ Random & -2.269 & $[-2.323, -2.215]$ & $< 10^{-7}$\\
    Degree-preserving $-$ Random & -0.972 & $[-1.027, -0.918]$ & $< 10^{-7}$\\
    Degree-preserving $-$ Optimized & 1.297 & $[1.242, 1.351]$ & $< 10^{-7}$\\
    \bottomrule
  \end{tabular}
\end{table}

\clearpage
\newpage

\section{Sentinel node set optimization on directed networks\label{sec:SIdirected}}

In this section, we show that our sentinel node approximation also performs well for directed networks. For simplicity, we consider the largest weakly connected component of each network and do not consider networks with edge weights.

We collected ten directed networks from the Netzschleuder repository \cite{Netzschleuder}: a freshwater trophic network \cite{thompson2003} (Trophic: $N=109$, $M=717$), a social network of physicians \cite{coleman1957} (Physicians: $N=117$, $M=542$), an email network from a manufacturing company \cite{michalski2011} (Email (manufacturer): $N=167$, $M=5,783$), a dependency network of the Flamingo software \cite{subelj2011b} (Flamingo: $N=228$, $M=497$), a transcription network of the bacterium {\em Escherichia coli} \cite{shenorr2002} ({\em Escherichia coli}: $N=328$, $M=456$), a transcription network of the yeast {\em Saccharomyces cerevisiae} \cite{milo2002} ({\em Saccharomyces cerevisiae}: $N=664$, $M=1,066$), a network of US air carrier flights in 2020 \cite{BTS} (Flights: $N=806$, $M=11,924$), a dependency network of the JUNG software \cite{subelj2011b} (JUNG: $N=879$, $M=2,051$), the directed version of the university email network from the main text \cite{guimera2003} (Email (university): $N=1,133$, $M=10,902$), and a network of air routes preferred by the US Federal Aviation Administration \cite{FAA} (Air route: $N=1,226$, $M=2,613$). We discarded any self-loops and the multiplicity of any multi-edges.

\begin{figure}[b]
  \centering
  \includegraphics[width = 0.55\textwidth]{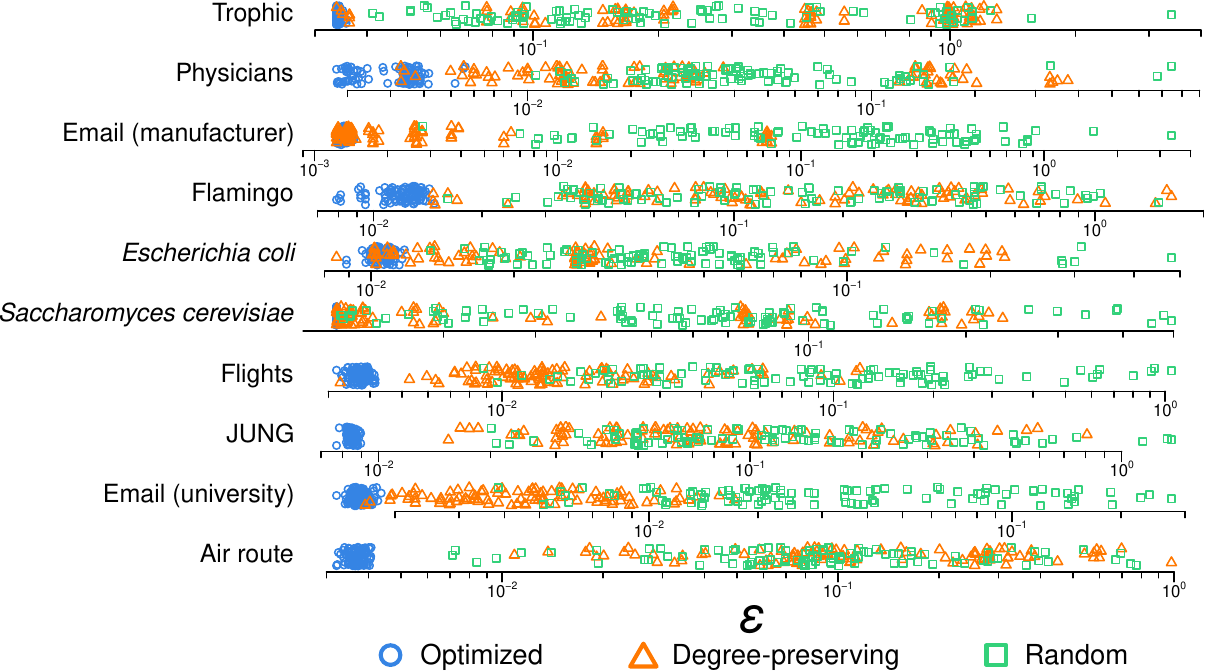}
  \caption{
    Approximation error for the coupled double-well dynamics on directed networks.
  }
  \label{fig:comp-nets-directed}
\end{figure}

We show the approximation error for the coupled double-well dynamics on the ten directed networks in Fig.~\ref{fig:comp-nets-directed}. The figure indicates that our optimization algorithm reliably identifies node sets that obtain relatively small approximation error in each of the ten directed networks.

We verified our observations with an ANOVA (model $R^2= 0.60$; dynamics: $df=3$, $F=5,048$, $p<10^{-7}$; network: $df=9$, $F=1,925$, $p<10^{-7}$; node set type: $df=6$, $F=1,541$, $p<10^{-7}$). On average, the approximation error for optimized node sets was 211 times smaller than completely random node sets ($b = -5.351$, $e^b=0.005$, $p<10^{-7}$) and 62.4 times smaller than degree-preserving node sets (difference in coefficients: $-4.134$, $e^b = 0.016$, $p < 10^{-7}$); see Table~\ref{tab:HSD-home-directed}.

\bigskip

\begin{table}
  \centering
  \caption{Differences between node set types in terms of average $\ln \varepsilon$ when the edges are directed and we know the test dynamics. CI: confidence interval.}
  \label{tab:HSD-home-directed}
  \begin{tabular}{lScc}
    \toprule
    & {Difference} & CI & $p$\\
    \midrule
    Optimized $-$ Random & -5.351 & $[-5.563, -5.139]$ & $< 10^{-7}$\\
    Degree-preserving $-$ Random & -1.217 & $[-1.429, -1.005]$ & $< 10^{-7}$\\
    $k$-constrained $-$ Random & -0.081 & $[-0.293, 0.131]$ & 0.921\\
    $k$-quantiled $-$ Random & -0.362 & $[-0.574, -0.150]$ & $1.00 \times 10^{-5}$\\
    $k_{\rm nn}$-constrained $-$ Random & -0.045 & $[-0.257,  0.167]$ & 0.996\\
    Community-based $-$ Random & 0.347 & $[0.134, 0.561]$ & $3.19 \times 10^{-5}$\\
    Degree-preserving $-$ Optimized & 4.134 & $[3.922, 4.346]$ & $< 10^{-7}$\\
    $k$-constrained $-$ Optimized & 5.270 & $[5.058, 5.482]$ & $< 10^{-7}$\\
    $k$-quantiled $-$ Optimized & 4.989 & $[4.777, 5.201]$ & $< 10^{-7}$\\
    $k_{\rm nn}$-constrained $-$ Optimized & 5.306 & $[5.094, 5.518]$ & $< 10^{-7}$\\
    Community-based $-$ Optimized & 5.699 & $[5.486, 5.912]$ & $< 10^{-7}$\\
    $k$-constrained $-$ Degree-preserving & 1.136 & $[0.924, 1.348]$ & $< 10^{-7}$\\
    $k$-quantiled $-$ Degree-preserving & 0.855 & $[0.643, 1.067]$ & $< 10^{-7}$\\
    $k_{\rm nn}$-constrained $-$ Degree-preserving & 1.172 & $[0.960, 1.384]$ & $< 10^{-7}$\\
    Community-based $-$ Degree-preserving & 1.564 & $[1.351, 1.778]$ & $< 10^{-7}$\\
    $k$-quantiled $-$ $k$-constrained & -0.281 & $[-0.493, -0.069]$ & $1.80 \times 10^{-3}$\\
    $k_{\rm nn}$-constrained $-$ $k$-constrained & 0.036 & -$[0.176, 0.248]$ & 0.999\\
    Community-based $-$ $k$-constrained & 0.428 & $[0.215, 0.641]$ & $10^{-7}$\\
    $k_{\rm nn}$-constrained $-$ $k$-quantiled & 0.317 & $[0.105, 0.529]$ & $2.15 \times 10^{-4}$\\
    Community-based $-$ $k$-quantiled & 0.709 & $[0.496, 0.922]$ & $< 10^{-7}$\\
    Community-based $-$ $k_{\rm nn}$-constrained & 0.393 & $[0.179, 0.606]$ & $1.2 \times 10^{-6}$\\
    \bottomrule
  \end{tabular}
\end{table}

The ANOVA for directed networks in the case when we do not know the test dynamics fits the data reasonably well (model $R^2= 0.52$; training dynamics: $df=3$, $F = 613$, $p<10^{-7}$; test dynamics: $df=3$, $F = 4,701$, $p<10^{-7}$; network: $df=9$, $F = 2,442$, $p<10^{-7}$; node set type: $df= 2$, $F = 353$, $p<10^{-7}$). As is the case for undirected networks, the performance of optimized node sets is worse when the test dynamics is different from the training dynamics, but optimized node sets still outperform all other node sets. Optimized node sets had on average 3.02 times lower approximation error than completely random node sets ($b=-1.104$, $e^b=0.332$, $p<10^{-7}$) and 1.81 times smaller error than degree-preserving node sets (difference in coefficients: $-0.595$, $e^b=0.552$, $p<10^{-7}$); see Table \ref{tab:HSD-other-directed}.

\begin{table}
 \centering
  \caption{Differences between node set types in terms of average $\ln \varepsilon$ when the edges are directed and we do not know the test dynamics. CI: confidence interval.}
  \label{tab:HSD-other-directed}
  \begin{tabular}{lScc}
    \toprule
    & {Difference} & CI & $p$\\
    \midrule
    Optimized $-$ Random & -1.104 & $[-1.201, -1.006]$ & $< 10^{-7}$\\
    Degree-preserving $-$ Random & -0.508 & $[-0.606, -0.411]$ & $< 10^{-7}$\\
    Degree-preserving $-$ Optimized & 0.595 & $[0.498, 0.693]$ & $< 10^{-7}$\\
    \bottomrule
  \end{tabular}
\end{table}

\clearpage

\section{Characterization of individual sentinel nodes\label{sec:node-feature}}

\subsection{Descriptive statistics\label{sub:node-feature-descriptive-statistics}}

We investigated whether or not features of individual nodes contribute to their possibility of being selected as a sentinel node.
We considered six node features. Three of them are conventional centrality measures, i.e., the node's degree, denoted by $k$, closeness centrality, and betweenness centrality \cite{newman2018}. A fourth node feature is
the average nearest neighbor degree of a node, which is given for an $i$th node by 
\begin{equation} \label{eq:knni}
  k_{{\rm nn}, i} = \frac{1}{k_i}\sum_{j=1; j \in \mathcal{N}_i}^{N} k_j
\end{equation}
where $\mathcal{N}_i$ is the set of the $k_i$ neighbors of the $i$th node. 
A fifth node feature is the 
local clustering coefficient of an $i$th node, given by \cite{newman2018}
\begin{equation} \label{eq:Ci}
C_i = \frac{\text{number of }i\text{'s neighbors that are adjacent to each other}}{\frac{1}{2}k_i(k_i-1)}.
\end{equation}
The last node feature is coreness. If a node belongs to the $k$-core but not to the $(k+1)$-core, the coreness of the node is equal to $k$ \cite{Sayama2015book}.

\begin{figure}
  \centering
  \includegraphics[width=\textwidth]{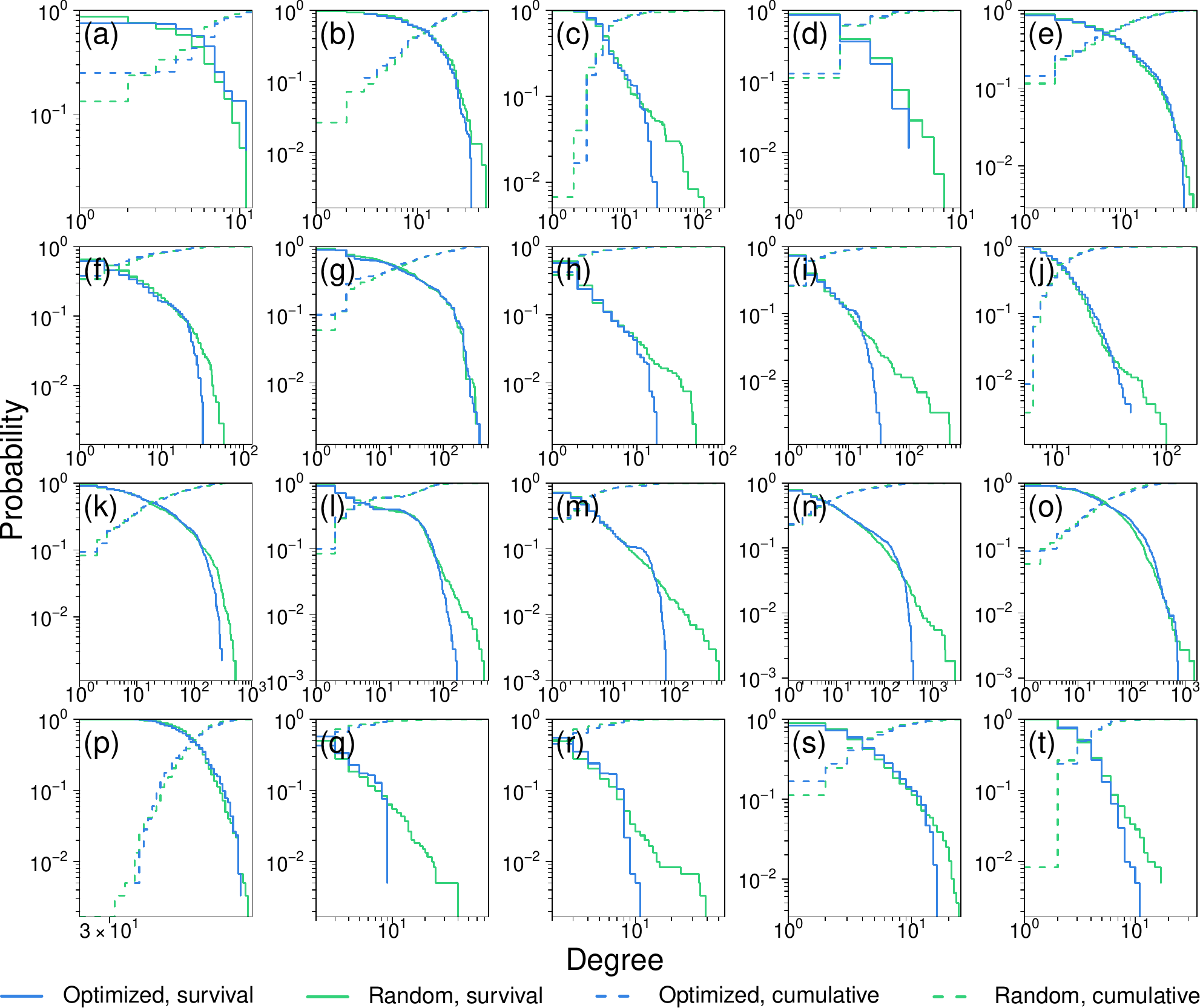}
  \caption{Distribution of the node's degree, $k$, in sentinel and completely random node sets under the coupled double-well dynamics. (a) Dolphin. (b) Proximity. (c) Metabolic. (d) Road. (e) Email. (f) FlyBi. (g) Reactome. (h) Route views. (i) Spanish words. (j) FOLDOC. (k) Tree of life. (\text{$\ell$}) English words. (m) Enron. (n) Marker Cafe. (o) Prosper. (p) ER. (q) BA. (r) HK. (s) GKK. (t) LFR.  The blue and green lines represent sentinel and completely random node sets, respectively. The solid lines represent the survival probability, i.e., the fraction of nodes having the degree larger than $k$. The dashed lines represent the cumulative distribution function, i.e., the fraction of nodes having the degree smaller than or equal to $k$. By definition, the sum of the survival probability and the cumulative distribution function is equal to $1$ at any $k$.
For each network, we generated 100 node sets of each type to generate each distribution shown.}
  \label{fig:dist-k-node}
\end{figure}

\begin{figure}
  \centering
  \includegraphics[width=\textwidth]{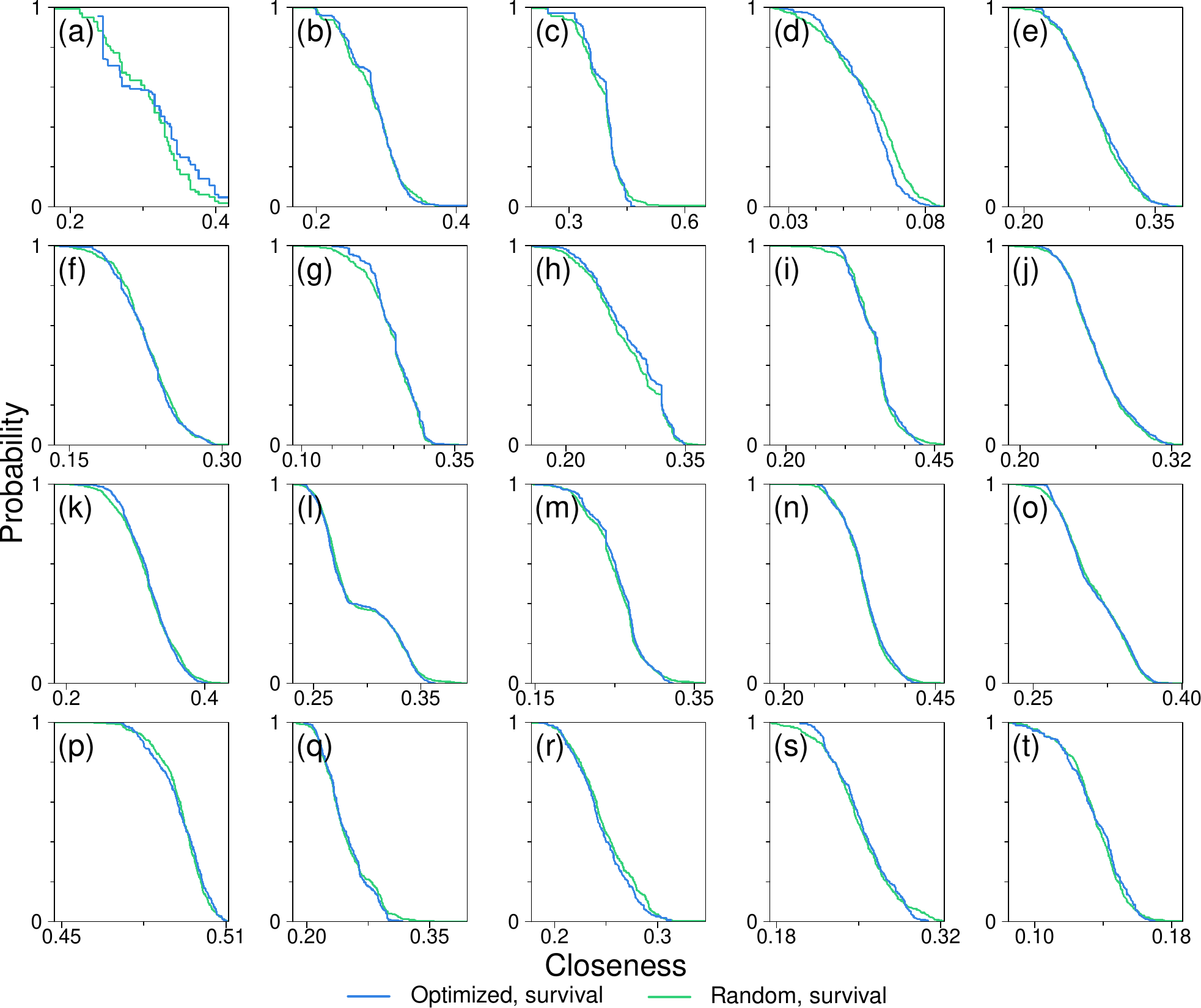}
  \caption{Distribution of the closeness centrality of the nodes in sentinel and completely random node sets under the coupled double-well dynamics. See the caption of Fig.~\ref{fig:dist-k-node} for the legends.}
  \label{fig:dist-closeness-node}
\end{figure}

\begin{figure}
  \centering
  \includegraphics[width=\textwidth]{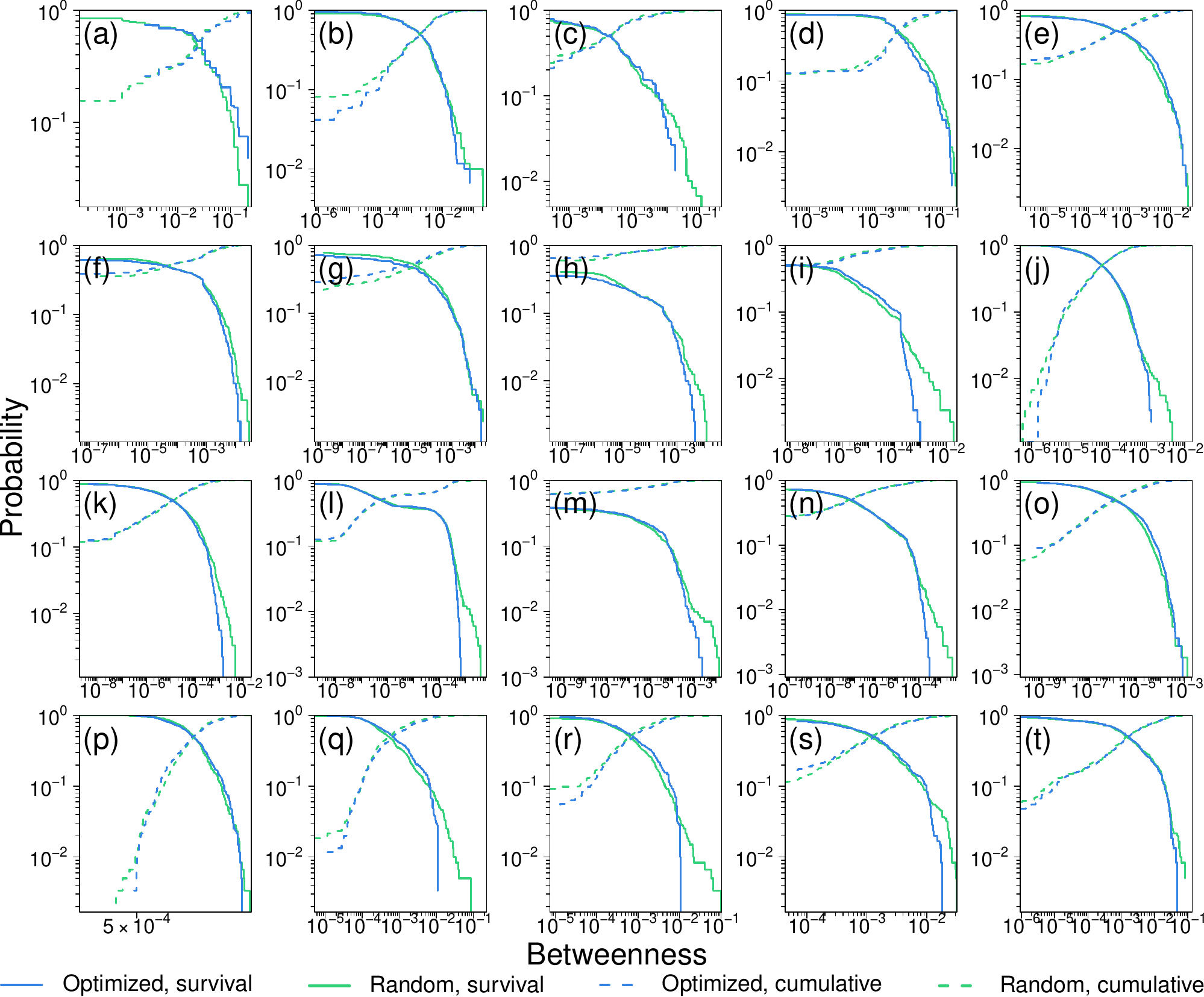}
  \caption{Distribution of the betweenness centrality of the nodes in sentinel and completely random node sets under the coupled double-well dynamics. See the caption of Fig.~\ref{fig:dist-k-node} for the legends.}
  \label{fig:dist-betweenness-node}
\end{figure}

\begin{figure}
  \centering
  \includegraphics[width=\textwidth]{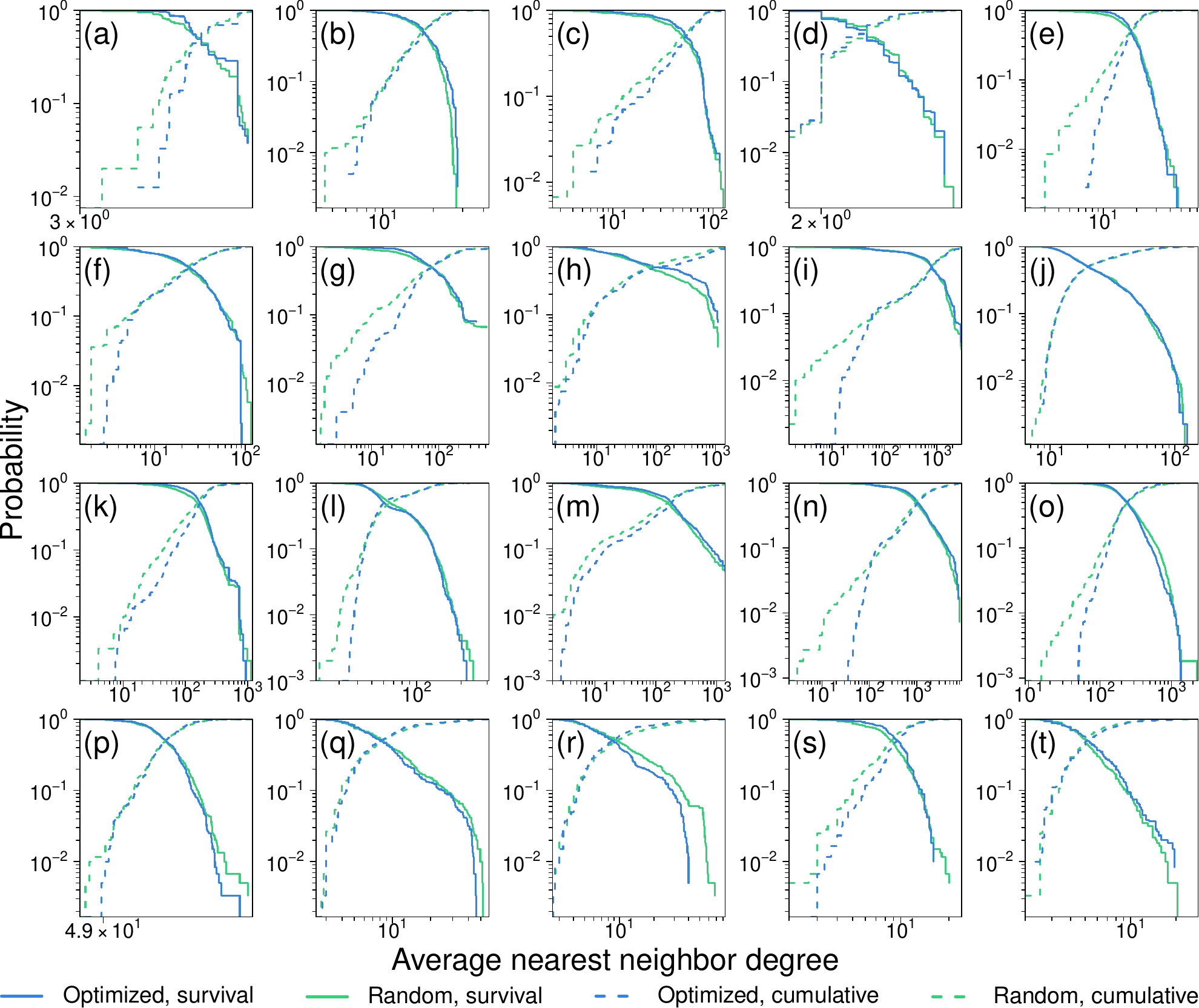}
  \caption{Distribution of the average nearest neighbor degree of the nodes in sentinel and completely random node sets under the coupled double-well dynamics. See the caption of Fig.~\ref{fig:dist-k-node} for the legends.}
  \label{fig:dist-knn-node}
\end{figure}

\begin{figure}
  \centering
  \includegraphics[width=\textwidth]{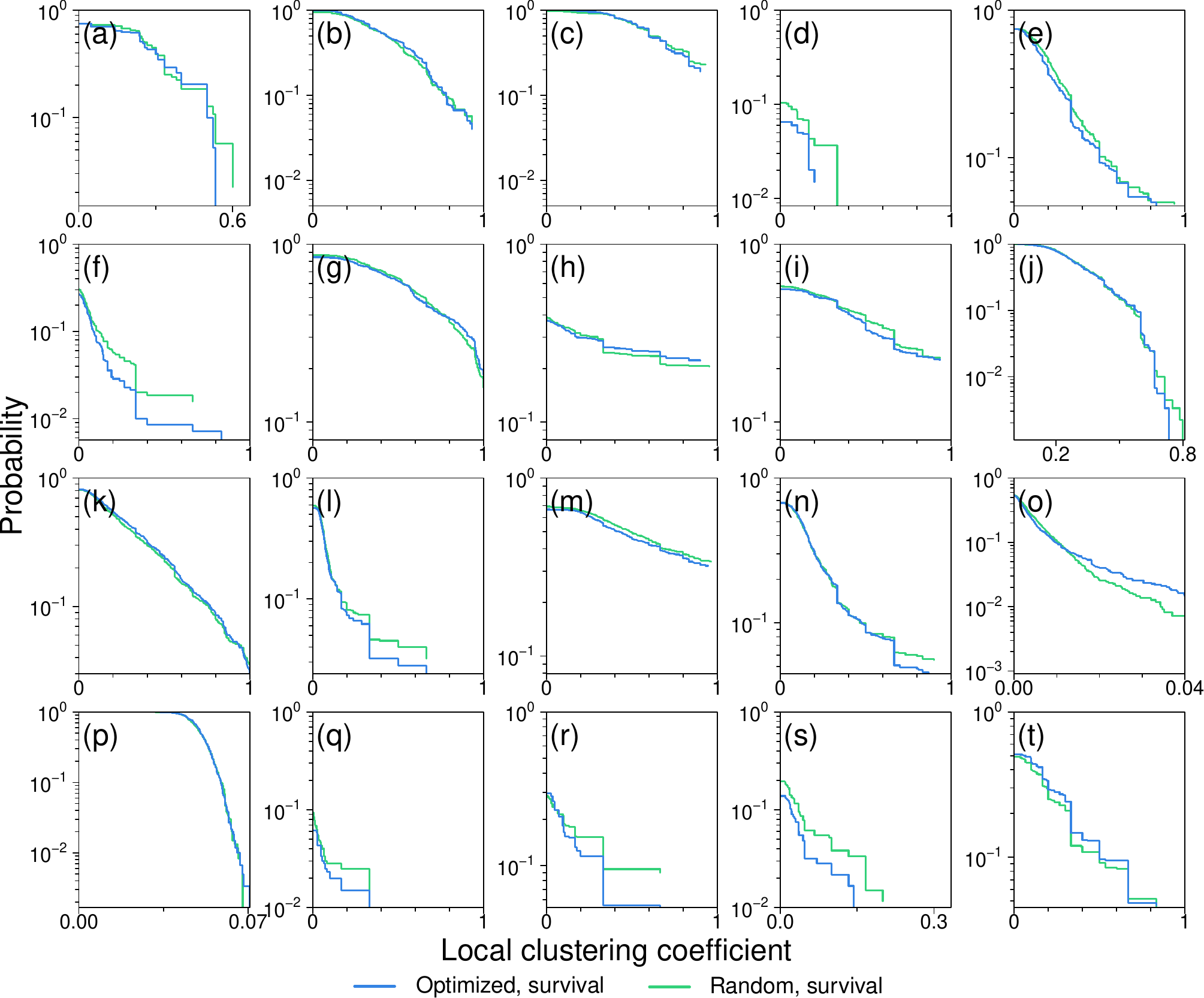}
  \caption{Distribution of the local clustering coefficient of the nodes in sentinel and completely random node sets under the coupled double-well dynamics. See the caption of Fig.~\ref{fig:dist-k-node} for the legends.}
  \label{fig:dist-Ci-node}
\end{figure}

\begin{figure}
  \centering
  \includegraphics[width=\textwidth]{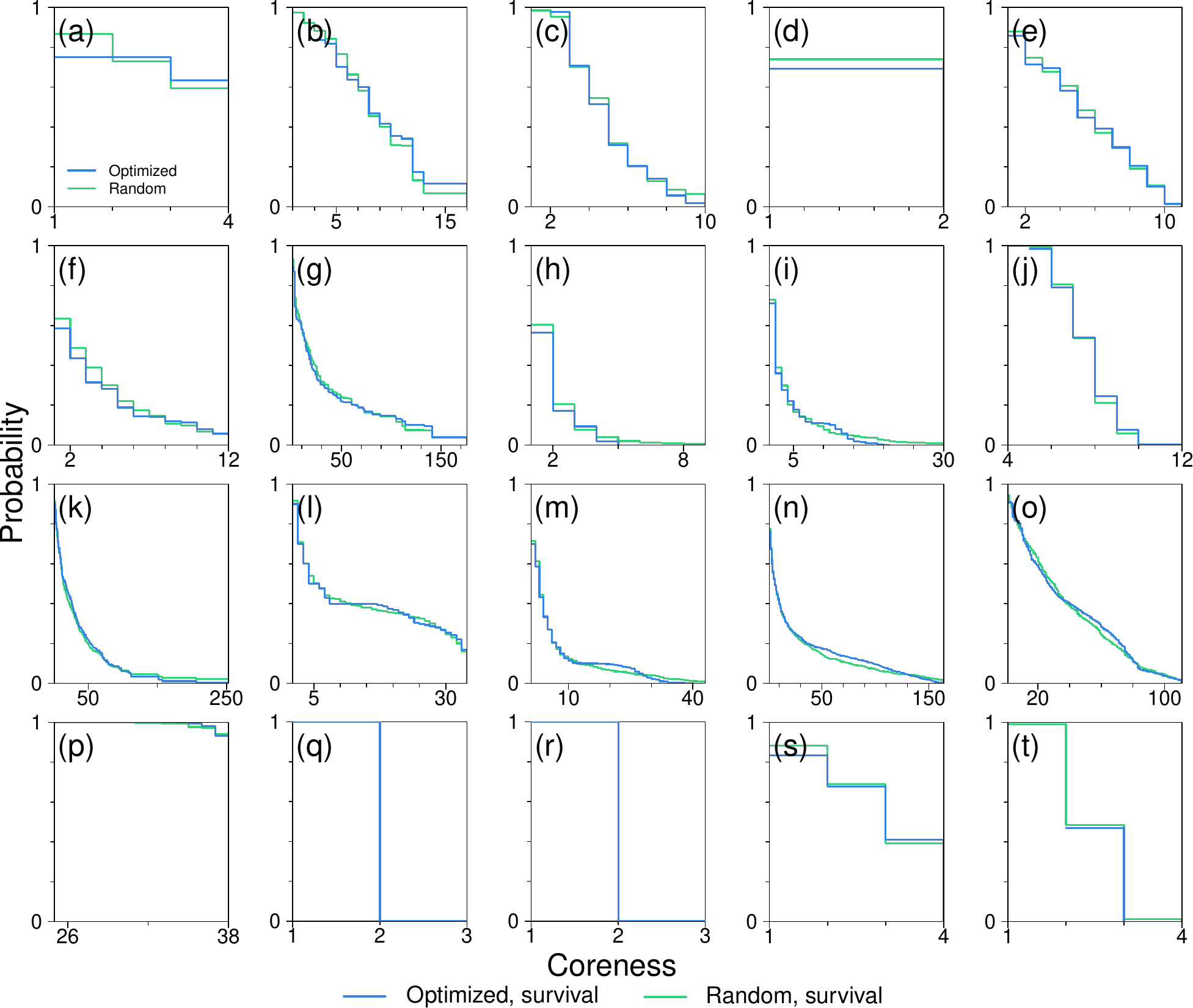}
  \caption{Distribution of the coreness of the nodes in sentinel and completely random node sets under the coupled double-well dynamics. In (q) and (r), all nodes have coreness equal to $2$, and therefore the blue and green lines completely overlap. See the caption of Fig.~\ref{fig:dist-k-node} for the legends.}
  \label{fig:dist-coreness-node}
\end{figure}

We produced 100 optimized node sets by running the optimization algorithm 100 times, for each combination of dynamics and network. For comparison, we also generated 100 completely random node sets for each network. Note that the completely random node sets do not depend on the dynamics. To first casually examine the difference between these two node sets in term of the six node features, for each network, we show the survival probability of $k$, closeness, betweenness, $k_{{\rm nn}, i}$, $C_i$, and coreness in
Figs.~\ref{fig:dist-k-node}, \ref{fig:dist-closeness-node}, \ref{fig:dist-betweenness-node}, \ref{fig:dist-knn-node}, \ref{fig:dist-Ci-node}, and \ref{fig:dist-coreness-node}, respectively. The solid lines show the survival probability, i.e., the fraction of nodes having the node feature values larger than the value shown on the horizontal axis. In some of these figures, we also show the cumulative distribution function by the dashed lines to inspect the difference between the distribution of two node sets at small values of the node feature. Our observations are as follows.
First, Fig.~\ref{fig:dist-k-node} suggests that sentinel nodes tend to avoid hubs for relatively large networks, albeit weakly. Second, Fig.~\ref{fig:dist-k-node} also suggests that sentinel nodes tend to avoid small-degree nodes for relatively small networks, albeit weakly. These results are consistent with the results we report in the main text. Third, Fig.~\ref{fig:dist-knn-node} suggests that $k_{{\rm nn}, i}$ also shows these two tendencies, albeit weakly. 
Qualitatively, the avoidance of small $k_{{\rm nn}, i}$ nodes suggests that our algorithm prefers nodes that ``see'' more of the network, i.e., nodes with neighbors receiving input from a larger portion of the network. 
Fourth, Fig.~\ref{fig:dist-Ci-node} suggests that sentinel node sets tend to avoid nodes with large $C_i$. However, all these results are weak tendencies.

\clearpage

\clearpage

\clearpage

\clearpage

\clearpage

To be more quantitative about the degree of the sentinel nodes, we carried out two additional analyses. First, given a pair of dynamics and network, we computed the $z$-score of the degree of the sentinel nodes relative to the average degree of the network. The $z$-score is given by
$z_1 = (\mu - \overline{k})/s_1$, where $\mu$ and $s_1$ are the mean and standard deviation, respectively, of the degree of the $100n$ sentinel nodes originating from $100$ sentinel node sets; $\overline{k}$ is the average degree of the network. We show the $z_1$ values and the associated $p$ values, denoted by $p_1$, for the coupled double-well, mutualistic species, SIS, and gene regulatory dynamics in Tables~\ref{tab:meandegree-doublewell}, \ref{tab:meandegree-mutualistic}, \ref{tab:meandegree-SIS}, and \ref{tab:meandegree-gene}, respectively. We find that $z_1$ tends to be negative except for the gene regulatory dynamics, for which $z_1$ is close to $0$ for all the networks. The negative $z_1$ values suggest that the sentinel nodes tend to have a smaller degree than the network average, $\overline{k}$. This result is consistent with the casual observation made earlier in this section that sentinel nodes tend to avoid hubs. However, this result is statistically insignificant for all pairs of dynamics and network.

We also compared the average degree of the sentinel nodes in a sentinel node set, denoted by $\overline{k}'$, and the average degree of the network using the $z$-score. The $z$-score is given by
$z_2 = (\mu - \overline{k})/s_2$, where $s_2$ is the standard deviation of $\overline{k}'$ calculated on the basis of the $100$ sentinel node sets. Note that the average of $\overline{k}'$ over the $100$ sentinel node sets is equal to $\mu$ used for the calculation of $z_1$.
We show the $z_2$ values and the associated $p$ values, denoted by $p_2$, in Tables~\ref{tab:meandegree-doublewell}--\ref{tab:meandegree-gene}. After Bonferroni correction, the $z_2$ values are significantly different from $0$ (i.e., $p_2 < 0.05/80 = 6.25 \times 10^{-4}$) for eight out of $80$ pairs of dynamics and network because $s_2$ is substantially smaller than $s_1$ partly due to the averaging over the $n$ sentinel nodes. We obtained $z_2 < 0$ for seven among the eight significant pairs of dynamics and network, supporting our claim that sentinel nodes tend not to be hubs. However, the result that only a small fraction of pairs of dynamics and network show significant differences between $\mu$ and $\overline{k}$ at the node set level suggests that the degree of sentinel nodes is only weakly different from average nodes.

\setlength{\tabcolsep}{3pt}


\begin{table}
\centering
  \caption{The comparison between the average degree of the sentinel (i.e., optimized) node sets, $\mu$, and the average degree of the network, $\overline{k}$, when one obtains the sentinel node sets with the coupled double-well dynamics. We denoted by $s_1$ the standard deviation of the degree based on $100 n$ sentinel nodes. The associated $z$-score and its $p$ value are denoted by $z_1$ and $p_1$, respectively. The quantities $s_2$, $z_2$, and $p_2$ are the same except that the standard deviation $s_2$ is computed on the average degrees of each of the 100 node sets.}
  \label{tab:meandegree-doublewell}
  \begin{tabular}{lSSSSSSSS}
    \toprule
    Network & {$\mu$} & {$\overline{k}$} & {$s_1$} & {$z_1$} & {$p_1$} & {$s_2$} & {$z_2$} & {$p_2$}\\
    \midrule
    Dolphin & 5.663 & 5.129 & 3.369 & 0.158 & 0.874 & 0.186 & 2.869 & 0.004\\
    Proximity & 12.852 & 13.488 & 7.677 & -0.083 & 0.934 & 0.430 & -1.481 & 0.139\\
    Metabolic & 7.192 & 8.940 & 5.290 & -0.331 & 0.741 & 0.561 & -3.116 & 0.002\\
    Road & 2.470 & 2.512 & 1.048 & -0.040 & 0.968 & 0.122 & -0.345 & 0.730\\
    Email & 9.491 & 9.622 & 8.810 & -0.015 & 0.988 & 0.508 & -0.257 & 0.797\\
    FlyBi & 5.626 & 6.254 & 7.220 & -0.087 & 0.931 & 0.544 & -1.154 & 0.248\\
    Reactome & 48.784 & 48.812 & 67.043 & 0.000 & 1.000 & 4.832 & -0.006 & 0.995\\
    Route views & 2.495 & 3.884 & 2.613 & -0.532 & 0.595 & 0.468 & -2.965 & 0.003\\
    Spanish words & 4.343 & 7.449 & 5.535 & -0.561 & 0.575 & 0.550 & -5.647 & {$< 10^{-7}$}\\
    FOLDOC & 13.070 & 13.697 & 6.813 & -0.092 & 0.927 & 0.491 & -1.278 & 0.201\\
    Tree of life & 47.799 & 53.624 & 61.831 & -0.094 & 0.925 & 3.906 & -1.492 & 0.136\\
    English words & 23.936 & 25.687 & 30.469 & -0.057 & 0.954 & 1.180 & -1.484 & 0.138\\
    Enron & 8.306 & 10.732 & 13.738 & -0.177 & 0.860 & 0.908 & -2.670 & 0.008\\
    Marker Cafe & 37.021 & 47.457 & 71.305 & -0.146 & 0.884 & 4.978 & -2.096 & 0.036\\
    Prosper & 75.814 & 74.687 & 107.632 & 0.010 & 0.992 & 8.892 & 0.127 & 0.899\\
    ER & 50.287 & 50.264 & 7.357 & 0.003 & 0.998 & 0.075 & 0.301 & 0.763\\
    BA & 3.683 & 3.992 & 2.248 & -0.137 & 0.891 & 0.135 & -2.286 & 0.022\\
    HK & 3.625 & 3.992 & 2.137 & -0.172 & 0.864 & 0.131 & -2.811 & 0.005\\
    GKK & 5.197 & 5.260 & 3.904 & -0.016 & 0.987 & 0.274 & -0.232 & 0.816\\
    LFR & 3.795 & 3.984 & 1.631 & -0.116 & 0.908 & 0.183 & -1.031 & 0.303\\
    \bottomrule
  \end{tabular}
\end{table}

\begin{table}
\centering
  \caption{The comparison between the average degree of the sentinel node sets and the average degree of the network when one obtains the sentinel node sets with the mutualistic species dynamics. See the caption of Table~\ref{tab:meandegree-doublewell} for the legends.}
  \label{tab:meandegree-mutualistic}
  \begin{tabular}{lSSSSSSSS}
    \toprule
    Network & {$\mu$} & {$\overline{k}$} & {$s_1$} & {$z_1$} & {$p_1$} & {$s_2$} & {$z_2$} & {$p_2$}\\
    \midrule
    Dolphin & 4.968 & 5.129 & 2.597 & -0.062 & 0.950 & 0.224 & -0.722 & 0.470\\
    Proximity & 13.580 & 13.488 & 8.689 & 0.011 & 0.992 & 0.693 & 0.133 & 0.894\\
    Metabolic & 7.213 & 8.940 & 4.003 & -0.431 & 0.666 & 0.320 & -5.403 & {$< 10^{-7}$}\\
    Road & 2.252 & 2.512 & 1.096 & -0.238 & 0.812 & 0.486 & -0.536 & 0.592\\
    Email & 9.203 & 9.622 & 7.589 & -0.055 & 0.956 & 0.580 & -0.723 & 0.469\\
    FlyBi & 5.477 & 6.254 & 6.357 & -0.122 & 0.903 & 0.575 & -1.350 & 0.177\\
    Reactome & 41.960 & 48.812 & 46.191 & -0.148 & 0.882 & 3.641 & -1.882 & 0.060\\
    Route views & 2.575 & 3.884 & 2.107 & -0.621 & 0.534 & 0.354 & -3.702 & {$2.14 \times 10^{-4}$}\\
    Spanish words & 4.536 & 7.449 & 5.982 & -0.487 & 0.626 & 0.600 & -4.860 & {$1.17 \times 10^{-6}$}\\
    FOLDOC & 12.943 & 13.697 & 6.095 & -0.124 & 0.902 & 0.437 & -1.725 & 0.085\\
    Tree of life & 50.164 & 53.624 & 71.893 & -0.048 & 0.962 & 4.650 & -0.744 & 0.457\\
    English words & 24.228 & 25.687 & 31.309 & -0.047 & 0.963 & 1.689 & -0.864 & 0.388\\
    Enron & 8.425 & 10.732 & 12.709 & -0.182 & 0.856 & 1.011 & -2.282 & 0.023\\
    Marker Cafe & 31.080 & 47.457 & 49.619 & -0.330 & 0.741 & 3.628 & -4.514 & {$6.36 \times 10^{-6}$}\\
    Prosper & 73.640 & 74.687 & 103.538 & -0.010 & 0.992 & 8.044 & -0.130 & 0.896\\
    ER & 50.297 & 50.264 & 7.390 & 0.004 & 0.996 & 0.069 & 0.471 & 0.638\\
    BA & 3.752 & 3.992 & 2.636 & -0.091 & 0.927 & 0.233 & -1.033 & 0.302\\
    HK & 3.678 & 3.992 & 2.294 & -0.137 & 0.891 & 0.200 & -1.569 & 0.117\\
    GKK & 5.182 & 5.260 & 4.131 & -0.019 & 0.985 & 0.356 & -0.221 & 0.825\\
    LFR & 3.858 & 3.984 & 1.812 & -0.069 & 0.945 & 0.208 & -0.603 & 0.547\\
    \bottomrule
  \end{tabular}
\end{table}

\begin{table}
\centering
  \caption{The comparison between the average degree of the sentinel node sets and the average degree of the network when one obtains the sentinel node sets with the SIS dynamics. See the caption of Table~\ref{tab:meandegree-doublewell} for the legends.}
  \label{tab:meandegree-SIS}
  \begin{tabular}{lSSSSSSSS}
    \toprule
    Network & {$\mu$} & {$\overline{k}$} & {$s_1$} & {$z_1$} & {$p_1$} & {$s_2$} & {$z_2$} & {$p_2$}\\
    \midrule
    Dolphin & 5.333 & 5.129 & 3.163 & 0.064 & 0.949 & 0.316 & 0.644 & 0.520\\
    Proximity & 13.377 & 13.488 & 8.342 & -0.013 & 0.989 & 1.156 & -0.096 & 0.923\\
    Metabolic & 6.318 & 8.940 & 4.400 & -0.596 & 0.551 & 0.915 & -2.865 & 0.004\\
    Road & 2.362 & 2.512 & 0.869 & -0.173 & 0.863 & 0.174 & -0.863 & 0.388\\
    Email & 9.193 & 9.622 & 8.279 & -0.052 & 0.959 & 1.035 & -0.415 & 0.678\\
    FlyBi & 4.741 & 6.254 & 5.757 & -0.263 & 0.793 & 0.964 & -1.569 & 0.117\\
    Reactome & 46.378 & 48.812 & 67.409 & -0.036 & 0.971 & 9.073 & -0.268 & 0.788\\
    Route views & 2.058 & 3.884 & 1.137 & -1.606 & 0.108 & 0.205 & -8.899 & {$< 10^{-7}$}\\
    Spanish words & 4.354 & 7.449 & 5.716 & -0.541 & 0.588 & 0.488 & -6.346 & {$< 10^{-7}$}\\
    FOLDOC & 12.567 & 13.697 & 6.613 & -0.171 & 0.864 & 1.033 & -1.094 & 0.274\\
    Tree of life & 50.078 & 53.624 & 74.450 & -0.048 & 0.962 & 9.098 & -0.390 & 0.697\\
    English words & 24.025 & 25.687 & 32.452 & -0.051 & 0.959 & 1.929 & -0.862 & 0.389\\
    Enron & 7.804 & 10.732 & 13.200 & -0.222 & 0.824 & 1.620 & -1.807 & 0.071\\
    Marker Cafe & 36.568 & 47.457 & 73.117 & -0.149 & 0.882 & 5.959 & -1.827 & 0.068\\
    Prosper & 71.658 & 74.687 & 101.309 & -0.030 & 0.976 & 13.350 & -0.227 & 0.821\\
    ER & 50.117 & 50.264 & 6.496 & -0.023 & 0.982 & 0.309 & -0.476 & 0.634\\
    BA & 3.410 & 3.992 & 2.110 & -0.276 & 0.783 & 0.351 & -1.659 & 0.097\\
    HK & 3.345 & 3.992 & 2.002 & -0.323 & 0.747 & 0.345 & -1.875 & 0.061\\
    GKK & 5.000 & 5.260 & 3.598 & -0.072 & 0.942 & 0.565 & -0.461 & 0.645\\
    LFR & 3.720 & 3.984 & 1.736 & -0.152 & 0.879 & 0.362 & -0.730 & 0.465\\
    \bottomrule
  \end{tabular}
\end{table}

\begin{table}
\centering
  \caption{The comparison between the average degree of the sentinel node sets and the average degree of the network when one obtains the sentinel node sets with the gene regulatory dynamics. See the caption of Table~\ref{tab:meandegree-doublewell} for the legends. For seven networks, all optimized node sets had the same average degree although the exact degree sequence is different for different optimized node sets in general. For these seven networks, we mark $z_2$ and $p_2$ ``NA'' because $s_2 = 0$ such that one cannot calculate $z_2$ or $p_2$. In these cases, $\mu$ is extremely close to $\overline{k}$, suggesting that choosing a node set whose $\mu$ is close to $\overline{k}$ may be a key to successful sentinel node approximation.}
  \label{tab:meandegree-gene}
  \begin{tabular}{lSSSSSSSS}
    \toprule
    Network & {$\mu$} & {$\overline{k}$} & {$s_1$} & {$z_1$} & {$p_1$} & {$s_2$} & {$z_2$} & {$p_2$}\\
    \midrule
    Dolphin & 5.088 & 5.129 & 2.858 & -0.015 & 0.988 & 0.120 & -0.347 & 0.729\\
    Proximity & 13.500 & 13.488 & 7.097 & 0.002 & 0.999 & 0.000 & {NA} & {NA}\\
    Metabolic & 9.002 & 8.940 & 8.728 & 0.007 & 0.994 & 0.017 & 3.676 & {$2.37 \times 10^{-4}$}\\
    Road & 2.497 & 2.512 & 0.977 & -0.016 & 0.987 & 0.125 & -0.123 & 0.902\\
    Email & 9.580 & 9.622 & 7.973 & -0.005 & 0.996 & 0.034 & -1.239 & 0.215\\
    FlyBi & 6.253 & 6.254 & 6.629 & 0.000 & 1.000 & 0.076 & -0.010 & 0.992\\
    Reactome & 48.796 & 48.812 & 61.184 & 0.000 & 1.000 & 0.061 & -0.265 & 0.791\\
    Route views & 3.880 & 3.884 & 4.609 & -0.001 & 0.999 & 0.056 & -0.069 & 0.945\\
    Spanish words & 7.436 & 7.449 & 14.098 & -0.001 & 0.999 & 0.034 & -0.405 & 0.686\\
    FOLDOC & 13.667 & 13.697 & 6.763 & -0.005 & 0.996 & 0.000 & {NA} & {NA}\\
    Tree of life & 53.640 & 53.624 & 68.758 & 0.000 & 1.000 & 0.048 & 0.328 & 0.743\\
    English words & 25.700 & 25.687 & 31.859 & 0.000 & 1.000 & 0.000 & {NA} & {NA}\\
    Enron & 10.642 & 10.732 & 17.052 & -0.005 & 0.996 & 0.059 & -1.526 & 0.127\\
    Marker Cafe & 47.455 & 47.457 & 81.439 & 0.000 & 1.000 & 0.020 & -0.084 & 0.933\\
    Prosper & 74.725 & 74.687 & 82.797 & 0.000 & 1.000 & 0.013 & 2.984 & 0.003\\
    ER & 50.333 & 50.264 & 6.762 & 0.010 & 0.992 & 0.000 & {NA} & {NA}\\
    BA & 4.000 & 3.992 & 2.417 & 0.003 & 0.997 & 0.000 & {NA} & {NA}\\
    HK & 4.000 & 3.992 & 2.563 & 0.003 & 0.998 & 0.000 & {NA} & {NA}\\
    GKK & 5.305 & 5.260 & 3.884 & 0.012 & 0.991 & 0.063 & 0.711 & 0.477\\
    LFR & 4.000 & 3.984 & 1.622 & 0.010 & 0.992 & 0.000 & {NA} & {NA}\\
    \bottomrule
  \end{tabular}
\end{table}
\setlength{\tabcolsep}{10pt}

\newpage
\clearpage

\begin{figure}
  \centering
  \includegraphics[width=\textwidth]{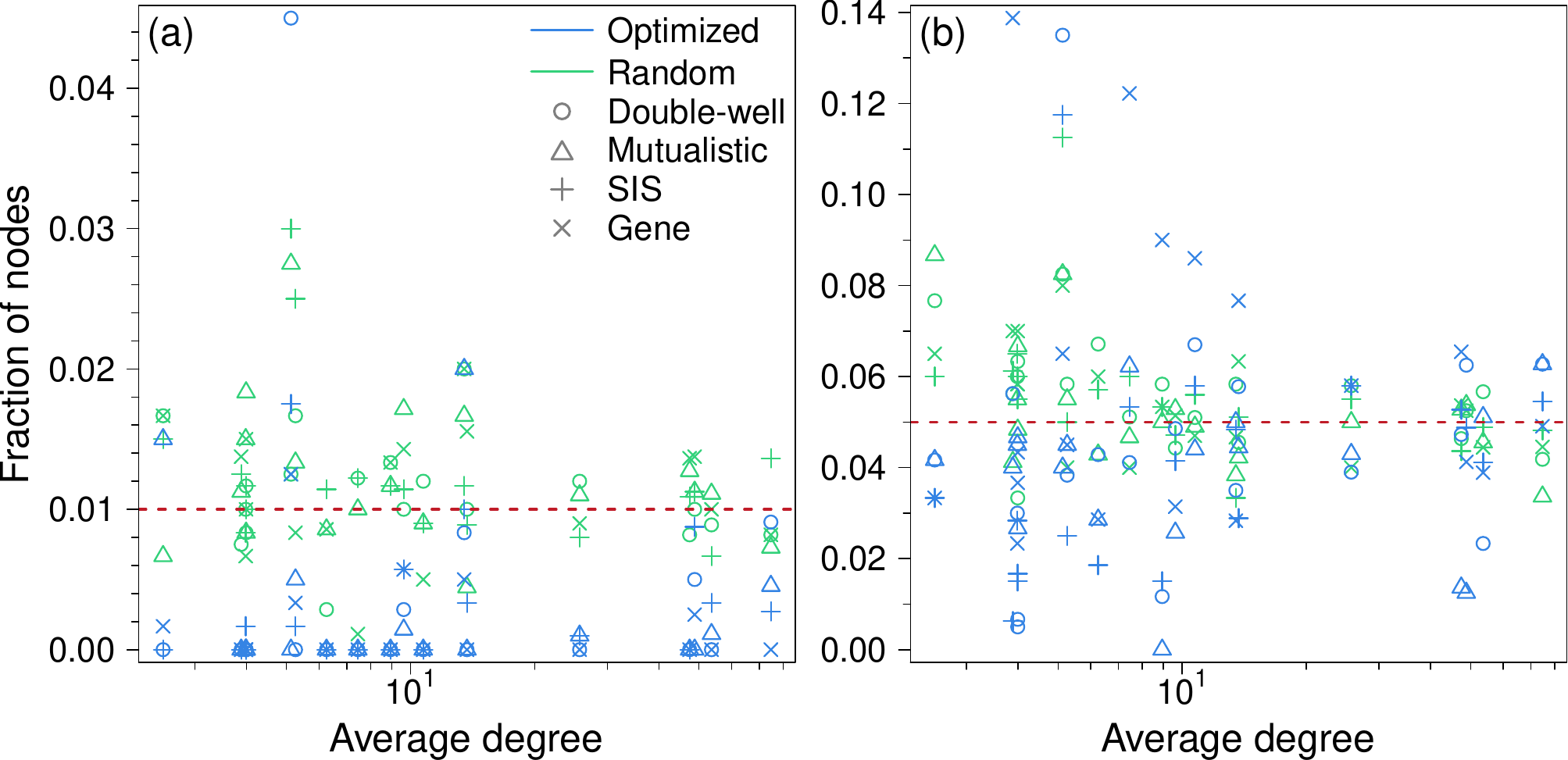}
  \caption{Fraction of sentinel nodes and completely randomly selected nodes that are hubs. Each symbol corresponds to a pair of dynamics and network, and represents the fraction of nodes that are within (a) top 1\% or (b) top 5\% in terms of the degree. We used the $100$ sentinel node sets or $100$ completely random node sets to compute the respective fraction of nodes. When the same node is selected multiple times as sentinel node, for example, in the $100$ node sets, we regarded the multiple appearances of the same node as different instances when computing the fraction. The dashed lines represent the expectation for the completely random node set. We excluded the ER network because it has a narrow degree distribution.}
  \label{fig:ns-count-extreme}
\end{figure}

The second additional analysis aims to quantify the avoidance of hubs by sentinel nodes. We counted the sentinel nodes and the nodes in completely random node sets that are in the top 1\% or 5\% in terms of the degree. We show the fraction of such nodes for each pair of dynamics and network, separately for sentinel nodes and completely randomly selected nodes in Fig.~\ref{fig:ns-count-extreme}.
The blue symbols in Fig.~\ref{fig:ns-count-extreme}a represent the fraction of the sentinel nodes in the top 1\% in terms of the degree. The green symbols represent the same fraction for completely randomly selected nodes. Figure~\ref{fig:ns-count-extreme}b shows the same fractions when the threshold is the top 5\%. As expected, we find that the fraction for completely randomly selected nodes is distributed around its expected value (i.e., 1\% and 5\% in Fig.~\ref{fig:ns-count-extreme}a and~\ref{fig:ns-count-extreme}b, respectively). In contrast, the sentinel nodes tend to be non-hubs across the pairs of dynamics and network, supporting our casual observations made in Fig.~3 in the main text and Fig.~\ref{fig:dist-k-node}. We also find that the tendency for sentinel nodes to avoid hubs is stronger when we consider the top 1\% nodes as hubs than the top 5\%. Quantitatively, there are more completely randomly selected nodes than sentinel nodes in the top 1\% group (i.e., the green symbol above the corresponding blue symbol in Fig.~\ref{fig:ns-count-extreme}a) in 93.4\% of the pairs of dynamics and network.
%
%
The same number is 61.8\% in the top 5\% group.
%
%

\newpage
\clearpage

\subsection{Search for node features that would make nodes sentinel}

Next, we built a random forest regression, i.e., the random forest version of a Poisson regression for count data.
The goal of this regression analysis is to find node features that contribute to the frequency of the node to be selected as sentinel.
The random forest model uses Poisson deviance as the criterion for finding splits. We used 100 trees, each with a random subset of three features. In each tree, nodes were split until leaves were homogeneous or contained at most two data points. 
For a given network, we used 90\% of the data as training set and the remaining 10\% as test set; each data point is a pair of a node and a dynamics. The dependent variable for each data point in the regression is the count of how often a node is selected as sentinel node in the 100 sentinel node sets for a given pair of dynamics and network. The independent variables are the six node features, which are numeric variables, and the dymamics, which is a categorical variable. We used the implementation in Python package scikit-learn version 1.6.0. We obtained the coefficient of determination, denoted by $R^2$, as the goodness of fit measure. It is calculated as $1 - (\text{sum of squared errors})/(\text{variance of the dependent variable})$ for the test data. The model perfectly fits the data if $R^2 = 1$. The model is guessing completely randomly if $R^2 = 0$. Note that $R^2$ can be negative, which contrasts to the case of conventional linear regressions. Another output of the algorithm is the variable importance~\cite{Breiman2001MachLearn, Hastie2009book}, which is a measure of the effect of an independent variable on model fit. We calculate it for each independent variable using the training data. We ran the analysis separately for each network because it is not valid to compare node feature values across networks with different numbers of nodes, $N$. We excluded the ER network because all the nodes are statistically the same and node feature values are narrowly distributed for the ER network. 

We show the results of the random forest regression in Table~\ref{tab:rf}. We find that $R^2$ is small, which supports the idea that these popular features of individual nodes are not helpful for indicating whether nodes tend to be selected as sentinel nodes. This result is consistent with those shown in Figs.~\ref{fig:dist-k-node}--\ref{fig:dist-coreness-node}, which reveal only modest differences between the nodes in the sentinel and completely random node sets. Table~\ref{tab:rf} indicates that the dynamics is a stronger determinant of sentinel node selection than node features in terms of the importance measure, especially for large networks. However, we should not over-interpret the importance values, including their differences across node features, because the model fit is poor.

\setlength{\tabcolsep}{2.5pt}
\begin{table}
\centering
  \caption{Model performance and variable importance for a random forest model fitted to 19 networks. Close: closeness centrality. Betw: betweenness centrality. LC: local clustering coefficient. Core: coreness. The variable importance of the coreness is $0$ for the HK and BA networks because all nodes have the same coreness value (i.e., $=2$) in these networks.}
  \label{tab:rf}
  \begin{tabular}{lS[table-format=5.0]S[table-format=3.3]*7{S[table-format=1.3]}}
    \toprule
    Network & {$N$} & {$R^2$} & {$k$} & {Close} & {Betw} & {$k_{\rm nn}$} & {LC} & {Core} & {Dynamics}\\
    \midrule
    Dolphin & 62 & -0.115 & 0.180 & 0.237 & 0.265 & 0.245 & 0.204 & 0.138 & 0.722\\
    Proximity & 410 & 0.030 & 0.435 & 0.269 & 0.257 & 0.276 & 0.238 & 0.268 & 0.477\\
    Metabolic & 453 & -0.383 & 0.197 & 0.229 & 0.278 & 0.234 & 0.161 & 0.111 & 0.354\\
    GKK & 949 & -0.114 & 0.242 & 0.402 & 0.347 & 0.380 & 0.064 & 0.149 & 0.442\\
    LFR & 998 & -0.058 & 0.240 & 0.359 & 0.398 & 0.300 & 0.189 & 0.108 & 0.396\\
    HK & 1000 & -0.007 & 0.321 & 0.369 & 0.400 & 0.334 & 0.163 & 0.000 & 0.482\\
    BA & 1000 & 0.063 & 0.314 & 0.393 & 0.432 & 0.388 & 0.042 & 0.000 & 0.485\\
    Road & 1039 & -0.111 & 0.249 & 0.510 & 0.408 & 0.353 & 0.034 & 0.121 & 0.460\\
    Email & 1133 & 0.029 & 0.229 & 0.349 & 0.289 & 0.426 & 0.219 & 0.222 & 0.457\\
    FlyBi & 2705 & 0.013 & 0.209 & 0.319 & 0.257 & 0.308 & 0.081 & 0.188 & 0.323\\
    Reactome & 5973 & -0.054 & 0.161 & 0.167 & 0.130 & 0.184 & 0.122 & 0.182 & 0.203\\
    Route views & 6474 & -0.030 & 0.111 & 0.247 & 0.124 & 0.300 & 0.102 & 0.091 & 0.199\\
    Spanish words & 11558 & -0.038 & 0.153 & 0.185 & 0.161 & 0.192 & 0.133 & 0.165 & 0.158\\
    FOLDOC & 13356 & -0.048 & 0.198 & 0.298 & 0.260 & 0.264 & 0.200 & 0.107 & 0.175\\
    Tree of Life & 16415 & -0.038 & 0.195 & 0.256 & 0.197 & 0.279 & 0.154 & 0.200 & 0.147\\
    English words & 23132 & -0.031 & 0.207 & 0.317 & 0.270 & 0.255 & 0.130 & 0.189 & 0.171\\
    Enron & 33696 & -0.015 & 0.117 & 0.150 & 0.100 & 0.143 & 0.123 & 0.130 & 0.115\\
    Marker Cafe & 69317 & -0.022 & 0.195 & 0.253 & 0.179 & 0.177 & 0.125 & 0.198 & 0.120\\
    Propser & 89171 & -0.011 & 0.248 & 0.211 & 0.208 & 0.154 & 0.084 & 0.242 & 0.122\\
    \bottomrule
  \end{tabular}
\end{table}
\setlength{\tabcolsep}{10pt}

We further investigated whether node features inform the approximation error directly.
To this end, for each pair of dynamics and network, we first generated $\lfloor 20 N/n \rfloor$ completely random node sets.
With this number, each node is contained in a completely random node set $20$ times on average.
Second, we measured the approximation error, $\varepsilon$, for each completely random node set.
Third, for each $i$th node, we averaged $\varepsilon$ over all the completely random node sets to which the $i$th node belongs.
Because we ensured that each node belongs to $20$ completely random node sets on average, each node belonged to at least some completely random node sets such that we were able compute to the average $\varepsilon$ for each node.
Finally, we regressed $\ln \varepsilon$ on the six node features we have been considering using the random forest model.
We separately built the random forest model for the different dynamics because the $\varepsilon$ value is not comparable across the dynamics.

We show the regression results for the coupled double-well, mutualistic species, SIS, gene-regulatory dynamics in Tables~\ref{tab:rf-error-doublewell}, \ref{tab:rf-error-mutualistic}, \ref{tab:rf-error-SIS}, and \ref{tab:rf-error-genereg}, respectively. We find that the $R^2$ values are overall small. There are only two pairs of dynamics and network among the 76 pairs that yielded $R^2 > 0.3$, and only five pairs that yielded $0.2 < R^2 \le 0.3$. The average of $R^2$ over the 76 pairs is $0.04$. The present results further strengthen our finding that features of individual nodes are not sufficiently informative for constructing high-quality sentinel node sets.


\newpage
\clearpage

\setlength{\tabcolsep}{2.5pt}
\begin{table}
\centering
  \caption{Model performance and variable importance for a random forest model predicting the approximation error for completely random node sets using the coupled double-well dynamics on 19 networks. The random forest model uses $\ln \varepsilon$ as the dependent variable and squared error as the criterion for finding splits. The regression was otherwise the same as described for Table \ref{tab:rf}. See the caption of Table \ref{tab:rf} for the legends.}
  \label{tab:rf-error-doublewell}
  \begin{tabular}{lS[table-format=5.0]S[table-format=3.3]*6{S[table-format=1.3]}}
    \toprule
    Network & {$N$} & {$R^2$} & {$k$} & {Close} & {Betw} & {$k_{\rm nn}$} & {LC} & {Core}\\
    \midrule
    Dolphin & 62 & -0.169 & 0.179 & 0.164 & 0.283 & 0.079 & 0.049 & 0.089\\
    Proximity & 410 & 0.354 & 0.334 & 0.227 & 0.255 & 0.233 & 0.157 & 0.368\\
    Metabolic & 453 & -0.230 & 0.203 & 0.226 & 0.261 & 0.206 & 0.267 & 0.113\\
    GKK & 949 & 0.163 & 0.382 & 0.392 & 0.396 & 0.314 & 0.060 & 0.189\\
    LFR & 998 & -0.121 & 0.460 & 0.438 & 0.411 & 0.442 & 0.237 & 0.104\\
    HK & 1000 & -0.024 & 0.315 & 0.388 & 0.579 & 0.347 & 0.051 & 0.000\\
    BA & 1000 & 0.174 & 0.327 & 0.348 & 0.461 & 0.313 & 0.130 & 0.000\\
    Road & 1039 & -0.009 & 0.389 & 0.610 & 0.545 & 0.463 & 0.076 & 0.121\\
    Email & 1133 & 0.122 & 0.326 & 0.433 & 0.317 & 0.373 & 0.237 & 0.216\\
    FlyBi & 2705 & -0.095 & 0.377 & 0.397 & 0.344 & 0.404 & 0.123 & 0.246\\
    Reactome & 5973 & 0.033 & 0.323 & 0.209 & 0.148 & 0.231 & 0.156 & 0.237\\
    Route views & 6474 & 0.111 & 0.219 & 0.340 & 0.244 & 0.347 & 0.114 & 0.166\\
    Spanish words & 11558 & 0.041 & 0.226 & 0.378 & 0.279 & 0.390 & 0.216 & 0.218\\
    FOLDOC & 13356 & 0.141 & 0.550 & 0.485 & 0.515 & 0.442 & 0.422 & 0.316\\
    Tree of Life & 16415 & 0.079 & 0.578 & 0.485 & 0.406 & 0.401 & 0.358 & 0.460\\
    English words & 23132 & 0.004 & 0.438 & 0.647 & 0.635 & 0.578 & 0.327 & 0.370\\
    Enron & 33696 & 0.036 & 0.364 & 0.390 & 0.220 & 0.374 & 0.276 & 0.279\\
    Marker Cafe & 69317 & 0.109 & 0.436 & 0.509 & 0.486 & 0.455 & 0.357 & 0.443\\
    Propser & 89171 & 0.052 & 0.635 & 0.607 & 0.620 & 0.526 & 0.310 & 0.576\\
    \bottomrule
  \end{tabular}
\end{table}

\newpage
\clearpage

\begin{table}
\centering
  \caption{Model performance and variable importance for a random forest model predicting the approximation error for completely random node sets using the mutualistic species dynamics on 19 networks. See the caption of Table \ref{tab:rf} for the legends.}
  \label{tab:rf-error-mutualistic}
  \begin{tabular}{lS[table-format=5.0]S[table-format=3.3]*6{S[table-format=1.3]}}
    \toprule
    Network & {$N$} & {$R^2$} & {$k$} & {Close} & {Betw} & {$k_{\rm nn}$} & {LC} & {Core}\\
    \midrule
    Dolphin & 62 & 0.425 & 0.214 & 0.146 & 0.345 & 0.169 & 0.107 & 0.072\\
    Proximity & 410 & 0.224 & 0.408 & 0.267 & 0.292 & 0.263 & 0.182 & 0.333\\
    Metabolic & 453 & -0.084 & 0.277 & 0.258 & 0.310 & 0.183 & 0.201 & 0.113\\
    GKK & 949 & 0.154 & 0.365 & 0.440 & 0.450 & 0.358 & 0.081 & 0.174\\
    LFR & 998 & -0.088 & 0.425 & 0.507 & 0.382 & 0.308 & 0.317 & 0.103\\
    HK & 1000 & -0.078 & 0.318 & 0.415 & 0.581 & 0.357 & 0.059 & 0.000\\
    BA & 1000 & 0.048 & 0.279 & 0.409 & 0.506 & 0.368 & 0.157 & 0.000\\
    Road & 1039 & -0.141 & 0.360 & 0.791 & 0.519 & 0.417 & 0.058 & 0.155\\
    Email & 1133 & 0.181 & 0.352 & 0.491 & 0.341 & 0.379 & 0.279 & 0.242\\
    FlyBi & 2705 & -0.021 & 0.442 & 0.362 & 0.327 & 0.364 & 0.113 & 0.238\\
    Reactome & 5973 & 0.049 & 0.298 & 0.218 & 0.154 & 0.235 & 0.146 & 0.228\\
    Route views & 6474 & 0.065 & 0.172 & 0.350 & 0.257 & 0.353 & 0.115 & 0.158\\
    Spanish words & 11558 & -0.009 & 0.238 & 0.393 & 0.293 & 0.385 & 0.224 & 0.232\\
    FOLDOC & 13356 & 0.126 & 0.545 & 0.495 & 0.535 & 0.457 & 0.422 & 0.302\\
    Tree of Life & 16415 & 0.079 & 0.564 & 0.479 & 0.413 & 0.404 & 0.351 & 0.471\\
    English words & 23132 & 0.021 & 0.454 & 0.651 & 0.655 & 0.569 & 0.328 & 0.375\\
    Enron & 33696 & 0.049 & 0.378 & 0.395 & 0.228 & 0.361 & 0.272 & 0.280\\
    Marker Cafe & 69317 & 0.112 & 0.460 & 0.514 & 0.503 & 0.451 & 0.365 & 0.466\\
    Propser & 89171 & 0.048 & 0.647 & 0.609 & 0.625 & 0.519 & 0.314 & 0.590\\
    \bottomrule
  \end{tabular}
\end{table}

\newpage
\clearpage

\begin{table}
\centering
  \caption{Model performance and variable importance for a random forest model predicting the approximation error for completely random node sets using the SIS dynamics on 19 networks. See the caption of Table \ref{tab:rf} for the legends.}
  \label{tab:rf-error-SIS}
  \begin{tabular}{lS[table-format=5.0]S[table-format=3.3]*6{S[table-format=1.3]}}
    \toprule
    Network & {$N$} & {$R^2$} & {$k$} & {Close} & {Betw} & {$k_{\rm nn}$} & {LC} & {Core}\\
    \midrule
    Dolphin & 62 & 0.160 & 0.225 & 0.193 & 0.342 & 0.169 & 0.099 & 0.085\\
    Proximity & 410 & 0.266 & 0.339 & 0.221 & 0.252 & 0.238 & 0.187 & 0.384\\
    Metabolic & 453 & -0.200 & 0.209 & 0.289 & 0.210 & 0.251 & 0.244 & 0.129\\
    GKK & 949 & 0.059 & 0.405 & 0.431 & 0.460 & 0.364 & 0.067 & 0.201\\
    LFR & 998 & -0.089 & 0.594 & 0.400 & 0.398 & 0.417 & 0.220 & 0.109\\
    HK & 1000 & -0.003 & 0.328 & 0.409 & 0.607 & 0.383 & 0.058 & 0.000\\
    BA & 1000 & 0.167 & 0.363 & 0.378 & 0.527 & 0.353 & 0.128 & 0.000\\
    Road & 1039 & 0.162 & 0.442 & 0.470 & 0.525 & 0.428 & 0.102 & 0.129\\
    Email & 1133 & 0.090 & 0.292 & 0.446 & 0.335 & 0.405 & 0.281 & 0.228\\
    FlyBi & 2705 & -0.132 & 0.370 & 0.414 & 0.378 & 0.417 & 0.115 & 0.300\\
    Reactome & 5973 & -0.006 & 0.300 & 0.238 & 0.158 & 0.238 & 0.181 & 0.239\\
    Route views & 6474 & 0.145 & 0.241 & 0.317 & 0.239 & 0.331 & 0.145 & 0.216\\
    Spanish words & 11558 & 0.027 & 0.259 & 0.391 & 0.300 & 0.362 & 0.240 & 0.288\\
    FOLDOC & 13356 & 0.045 & 0.621 & 0.507 & 0.542 & 0.483 & 0.433 & 0.347\\
    Tree of Life & 16415 & 0.037 & 0.525 & 0.513 & 0.443 & 0.438 & 0.356 & 0.448\\
    English words & 23132 & -0.062 & 0.443 & 0.716 & 0.669 & 0.640 & 0.345 & 0.405\\
    Enron & 33696 & -0.063 & 0.353 & 0.410 & 0.253 & 0.404 & 0.305 & 0.319\\
    Marker Cafe & 69317 & -0.058 & 0.456 & 0.568 & 0.505 & 0.524 & 0.385 & 0.480\\
    Propser & 89171 & 0.066 & 0.586 & 0.616 & 0.622 & 0.527 & 0.291 & 0.596\\
    \bottomrule
  \end{tabular}
\end{table}

\newpage
\clearpage

\begin{table}
\centering
  \caption{Model performance and variable importance for a random forest model predicting the approximation error for completely random node sets using the gene-regulatory dynamics on 19 networks. See the caption of Table \ref{tab:rf} for the legends.}
  \label{tab:rf-error-genereg}
  \begin{tabular}{lS[table-format=5.0]S[table-format=3.3]*6{S[table-format=1.3]}}
    \toprule
    Network & {$N$} & {$R^2$} & {$k$} & {Close} & {Betw} & {$k_{\rm nn}$} & {LC} & {Core}\\
    \midrule
    Dolphin & 62 & 0.318 & 0.245 & 0.160 & 0.283 & 0.224 & 0.106 & 0.058\\
    Proximity & 410 & -0.127 & 0.505 & 0.241 & 0.284 & 0.254 & 0.198 & 0.278\\
    Metabolic & 453 & -0.063 & 0.237 & 0.296 & 0.353 & 0.278 & 0.257 & 0.134\\
    GKK & 949 & 0.209 & 0.328 & 0.378 & 0.532 & 0.323 & 0.089 & 0.146\\
    LFR & 998 & -0.003 & 0.459 & 0.434 & 0.401 & 0.381 & 0.263 & 0.088\\
    HK & 1000 & -0.210 & 0.375 & 0.535 & 0.599 & 0.454 & 0.066 & 0.000\\
    BA & 1000 & -0.167 & 0.275 & 0.482 & 0.530 & 0.464 & 0.220 & 0.000\\
    Road & 1039 & 0.157 & 0.520 & 0.438 & 0.479 & 0.341 & 0.114 & 0.133\\
    Email & 1133 & 0.218 & 0.453 & 0.379 & 0.344 & 0.322 & 0.273 & 0.222\\
    FlyBi & 2705 & 0.028 & 0.411 & 0.393 & 0.321 & 0.368 & 0.131 & 0.241\\
    Reactome & 5973 & 0.240 & 0.339 & 0.196 & 0.164 & 0.194 & 0.128 & 0.194\\
    Route views & 6474 & -0.138 & 0.164 & 0.384 & 0.324 & 0.408 & 0.163 & 0.117\\
    Spanish words & 11558 & -0.114 & 0.193 & 0.466 & 0.365 & 0.448 & 0.274 & 0.207\\
    FOLDOC & 13356 & 0.013 & 0.501 & 0.596 & 0.595 & 0.567 & 0.525 & 0.300\\
    Tree of Life & 16415 & 0.157 & 0.574 & 0.471 & 0.410 & 0.418 & 0.342 & 0.405\\
    English words & 23132 & 0.057 & 0.439 & 0.613 & 0.600 & 0.555 & 0.365 & 0.375\\
    Enron & 33696 & -0.078 & 0.335 & 0.444 & 0.289 & 0.439 & 0.296 & 0.271\\
    Marker Cafe & 69317 & 0.025 & 0.444 & 0.591 & 0.557 & 0.550 & 0.432 & 0.431\\
    Propser & 89171 & 0.031 & 0.644 & 0.652 & 0.665 & 0.550 & 0.368 & 0.588\\
    \bottomrule
  \end{tabular}
\end{table}

\setlength{\tabcolsep}{10pt}

\newpage
\clearpage

\section{$x_i^*$ of sentinel nodes}

To demonstrate that sentinel nodes do not particularly tend to be outliers in terms of the $x_i^*$ value, we have carried out the following analysis. For each pair of dynamics and network, we observe the node states in the equilibrium for the largest value of the control parameter (i.e., closest to the regime shift), $x_1^*$, $\ldots$, $x_N^*$. This choice of the control parameter value is arbitrary but tends to yield a spread-out distribution of $x_1^*$, $\ldots$, $x_N^*$, making outlier analyses easier.
Then, we calculate a modified $z$-score for each node by
$z = (x_i^* - \text{median}(x_1^*, \ldots, x_N^*)) / \text{MAD}$, where ``median'' represents the median of the arguments, and $\text{MAD}$ is the median absolute deviation, i.e., the median of $\left| x_1^* - \text{median}(x_1^*, \ldots, x_N^*) \right|$, $\ldots$,
$\left| x_N^* - \text{median}(x_1^*, \ldots, x_N^*) \right|$. We use the modified $z$-score instead of the ordinary $z$-score because the former is more suitable for identifying outliers. Then, for the set of all the $N$ nodes, we compute the 95\% quantile of the $|z|$ value, denoted by $|z|_{\text{c, all}}$. In other words, just 5\% of the nodes have $|z|$ values larger than $|z|_{\text{c, all}}$; these nodes are 5\%-level outliers in terms of $x_i^*$. We also compute $100$ sentinel node sets to obtain $100 n$ sentinel nodes, counting multiplicities. The same node may appear multiple times in the set of $100 n$ sentinel nodes. For these $100 n$ nodes, we obtain the 95\% quantile of the $|z|$ value, denoted by $|z|_{\text{c, sentinel}}$. If $|z|_{\text{c, sentinel}}$ is significantly larger than $|z|_{\text{c, all}}$, then we conclude that sentinel nodes tend to be outliers in terms of $x_i^*$. In this case, outlier sentinel nodes tend to take extreme values of $x_i^*$ compared to outlier nodes in general.

For each of the four dynamics, we thus compute and $|z|_{\text{c, all}}$ and $|z|_{\text{c, sentinel}}$ for each of the 20 networks and run a paired $t$-test. A positive $t$ statistic signifies that $|z|_{\text{c, all}} > |z|_{\text{c, sentinel}}$. We find that $|z|_{\text{c, all}}$ and $|z|_{\text{c, sentinel}}$ are not significantly different for all the four dynamics after Bonferroni correction (coupled double-well: $t=2.07$, $p = 0.052$; mutualistic interaction: $t=2.06$, $p=0.054$; SIS: $t=2.10$, $p=0.049$; gene-regulatory: $t=-1.06$, $p=0.301$; $df=19$ and the $p$ value is not corrected for multiple comparison for each dynamics).
%
%
In all the marginally insignificant cases (i.e., coupled double-well, mutualistic interaction, SIS), the $t$ values are positive, implying that sentinel nodes tend to avoid outlier nodes in terms of $x_i^*$. Therefore, we conclude that sentinel nodes are not outliers in terms of $x_i^*$. This result is consistent with the result shown in section~\ref{sec:node-feature}\ref{sub:node-feature-descriptive-statistics} that sentinel nodes weakly tend to avoid hubs.

\newpage
\clearpage

\section{Kullback-Leibler divergence\label{sec:kld}}

We measure how different a sampled degree distribution is from the original degree distribution of the network with the Kullback-Leibler divergence, $D_{\rm KL}$.  The Kullback-Leibler divergence is defined as
\begin{equation} \label{eq:kld}
  D_{\rm KL}(Q \| P) = \sum_{k \in \mathcal{K}} Q(k) \log \left[ \frac{Q(k)}{P(k)} \right],
\end{equation}
where $P(k)$ is the degree distribution of the original network (i.e., the proportion of nodes having degree $k$), $Q(k)$ is the degree distribution based on the set of sampled nodes,
and $\mathcal{K}$ is the set of unique degree values in the original network.
To compute $Q(k)$, we first generated 100 optimized node sets with a given $n$. Then, we calculated the proportion of nodes, across the 100 node sets, which had each degree in $\mathcal{K}$.

\clearpage

\section{Approximation error as a function of $n$\label{sec:increasing-n}}

\begin{figure}[h]
\centering
  \includegraphics[width = 0.8\textwidth]{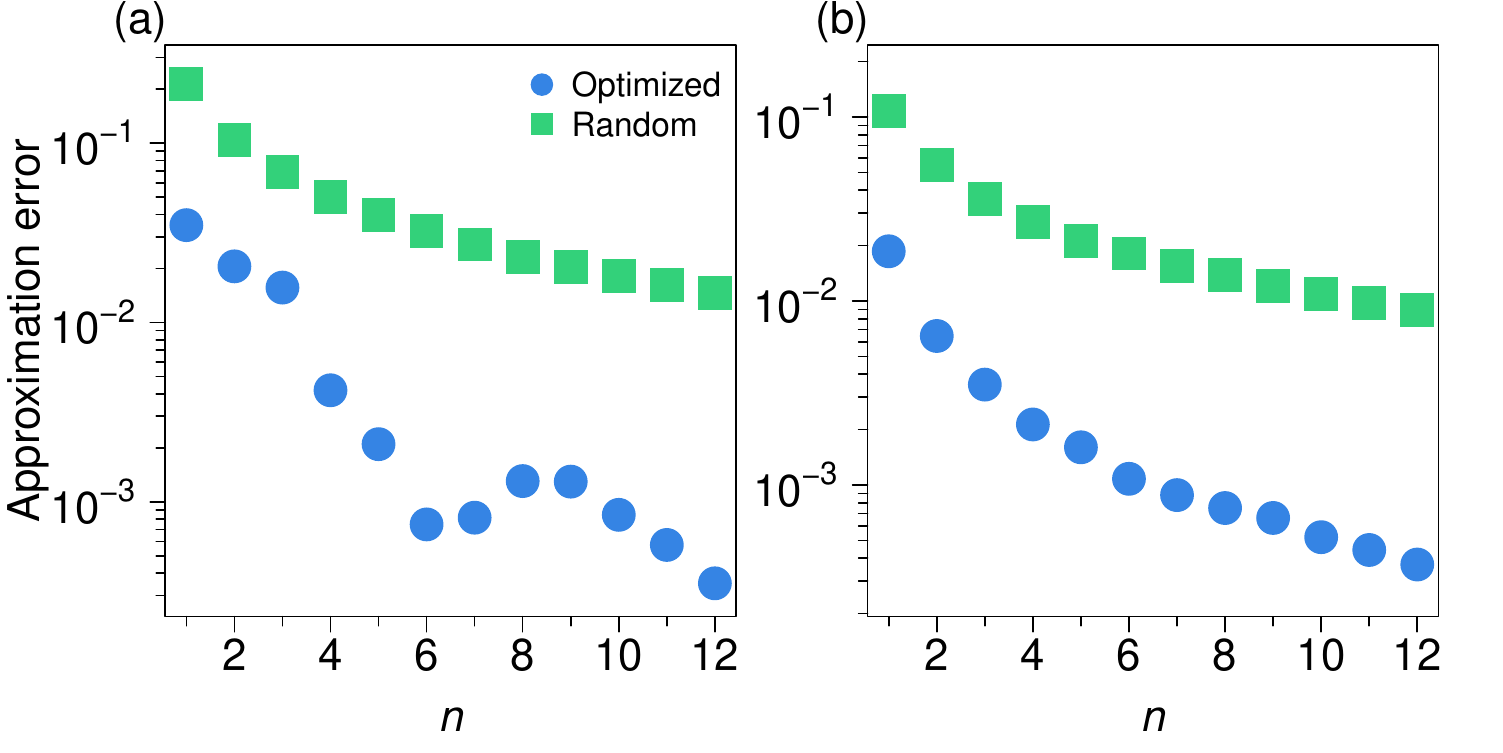}
  \caption{
    Approximation error as a function of node set size for two networks. The blue circles show the average approximation error over the 100 optimized node sets. The green squares show the average approximation error over 100 completely random node sets. The confidence interval for each average is smaller than the corresponding marker and is therefore not shown. (a) Dolphin network. (b) BA network.
  }
  \label{fig:errorfig}
\end{figure}

In the main text (section IIB), 
we showed that the degree distributions of optimized node sets remain distinct from those of completely random node sets up to $n=12$ for the dolphin and BA networks. Figure~\ref{fig:errorfig} indicates that the approximation errors of optimized node sets also remain distinct from and lower than those of completely random node sets. We note that the approximation error of both optimized and completely random node sets decreases as $n$ increases, which is expected.

\newpage
\clearpage

\section{Characterization of sentinel node sets\label{sec:structural-features}}

In the main text, we analyzed the degree distributions of optimized sentinel node sets in comparison with the degree distribution of completely randomly node sets. 
In section~\ref{sec:node-feature}, we showed that six features of individual nodes are not predictive of whether or not they are sentinel nodes. However, other features of node sets that are not ascribed to individual nodes' features may play roles in making node sets suitable as optimized node sets. In this section, we investigate whether or not two such features as well as averages of individual node features over the $n$ nodes in the node set are helpful for constructing good sentinel node sets without resorting to our optimization algorithm.

\subsection{Two node set features that cannot be decomposed into single-node features}

\begin{figure}
  \centering
  \includegraphics[width = 0.8\textwidth]{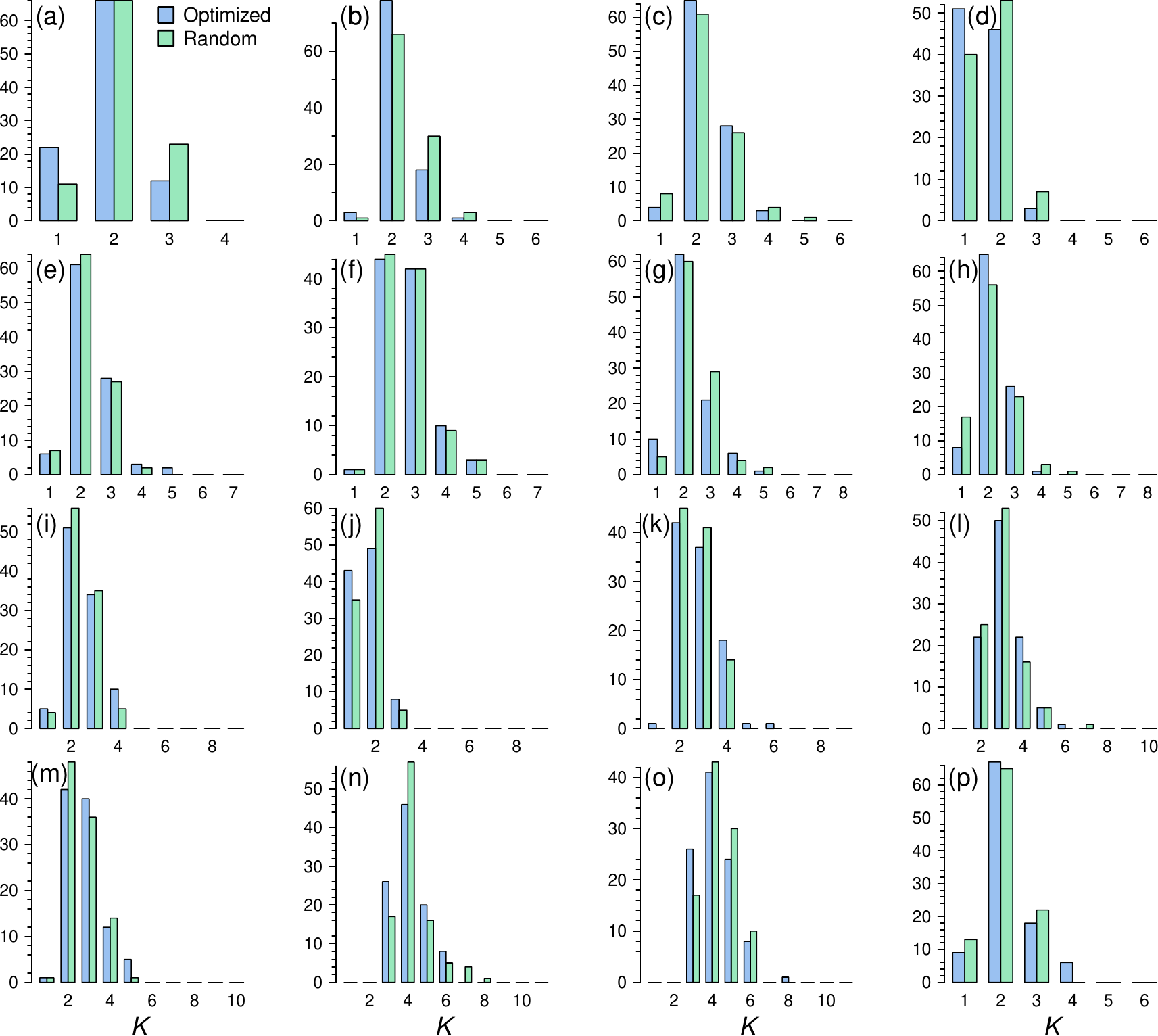}
  \caption{The maximum number of nodes in each node set coming from the same community. (a) Dolphin. (b) Proximity. (c) Metabolic. (d) Road. (e) Email. (f) FlyBi. (g) Reactome. (h) Route views. (i) Spanish words. (j) FOLDOC. (k) Tree of life. (\text{$\ell$}) English words. (m) Enron. (n) Marker Cafe. (o) Prosper. (p) LFR.}
  \label{fig:communities}
\end{figure}

First, we hypothesized that nodes in optimized node sets are more likely to come from different communities than those in completely random node sets. This hypothesis is intuitive because nodes in the same community are expected to show relatively similar dynamics  \cite{arenas2006, laurence2019, schaub2020, vegue2023}, and therefore it is probably redundant to devote multiple sentinel nodes to a single community when one can only use a small number of sentinel nodes. To test the hypothesis, we confined ourselves to sixteen networks, i.e., the 15 empirical networks and the LFR network, which have relatively strong community structure. We exclude the other four model networks because they do not have community structure by construction. We obtained a partition of each of the 16 networks into communities using the Louvain algorithm \cite{blondel2008}. Then, for each node set, we counted the number of nodes in the same community and recorded the maximum such number over all the communities, which we denote by $K$. In Fig.~\ref{fig:communities}, we show the distribution of $K$ for the optimized and completely random node sets. We find that the two node sets are not notably different from each other across the different networks.
%

\begin{figure}
  \centering
  \includegraphics[width = 0.8\textwidth]{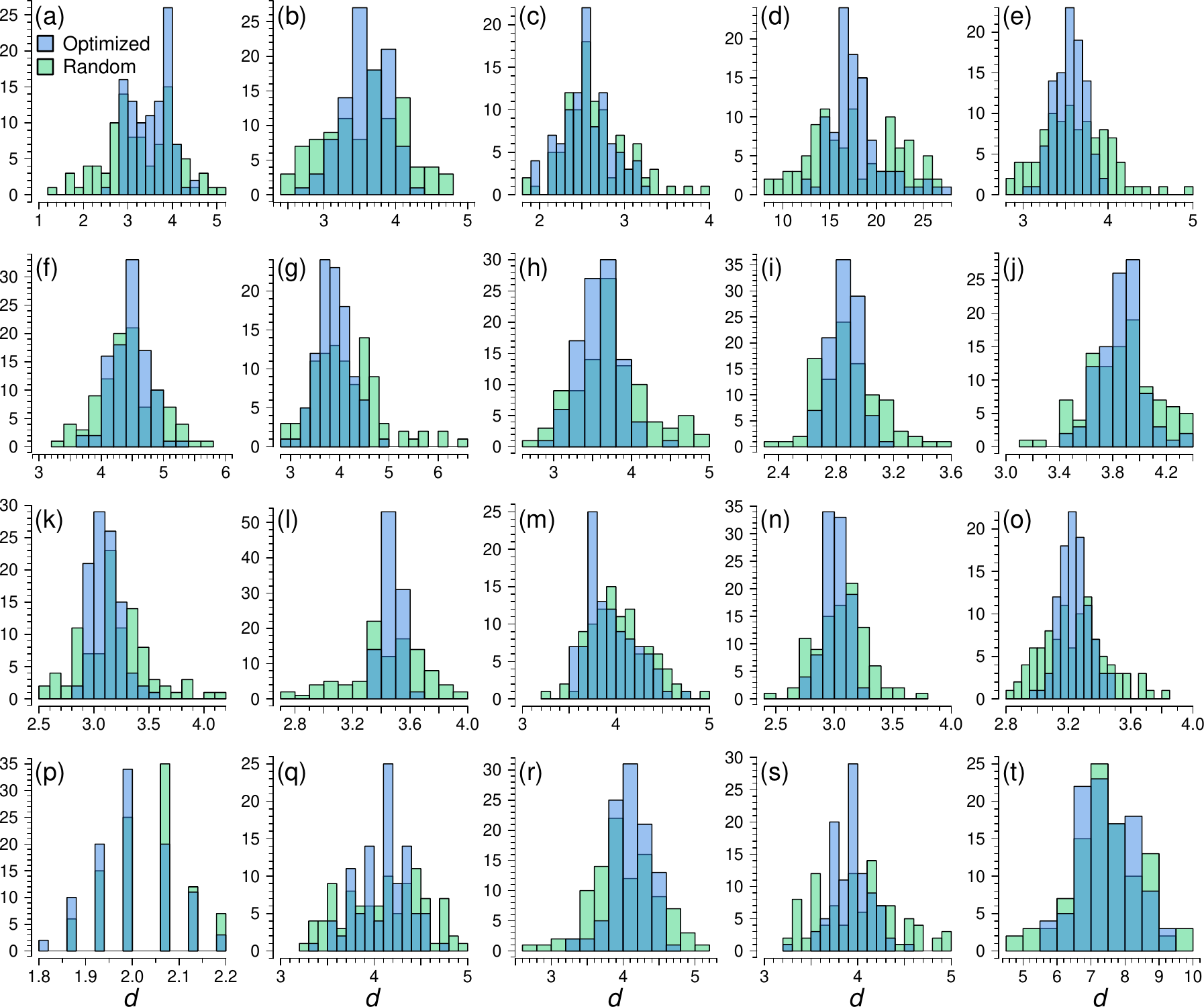}
  \caption{The average distance between node pairs in the sentinel or completely random node set.  (a) Dolphin. (b) Proximity. (c) Metabolic. (d) Road. (e) Email. (f) FlyBi. (g) Reactome. (h) Route views. (i) Spanish words. (j) FOLDOC. (k) Tree of life. (\text{$\ell$}) English words. (m) Enron. (n) Marker Cafe. (o) Prosper. (p) ER. (q) BA. (r) HK. (s) GKK. (t) LFR.}
  \label{fig:d-node-set}
\end{figure}

Second, regardless of the community structure, the sentinel node approximation may work better when the members of a sentinel node set are well separated from each other because the different sentinel nodes carry relatively independent information in this case. Therefore, we calculated the distance between two nodes (i.e., the smallest number of hops necessary to reach from one node to the other), which we averaged over all the $n(n-1)/2$ node pairs in a node set. We show this average node-to-node distance, denoted by $d$, for all the 20 networks in Fig.~\ref{fig:d-node-set}. We find that sentinel node sets tend to avoid small and large $d$ compared to completely random node sets. Avoidance of small $d$ supports the idea that it may be better for sentinel nodes in the same sentinel node set to be separated from each other. Avoidance of large $d$ may be because sentinel node sets may tend to exclude nodes that are located far from the remainder of the network.

\subsection{Search for node set features that would make node sets sentinel}

To formally examine whether or not the $K$ or $d$ value of a node set informs us of the quality of the node set in approximating $\overline{x}$, we first attempted to classify each node set as either optimized or completely random based only on features of the node set. A successful classification would imply that one can likely construct a high-performance node set by relying on these features and avoid running our optimization algorithm. It should be noted that we examined the same question in section~\ref{sec:node-feature}, but only with features of individual nodes. Because the classification performance was poor when we only use these individual-node features, here we assess the effectiveness of features of node \textit{sets}. In addition to $K$ and $d$, we used the average of each of the six node features used in section~\ref{sec:node-feature} (i.e., $k$, closeness centrality, betweenness centrality, $k_{\text{nn}, i}$, $C_i$, and coreness) over the $n$ nodes in a node set as additional features of the node set. 

We started by using logistic regression for this classification task. Specifically, we use a generalized linear model with binomial errors and a logit link function in which the dependent variable is the binary outcome of a trial (i.e., either a node set is completely random, which is the reference, or it is optimized). The independent variables are the simulation conditions (i.e., dynamics and network); the six node features, each averaged over the $n$ nodes in a node set; $K$, the maximum number of nodes in a node set assigned to the same community; and $d$, the average node-to-node distance in a node set. We analyzed the same number of node sets of each type: 100 optimized node sets for each combination of dynamics and network, and an equal number of completely random node sets (i.e., 400 random node sets for each network). We again exclude the ER network because of its relatively homogeneous degree distribution.

We show the results of this analysis in Table \ref{tab:struct}. We find that the classification is unsuccessful; our model explains less than 2\% of the variance in classification of a node set as random or optimized (McFadden's pseudo-$R^2 = 0.016$). The $p$ value is small for many independent variables due to a large sample size. Additionally, the controls for dynamics in this logistic regression are not significant, suggesting that the composition of optimized node sets may be more strongly related to network structure than dynamics.

\renewcommand{\arraystretch}{0.75}
\setlength{\tabcolsep}{3.5pt}
\begin{table}
\centering
  \caption{Regression results in search for features of the optimized node set. The dependent variable is whether the node set is completely random (the reference) or optimized. The model is a logistic regression, i.e., a generalized linear model with binomial errors and a logit link function. SE: standard error. $z$: coefficient estimate divided by the standard error. LC: local clustering coefficient. The single-node features (i.e., $k$, closeness centrality, betweenness centrality, $k_{\text{nn}}$, LC, and coreness) are averages over the $n$ nodes in the node set. The null deviance is $21,072$ ($df=15,199$) and the residual deviance is $20,726$ ($df=15,170$). The pseudo-$R^2$ is $0.016$.}
  \label{tab:struct}
  \begin{tabular}{l*4{S[table-format=4.3]}}
    \toprule
    & {Estimate} & {SE} & {$z$} & {$p$}\\
    \midrule
    Intercept & -6.971 & 0.596 & -11.698 & {$< 10^{-7}$}\\
    Dynamics &  &  &  & \\
    \hspace{1 em} Mutualistic species & 0.011 & 0.046 & 0.242 & 0.809\\
    \hspace{1 em} SIS & -0.005 & 0.046 & -0.115 & 0.909\\
    \hspace{1 em} Gene-regulatory & -0.014 & 0.046 & -0.306 & 0.760\\
    Network &  &  &  & \\
    \hspace{1 em} Proximity & 0.269 & 0.139 & 1.941 & 0.052\\
    \hspace{1 em} Metabolic & -2.452 & 0.231 & -10.625 & {$< 10^{-7}$}\\
    \hspace{1 em} Road & 5.974 & 0.455 & 13.131 & {$< 10^{-7}$}\\
    \hspace{1 em} Email & 0.130 & 0.140 & 0.928 & 0.353\\
    \hspace{1 em} FlyBi & 1.558 & 0.189 & 8.246 & {$< 10^{-7}$}\\
    \hspace{1 em} Reactome & 1.119 & 0.224 & 4.998 & {$5.79 \times 10^{-7}$}\\
    \hspace{1 em} Route views & 0.144 & 0.148 & 0.975 & 0.329\\
    \hspace{1 em} Spanish words & -1.556 & 0.202 & -7.691 & {$< 10^{-7}$}\\
    \hspace{1 em} FOLDOC & 0.654 & 0.155 & 4.231 & {$2.33 \times 10^{-5}$}\\
    \hspace{1 em} Tree of life & -0.560 & 0.182 & -3.069 & {$2.15 \times 10^{-3}$}\\
    \hspace{1 em} English words & 0.077 & 0.156 & 0.493 & 0.622\\
    \hspace{1 em} Enron & 0.966 & 0.170 & 5.680 & {$< 10^{-7}$}\\
    \hspace{1 em} Marker Cafe & -0.712 & 0.273 & -2.610 & {$9.06 \times 10^{-3}$}\\
    \hspace{1 em} Prosper & 0.074 & 0.218 & 0.340 & 0.734\\
    \hspace{1 em} BA & 0.819 & 0.163 & 5.026 & {$5.00 \times 10^{-7}$}\\
    \hspace{1 em} HK & 0.844 & 0.159 & 5.300 & {$1.16 \times 10^{-7}$}\\
    \hspace{1 em} GKK & 0.812 & 0.162 & 5.003 & {$5.65 \times 10^{-7}$}\\
    \hspace{1 em} LFR & 3.837 & 0.312 & 12.285 & {$< 10^{-7}$}\\
    $k$ & -0.025 & 0.003 & -9.659 & {$< 10^{-7}$}\\
    Closeness & 25.820 & 1.932 & 13.365 & {$< 10^{-7}$}\\
    Betweenness & -15.740 & 2.531 & -6.220 & {$< 10^{-7}$}\\
    $k_{\rm nn}$ & {$2.04 \times 10^{-5}$} & {$1.27 \times 10^{-4}$} & -0.161 & 0.872\\
    LC & -0.255 & 0.196 & -1.302 & 0.193\\
    Coreness & 0.030 & 0.005 & 5.569 & {$< 10^{-7}$}\\
    $K$ & -0.124 & 0.023 & -5.283 & {$1.27 \times 10^{-7}$}\\
    $d$ & -0.003 & 0.018 & -0.191 & 0.848\\
    \bottomrule
  \end{tabular}
\end{table}
\renewcommand{\arraystretch}{1}

\begin{figure}
  \centering
  \includegraphics[width = 0.8\textwidth]{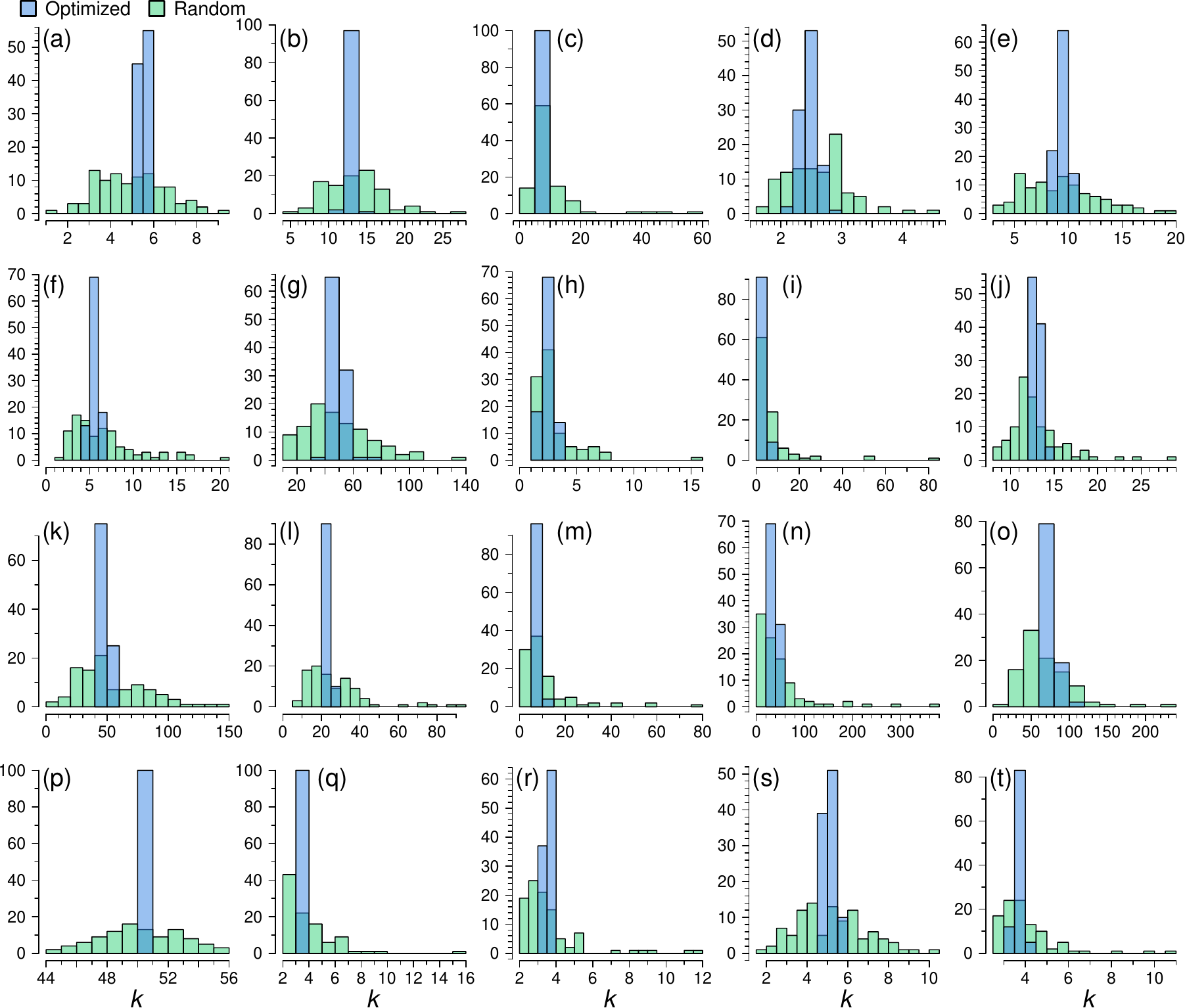}
  \caption{The average degree of the nodes in the sentinel or completely random node set.  (a) Dolphin. (b) Proximity. (c) Metabolic. (d) Road. (e) Email. (f) FlyBi. (g) Reactome. (h) Route views. (i) Spanish words. (j) FOLDOC. (k) Tree of life. (\text{$\ell$}) English words. (m) Enron. (n) Marker Cafe. (o) Prosper. (p) ER. (q) BA. (r) HK. (s) GKK. (t) LFR.}
  \label{fig:average-k-node-set}
\end{figure}

Although linear regression showed a poor classification result, nonlinear effects of node set features (as opposed to single-node features as investigated in section~\ref{sec:node-feature}) may make high-quality sentinel node sets. Therefore, we constructed the random forest classifier for the binary classification of node sets using the same data as those used Table~\ref{tab:struct}. We show the classification performance and the variable importance in Table~\ref{tab:rf-ns-class}. We find a large $R^2$ value (i.e., 0.865), indicating that the binary classification is successful. This result is in stark contrast with that with the linear regression shown in Table~\ref{tab:struct} and implies that nonlinear or combinatorial effects of some independent variables contribute to improving the approximation accuracy of node sets. Table~\ref{tab:rf-ns-class} also shows that the average degree over the $n$ nodes, $k$, has the highest variable importance. Therefore, we compare the distribution of $k$ between the optimized and completely random node set in Fig.~\ref{fig:average-k-node-set}. The figure indicates that sentinel node sets tend to avoid both large $k$ and small $k$. This tendency is a nonlinear effect in terms of $k$, which likely accounts for our result that the random forest, but not the linear regression, performs well at the binary classification. We emphasize that this result on $k$ does not imply that the individual nodes in a good node set are all intermediate-degree nodes. We showed that, while sentinel nodes tend to be non-hubs (see Fig.~3 in the main text and section~\ref{sec:node-feature}\ref{sub:node-feature-descriptive-statistics}), different sentinel nodes in an optimized sentinel node set tend to have their degrees scattered (see Fig.~3).

\begin{table}
\centering
  \caption{Model performance ($R^2$) and variable importance for classifying node sets into two categories, completely random or optimized, using a random forest classifier. Close: closeness centrality. Betw: betweenness centrality. LC: local clustering coefficient. Core: coreness. The single-node features (i.e., $k$, closeness centrality, betweenness centrality, $k_{\text{nn}}$, LC, and coreness) are averages over the $n$ nodes in the node set. We used 100 trees and a random subset of three features per node. We used the Gini index as the classification criterion.}
  \label{tab:rf-ns-class}
  \begin{tabular}{*2{S[table-format=2.3]}*9{S[table-format=2.3]}}
    \toprule										
    {$R^2$} & {Dynamics} & {Network} & {$k$} & {Close} & {Betw} & {$k_{\rm nn}$} & {LC} & {Core} & {$K$} & {$d$}\\
    \midrule
    0.865 & 0.031 & 0.020 & 0.420 & 0.232 & 0.142 & 0.140 & 0.052 & 0.349 & 0.004 & 0.065\\
    \bottomrule
  \end{tabular}
\end{table}

Motivated by this random forest result, we constructed a different random forest model for explaining the approximation error. Because the value of the approximation error in different dynamics are not comparable to each other, we built a random forest model for each dynamics. We show the results in Table~\ref{tab:rf-ns} for the completely random node sets, merger of the completely random and optimized node sets, and the merge of the completely random, optimized, and degree-preserving random node sets. The results are qualitatively the same as those for the binary classification task shown in Table~\ref{tab:rf-ns-class}. In other words, in all the three combinations of node sets and all the four dynamics, the random forest model is at least reasonably good at predicting the approximation error of the node set. Furthermore, $k$ remains to be the most important independent variable among the nine independent variables for explaining the approximation error in nine out of twelve random forests shown in the table. Overall, the coreness, closeness centrality, and betweenness centrality are the next most important independent variables after $k$. The last result is also consistent with that for the binary classifier (see Table~\ref{tab:rf-ns-class}).

\begin{table}
\centering
  \caption{Model performance ($R^2$) and variable importance for random forest models predicting approximation error for different types of node sets. We built separate random forest models for each dynamics, including all networks except the ER network in each random forest model. We used $\ln \varepsilon$ as the dependent variable and squared error as the criterion for finding splits. See the caption of Table~\ref{tab:rf-ns-class} for the legends.}
  \label{tab:rf-ns}
  \begin{tabular}{lS[table-format=2.3]|*9{S[table-format=2.3]}}
    \toprule
    \multicolumn{11}{l}{Completely random only}\\
    Dynamics & {$R^2$} & {Network} & {$k$} & {Close} & {Betw} & {$k_{\rm nn}$} & {LC} & {Core} & {$K$} & {$d$}\\
    \midrule
    Double-well & 0.763 & 0.262 & 0.398 & 0.234 & 0.204 & 0.233 & 0.100 & 0.426 & 0.029 & 0.144\\
    Mutualistic & 0.736 & 0.099 & 0.469 & 0.250 & 0.191 & 0.198 & 0.084 & 0.388 & 0.063 & 0.215\\
    SIS & 0.509 & 0.144 & 0.355 & 0.324 & 0.240 & 0.255 & 0.177 & 0.328 & 0.047 & 0.263\\
    Gene-regulatory & 0.731 & 0.117 & 0.805 & 0.115 & 0.265 & 0.308 & 0.063 & 0.184 & 0.019 & 0.057\\
    \midrule
    \multicolumn{11}{l}{Completely random and optimized}\\
    Dynamics & {$R^2$} & {Network} & {$k$} & {Close} & {Betw} & {$k_{\rm nn}$} & {LC} & {Core} & {$K$} & {$d$}\\
    \midrule
    Double-well & 0.786 & 0.202 & 0.869 & 0.391 & 0.485 & 0.253 & 0.088 & 0.858 & 0.009 & 0.234\\
    Mutualistic & 0.792 & 0.121 & 0.816 & 0.524 & 0.349 & 0.261 & 0.112 & 0.704 & 0.011 & 0.301\\
    SIS & 0.638 & 0.108 & 0.526 & 0.621 & 0.428 & 0.355 & 0.177 & 0.697 & 0.031 & 0.257\\
    Gene-regulatory & 0.917 & 0.086 & 1.118 & 0.181 & 0.444 & 0.209 & 0.075 & 0.349 & 0.010 & 0.083\\
    \midrule
    \multicolumn{11}{l}{Completely random, optimized, and degree-preserving random}\\
    Dynamics & {$R^2$} & {Network} & {$k$} & {Close} & {Betw} & {$k_{\rm nn}$} & {LC} & {Core} & {$K$} & {$d$}\\
    \midrule
    Double-well & 0.793 & 0.295 & 1.144 & 0.471 & 0.547 & 0.318 & 0.129 & 0.986 & 0.010 & 0.248\\
    Mutualistic & 0.791 & 0.191 & 0.717 & 0.629 & 0.267 & 0.363 & 0.099 & 0.615 & 0.018 & 0.308\\
    SIS & 0.626 & 0.141 & 0.569 & 0.605 & 0.431 & 0.351 & 0.162 & 0.612 & 0.028 & 0.285\\
    Gene-regulatory & 0.829 & 0.142 & 1.478 & 0.280 & 0.574 & 0.228 & 0.107 & 0.547 & 0.015 & 0.098\\
    \bottomrule
  \end{tabular}
\end{table}

\clearpage

\subsection{Heuristic algorithms for node set selection\label{sub:SI-heuristic}}

To further explore the possibility of using the information obtained from these analyses to eliminate the need for our optimization algorithm, we investigated four heuristic algorithms for node set selection based on our observations. These algorithms do not depend on the dynamics and only use the information on the network structure to different extents. 

For the first algorithm, which we call ``$k$-constrained,'' we reject the largest 5\% of nodes in terms of degree and select $n$ nodes uniformly at random without replacement from the remaining 95\% of nodes. We also reject the 5\% largest-degree nodes before selecting $n$ nodes without replacement in each of the following algorithms. 

In the ``$k$-quantiled'' algorithm, we select one node uniformly at random from each of $n$ divisons of the degree distribution. For example, if $n = 4$, then we select one node from the smallest 25\% of nodes in terms of degree, one node from the 25th--50th percent of nodes, and so on, among the 95\% of nodes with the smallest degrees. 

In the ``$k_{\rm nn}$-constrained'' algorithm, we reject the bottom 5\% of nodes in terms of $k_{\rm nn}$ and select nodes uniformly at random and without replacement from the remaining nodes. 

Finally, the ``community-based'' algorithm runs as follows. First, we partition the network into communities using the Louvain algorithm \cite{blondel2008}. Then, we select nodes softly avoiding multiple nodes from the same community. Specifically, we select the $i$th community with probability $\sqrt{c_i}/\sum_{j=1}^{n_{\text{c}}} \sqrt{c_j}$, where $c_i$ is the number of nodes in the $i$th community and $n_{\text{c}}$ is the number of communities in the network, then select one node uniformly at random without replacement from the $i$th community. We repeat this procedure $n$ times.
Note that the unbiased sampling of nodes would select each community with probability $c_i / \sum_{j=1}^{n_{\text{c}}} c_j$. Therefore, nodes in large communities are underrepresented in our community-based node set such that selection of multiple nodes from the same community is discouraged.

To compare different node selection algorithms, we generated 100 optimized node sets by running the optimization algorithm 100 times, for each combination of dynamics and network. We also generated 100 degree-preserving node sets, 100 completely random node sets for each network, and 100 node sets from each heuristic algorithm (i.e., $k$-constrained, $k$-quantiled, $k_{\rm nn}$-constrained, and community-based) for each network. Note that neither the completely random nor heuristic algorithm node sets depend on the dynamics.

We computed the approximation error for each node set obtained by each of the seven algorithms, each of the four dynamics, and each of the 19 networks, i.e., all but the ER network. We exclude the ER network 
for the same reason as that for the ANOVA analysis shown in section~\ref{sec:compare-networks}. Then, we carried out a multi-way ANOVA in the same manner as in section~\ref{sec:compare-networks}. The only difference between the two ANOVA analyses is that the current analysis involves seven node sets, whereas the analysis shown in section~\ref{sec:compare-networks} involves three node set types (i.e., optimized, degree-preserving random, and completely random) that are among the seven node sets.

Overall, our ANOVA model predicts $\ln \varepsilon$ well ($R^2 = 0.72$). In part due to the large sample size (i.e., $53,200$ observations), all three independent variables are highly significant (dynamics: $df = 3$, $F = 14,278$, $p < 10^{-7}$, where the $F$ statistic is the mean-square error of the factor divided by the mean-square error of the residuals; network: $df = 18$, $F = 245.29$, $p < 10^{-7}$; node set type: $df = 6$, $F = 14,579$, $p < 10^{-7}$). 

We show the differences between the average error for each pair of node set types, as computed by a Tukey's honestly significant difference test, in Table \ref{tab:HSD-home}. As in Table~\ref{tab:HSD-short}, each row of Table \ref{tab:HSD-home} specifies the estimated difference in average $\ln \varepsilon$ between the two node set types, the 95\% confidence interval of the estimated mean difference, and a $p$ value for the difference adjusted for multiple comparisons. For example, the first row states that, across dynamics and networks, optimized node sets have an average $\ln \varepsilon$ value that is 6.892 smaller than that of completely random node sets. On the natural scale, the average $\varepsilon$ for optimized node sets is $1/e^{-6.892} = 984.4$ times smaller than the average $\varepsilon$ for completely random node sets. 

\begin{table}
\centering
  \caption{
    Differences in average $\ln \varepsilon$ between different node set types when we use the test dynamics for optimization. The differences are computed by a Tukey's honestly significant difference test. The 95\% confidence intervals (CI) of the differences and associated $p$ values, adjusted for multiple comparisons, are also shown.}
  \label{tab:HSD-home}
  \begin{tabular}{lScc}
    \toprule
    & {Difference} & CI & $p$\\
    \midrule
    Optimized $-$ Random & -6.892 & $[-6.983, -6.801]$ & $< 10^{-7}$\\
    Degree-preserving $-$ Random & -3.149 & $[-3.240, -3.058]$ & $< 10^{-7}$\\
    $k$-constrained $-$ Random & -0.129 & $[-0.219, -0.038]$ & $5.77 \times 10^{-4}$\\
    $k$-quantiled $-$ Random & -0.259 & $[-0.350, -0.168]$ & $< 10^{-7}$\\
    $k_{\rm nn}$-constrained $-$ Random & -0.220 & $[-0.310, -0.129]$ & $< 10^{-7}$\\
    Community-based $-$ Random & 0.024 & $[-0.067, 0.114]$ & $0.988$\\
    Degree-preserving $-$ Optimized & 3.743 & $[3.652, 3.834]$ & $< 10^{-7}$\\
    $k$-constrained $-$ Optimized & 6.763 & $[6.672, 6.854]$ & $< 10^{-7}$\\
    $k$-quantiled $-$ Optimized & 6.633 & $[6.542, 6.724]$ & $< 10^{-7}$\\
    $k_{\rm nn}$-constrained $-$ Optimized & 6.672 & $[6.582, 6.763]$ & $< 10^{-7}$\\
    Community-based $-$ Optimized & 6.916 & $[6.825, 7.006]$ & $< 10^{-7}$\\
    $k$-constrained $-$ Degree-preserving & 3.020 & $[2.929, 3.111]$ & $< 10^{-7}$\\
    $k$-quantiled $-$ Degree-preserving & 2.890 & $[2.799, 2.981]$ & $< 10^{-7}$\\
    $k_{\rm nn}$-constrained $-$ Degree-preserving & 2.929 & $[2.839, 3.020]$ & $< 10^{-7}$\\
    Community-based $-$ Degree-preserving & 3.173 & $[3.082, 3.263]$ & $< 10^{-7}$\\
    $k$-quantiled $-$ $k$-constrained & -0.130 & $[-0.221, -0.039]$ &  $4.68 \times 10^{-4}$\\
    $k_{\rm nn}$-constrained $-$ $k$-constrained & -0.091 & $[-0.182, 1.60 \times 10^{-4}]$ & $0.049$\\
    Community-based $-$ $k$-constrained & 0.152 & $[0.062, 0.243]$ & $1.52 \times 10^{-5}$\\
    $k_{\rm nn}$-constrained $-$ $k$-quantiled & 0.039 & $[-0.051, 0.13]$ & $0.863$\\
    Community-based $-$ $k$-quantiled & 0.283 & $[0.192, 0.373]$ & $< 10^{-7}$\\
    Community-based $-$ $k_{\rm nn}$-constrained & 0.243 & $[0.153, 0.334]$ & $< 10^{-7}$\\
    \bottomrule
  \end{tabular}
\end{table}

The average $\ln \varepsilon$ of any node set type is smaller than that of completely random node sets, except that the community-based algorithm performs poorly, on par with completely random node sets (1st--6th rows of Table \ref{tab:HSD-home}). In fact, optimized node sets have the smallest error (1st and 7th--11th rows), followed by the degree-preserving node sets (2nd, 7th, and 12th--15th rows), followed by the node sets generated by the heuristic algorithms (remaining rows). 
The heuristic algorithms, except the community-based algorithm, yielded smaller approximation error than with completely random node sets. Specifically, the $k$-constrained, $k$-quantiled, $k_{\rm nn}$-constrained, and community-based node sets have $1.14$, $1.30$, $1.25$, and $0.98$ times smaller approximation error, respectively, than the completely random node sets (see Table \ref{tab:HSD-home}). This result verifies that the structural properties that our optimization algorithm emphasizes, i.e., degree, average nearest-neighbor degree, and community membership, characterize good sentinel nodes. However, their contribution to suppressing the approximation error is modest. These observations also hold true for weighted (Table~\ref{tab:HSD-home-weighted}) and directed (Table~\ref{tab:HSD-home-directed}) networks. 

\clearpage

\section{Evaluating approximation error on an alternative dynamics}
\label{sec:alternate-dynamics}

We have assumed that we can use the equations of the actual dynamical system in order to optimize node set selection. This condition is unlikely to hold in practice. In this section, we assess the consequences of optimizing with a proposed, or training, dynamics that is not the actual, or test, dynamics. Our procedure works as follows. 
For a given network, we pretend that we do not know the test dynamics (e.g., SIS dynamics) and therefore run a training dynamics (e.g., coupled double-well dynamics). Then, we analyze the approximation error obtained by the node sets we optimized with the training dynamics when the actual dynamics are the test dynamics. See Fig.\,5a  
in the main text. We further show the corresponding results for a weighted mobility-based human network and a directed air-traffic network in Figs.~\ref{fig:dyn-com-Proxw} and \ref{fig:dyn-com-Flights}, respectively. The results are similar to those shown in Fig.\,5 in the main text, suggesting that our method can work for weighted networks and directed networks and that it is potentially useful for approximating real epidemic outbreaks for which a reasonable dynamics may be unknown.

\clearpage
\includegraphics[width=0.8\textwidth]{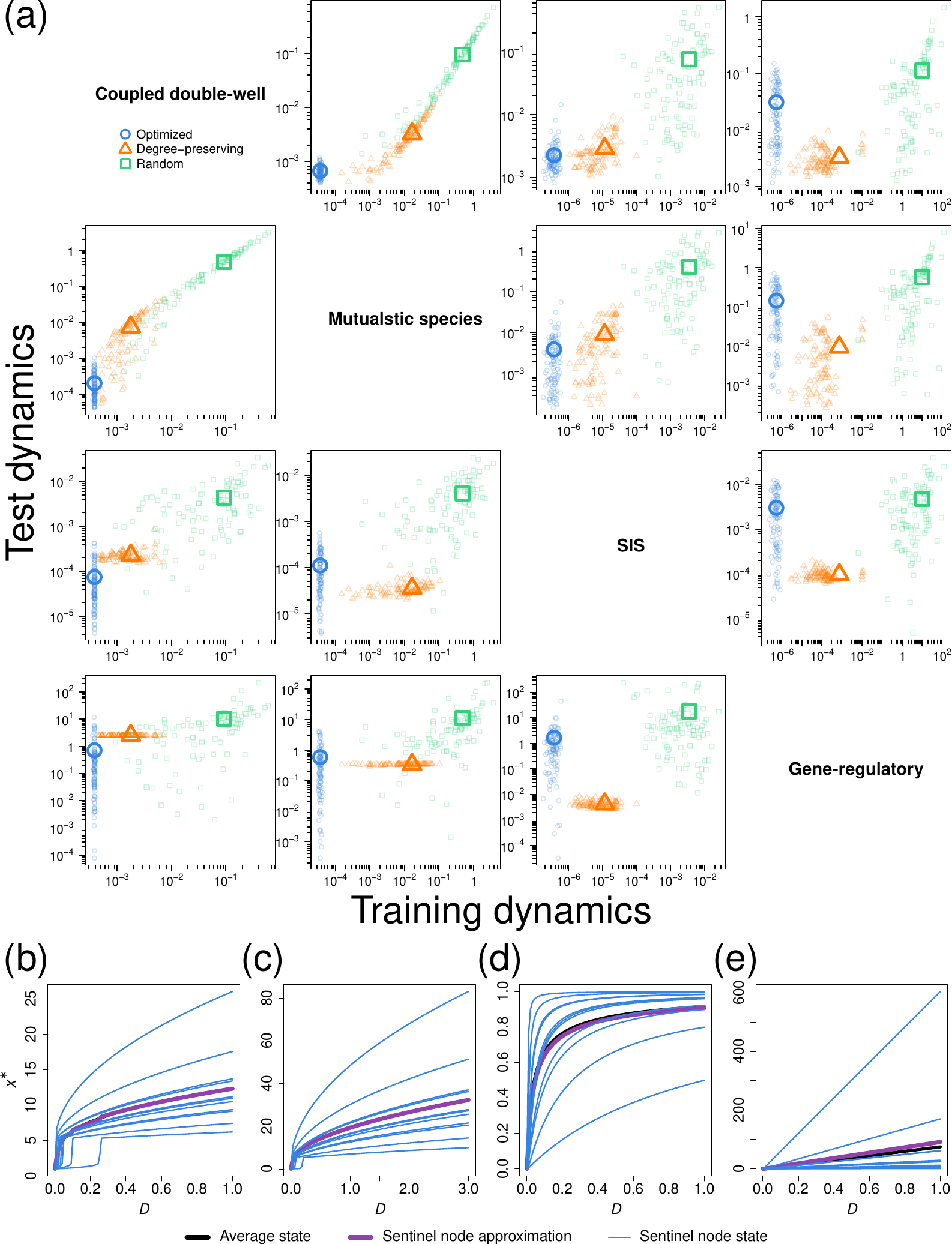}
\clearpage

\begin{figure}[b]
  \centering
  \caption{
    Transfer learnability of the sentinel node approximation on the Prosper network with $N = 89,171$ nodes.
    (a) Approximation error for optimized node sets evaluated on training and test dynamics. The large markers show the average approximation error over the 100 node sets for each type of node set.
    (b) Sentinel node approximation for the coupled double-well dynamics on the dolphin network. The states of the 
    $n = \lfloor \ln N \rfloor = 11$ sentinel nodes in a sentinel node set are shown in blue. We omit the states of the remaining nodes because there are too many of them. The black line represents the unweighted average state of all nodes, $\overline{x}$. The purple line represents the sentinel node approximation. 
    We show the sentinel node approximation for the (c) mutualistic species, (d) SIS, and (e) gene-regulatory dynamics on the same network when the node set is optimized for the coupled double-well dynamics. The blue lines show $x_i^*$ for the sentinel nodes used in (b). In (b) and (c), the purple lines cover the black lines.
}
  \label{fig:dyn-comp-Prosper}
\end{figure}

\clearpage
\newpage

\begin{figure}[b]
  \centering
  \includegraphics[width = 0.8\textwidth]{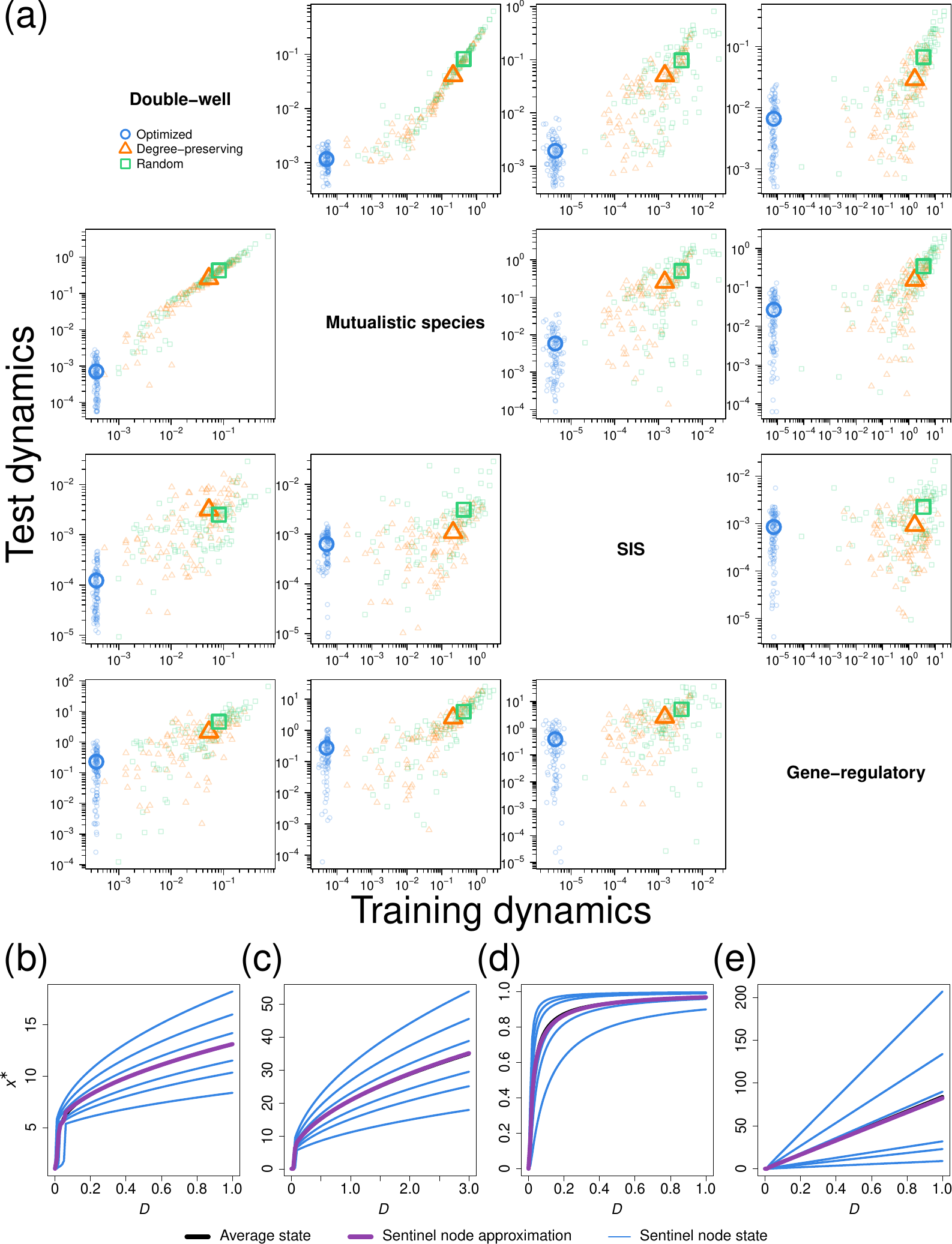}
  \caption{Transfer learnability of the sentinel node approximation on the weighted version of the Proximity network with $N = 410$ nodes. In (b)--(e), we used  $n = \lfloor \ln N \rfloor = 6$ sentinel nodes. See the caption of Fig.~\ref{fig:dyn-comp-Prosper} for details.}
  \label{fig:dyn-com-Proxw}
\end{figure}

\clearpage
\newpage

\begin{figure}[b]
  \centering
  \includegraphics[width = 0.8\textwidth]{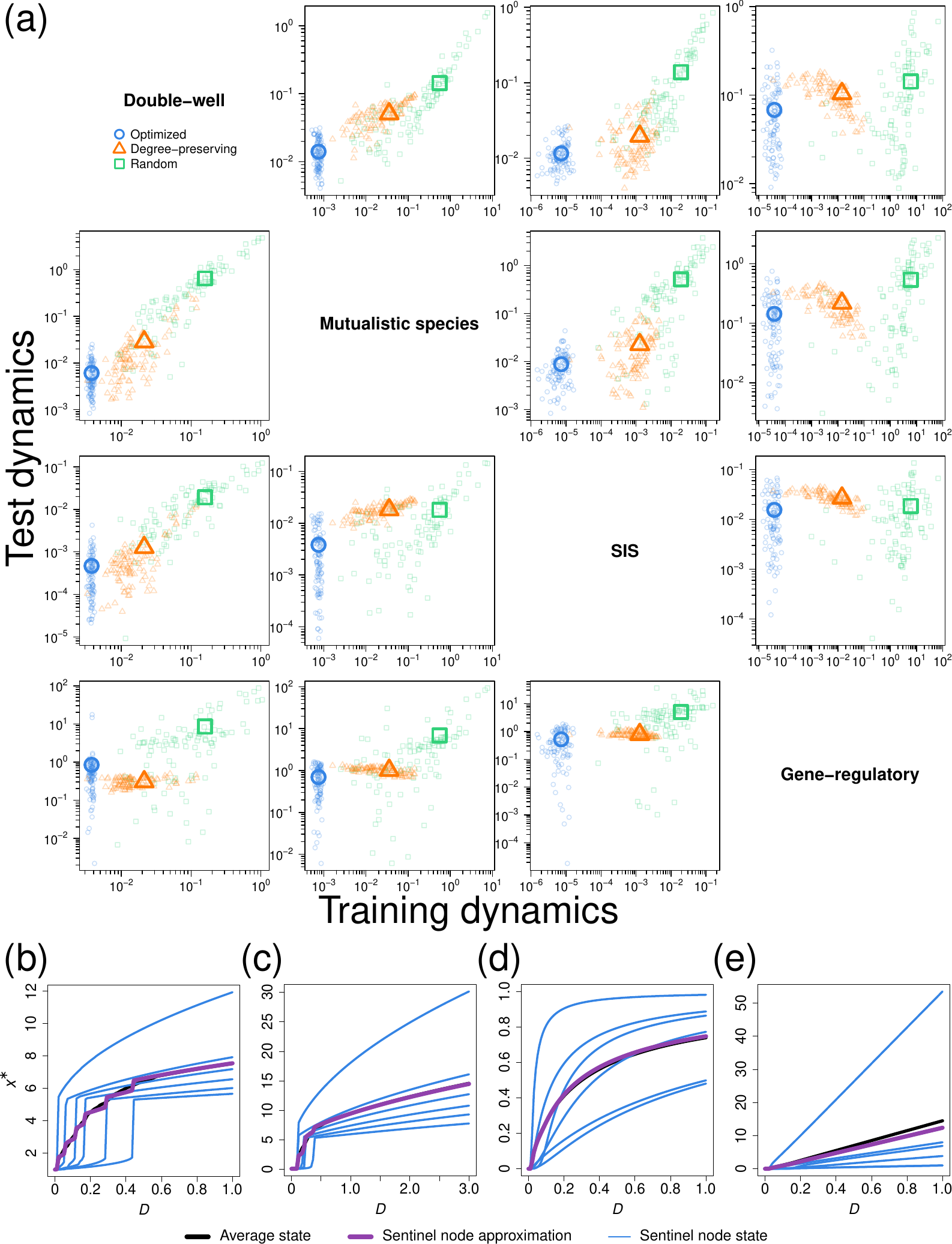}
  \caption{Transfer learnability of the sentinel node approximation on the Flights network with $N = 806$ nodes. In (b)--(e), we used  $n = \lfloor \ln N \rfloor = 6$ sentinel nodes. See the caption of Fig.~\ref{fig:dyn-comp-Prosper} for details.}
  \label{fig:dyn-com-Flights}
\end{figure}

\clearpage
\newpage

We followed this procedure for each combination of training dynamics, test dynamics, undirected and unweighted network, and node set type. Specifically, for each network, we used the 100 node sets optimized on the training dynamics (e.g., coupled double-well dynamics) to calculate approximation errors on each test dynamics (i.e., SIS dynamics, gene-regulatory dynamics, and mutualistic species dynamics if the training dynamics is the coupled double-well dynamics). We carry out this procedure for each pair of different training and test dynamics, each network, and each node set type. We omit the heuristic algorithms for node set selection considered in section~\ref{sec:structural-features}\ref{sub:SI-heuristic} because they do not perform well even when we assume to know the test dynamics (see section~\ref{sec:structural-features}\ref{sub:SI-heuristic} for the results). Note that only the optimized and degree-preserving node sets use information on the dynamics. Therefore, there are 100 optimized and degree-preserving node sets for each pair of training dynamics and network, but 100 node sets for each network for completely random node sets. 

As in section \ref{sec:compare-networks}, we analyzed $\ln \varepsilon$ as the dependent variable in a multi-way ANOVA. We excluded data from the ER network for the same reason as that stated in section~\ref{sec:compare-networks}. Even without knowledge of the test dynamics, our model explains a substantial portion of the variance in $\ln \varepsilon$ ($R^2 = 0.64$), although it is less than when we do know the test dynamics (see section~\ref{sec:compare-networks}). Each of the four independent variables is highly significant (training dynamics: $df = 3$, $F = 2,355$, $p < 10^{-7}$; test dynamics: $df = 3$, $F = 27,863$, $p < 10^{-7}$; network: $df = 18$, $F = 605$, $p < 10^{-7}$; node set type: $df = 2$, $F = 9,561$, $p < 10^{-7}$). Note that the large number (i.e., $68,400$) of observations in part explains the small $p$ values.

Table \ref{tab:HSD-other} shows the differences in average $\ln \varepsilon$, estimated by Tukey's honestly significant difference test. Despite not knowing the test dynamics, optimized node sets are still associated with the smallest $\varepsilon$, followed by the degree-preserving node sets and then by the completely random node sets.

\begin{table}[b]
\centering
    \caption{Differences in average $\ln \varepsilon$ between different node set types when we do not know the test dynamics. The differences are computed by a Tukey's honestly significant difference test. The 95\% confidence intervals (CI) of the differences and associated $p$ values, adjusted for multiple comparisons, are also shown.}
  \label{tab:HSD-other}
  \begin{tabular}{lScc}
    \toprule
    & {Difference} & CI & $p$\\
    \midrule
    Optimized $-$ Random & -2.006 &  $[-2.042, -1.970]$ & $< 10^{-7}$\\
    Degree-preserving $-$ Random & -1.587 & $[-1.623, -1.551]$ & $< 10^{-7}$\\
    Degree-preserving $-$ Optimized & 0.419 & $[0.383, 0.455]$ & $< 10^{-7}$\\
    \bottomrule
  \end{tabular}
\end{table}

\begin{figure}
  \centering
  \includegraphics[width = 0.9\textwidth]{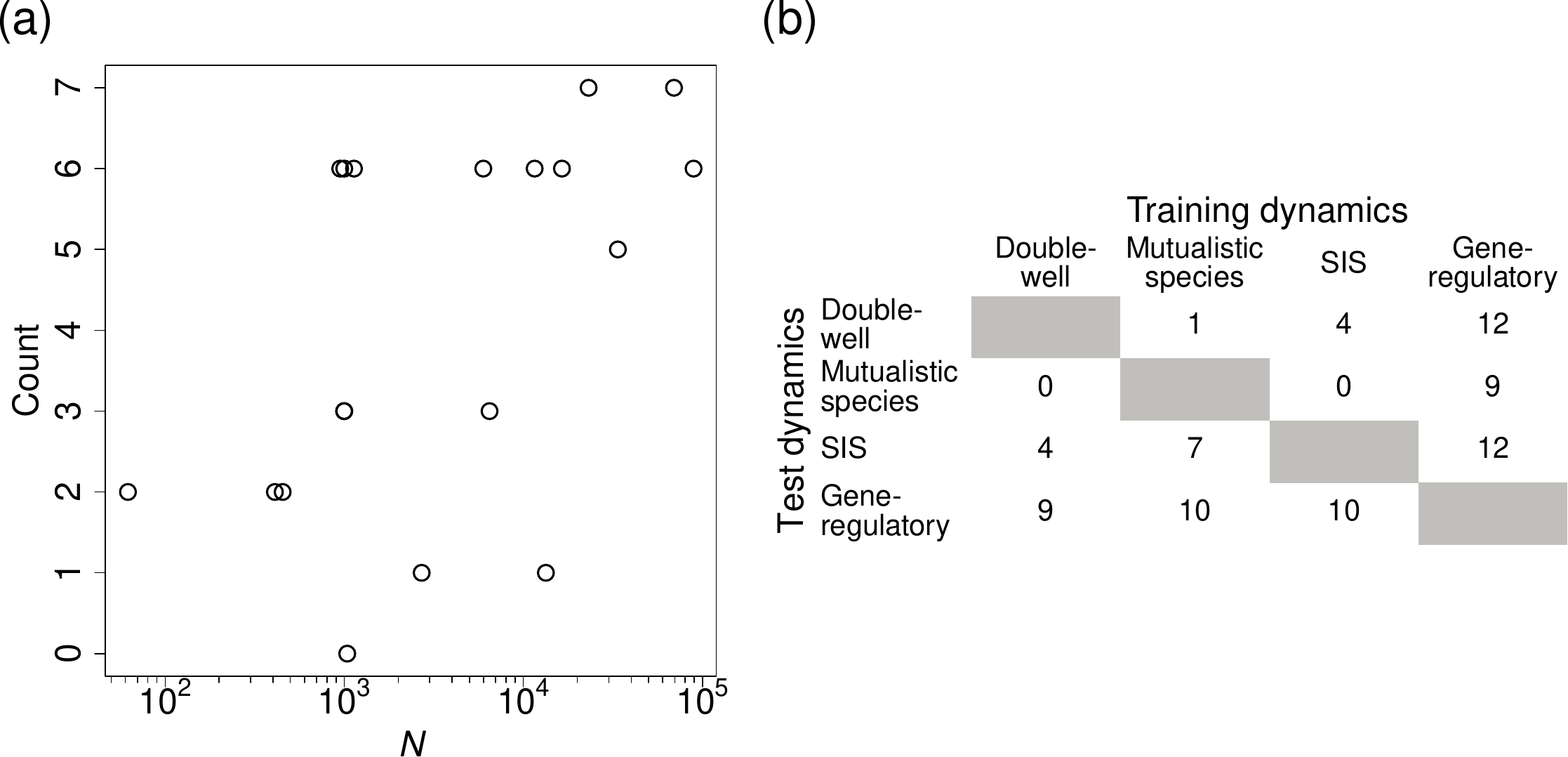}
  \caption{The number of cases in which the degree-preserving node sets outperform the optimized node sets in transfer learning. (a) Count of pairs of training and test dynamics for which the degree-preserving node sets are better. A circle represents a network. (b) Count of networks for which the degree-preserving node sets are better, given the pair of training and test dynamics. In both (a) and (b), we excluded the ER network because it has limited heterogeneity in the degree distribution.}
  \label{fig:degree-preserving-outperform}
\end{figure}

Despite these statistical results, an analysis of the Prosper network, which is the largest network used in the present study, reveals a different picture. On average, Fig.~\ref{fig:dyn-comp-Prosper} indicates that the degree-preserving node sets outperform the optimized node sets on the test dynamics in six out of the twelve pairs of training and test dynamics (i.e., the large triangle is below the large circle in six panels in Fig.~\ref{fig:dyn-comp-Prosper}a). This result contrasts to the case of the dolphin network shown in Fig.~5a in the main text, in which the degree-preserving node sets outperform the optimized node sets on the test dynamics only in two pairs of training and test dynamics. A major difference between the two networks used is the size of the network. Therefore, to explore generality of this observation, for each of the 19 networks, we counted the number of pairs of training and test dynamics for which the degree-preserving node sets perform better than the optimized node sets on average. We show the results as a function of the number of nodes, $N$, in Fig.~\ref{fig:degree-preserving-outperform}a. The figure suggests that the degree-preserving node sets perform relatively better for larger $N$.
The Pearson's correlation coefficient between $\log N$ and the count is $r = 0.52$ with $p = 0.021$.

We also found that the degree-preserving node sets tend to perform better when either the training or test dynamics is the gene-regulatory dynamics. To quantify this observation, for each pair of training and test dynamics, we counted the number of networks for which the degree-preserving node sets outperform the optimized node sets. We show the results in Fig.~\ref{fig:degree-preserving-outperform}b.

\newpage
\clearpage

\section{Incomplete observation of the network\label{sec:incomplete-observation}}

To investigate the performance of optimized sentinel node sets when some nodes or edges are not observable, we carry out the following stress test. For a given network, we first remove a given fraction of edges, $f$, selected uniformly at random. Then, we run our optimization algorithm on the reduced network to determine a sentinel node set. Then, we test its performance on the original network. This scenario corresponds to the case in which some edges are unobservable but contribute to the ongoing network dynamics. We use the coupled double-well dynamics for demonstration.

\begin{figure}
\centering
\includegraphics[width = \textwidth]{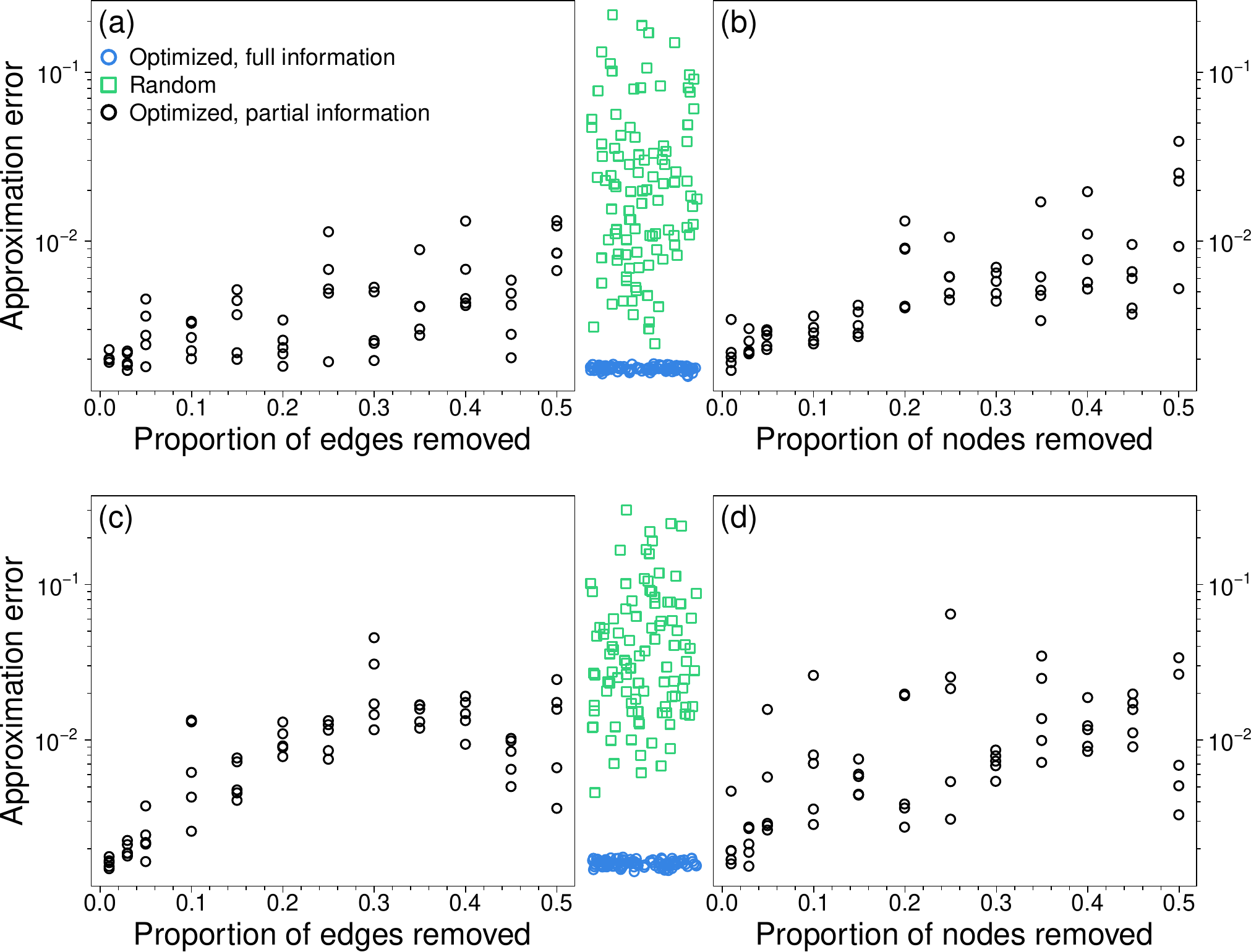}
  \caption{Sentinel node approximation under incomplete observation of networks. (a) Proximity network under edge removal. (b) Proximity network under node removal. (c) Email network under edge removal. (d) Email network under node removal. We use the coupled double-well dynamics and show the approximation error, $\varepsilon$, with $n = \lfloor \ln N \rfloor$ sentinel nodes. Each marker in the main plot represents the approximation error, $\varepsilon$, for a sentinel node set optimized on the reduced network in which some edges or nodes are unobserved. The obtained sentinel node set is tested against the dynamics on the original network. For each network, we also show the results for $100$ optimized node sets (blue circles) and 100 completely random node sets (green squares) for the original network. We show $\varepsilon$ for the node sets using the original network between panels (a) and (b) and between panels (c) and (d) to ease the comparison between the reduced and original networks.}
  \label{fig:incomplete-observation}
\end{figure}

We show in Fig.~\ref{fig:incomplete-observation}a results when some edges in the Proximity network are missing. For each $f$ value used in the figure, we generate one reduced network and run the optimization algorithm five times per reduced network. A circle in the figure corresponds to a node set and shows the approximation error measured against $\{x_1^*, \ldots, x_N^*\}$ for the dynamics run on the original network. The blue and green symbols represents the approximation error for node sets optimized for the original network and that for completely random node sets, respectively, for reference. By noting the logarithmic scale on the vertical axis, we suggest that the approximation error remains overall smaller than that for the completely random node sets roughly when $f \le 0.3$. 

We also run similar experiments removing a fraction $f$ of nodes, rather than edges, uniformly selected at random. In this scenario, we also remove all the edges incident to the removed nodes. We show the approximation error for the node set optimized for reduced networks, where the approximation error is measured for the dynamics run on the original network, in Fig.~\ref{fig:incomplete-observation}b. Qualitatively similar to the case of edge removal, the approximation error remains smaller than the case of completely random node sets roughly when $f \le 0.15$.

Results for a larger network are similar (see Fig.~\ref{fig:incomplete-observation}c and \ref{fig:incomplete-observation}d), whereas the approximation error grows faster as we remove edges in this network (see Fig.~\ref{fig:incomplete-observation}c) compared to the first network (see Fig.~\ref{fig:incomplete-observation}a). We conclude that our framework for selecting good sentinel node sets is tolerant to missing observation of up to a nonnegligible fraction (such as $f=0.1$ or $0.2$) of nodes or edges.

\newpage
\clearpage

\section{When the coupling strength of only some edges vary\label{sec:some-D}}

In the main text, we assumed that the coupling strength between nodes, $D$, is globally shared and tuned. In this section, we investigate the extent to which the performance of optimized sentinel node sets is robust when this assumption is violated. We assume that the coupling strength for a fraction $f$ of edges remains fixed at the initial value of $D$ and that the $D$ value for the other edges varies as it does in the analysis in the main text. If $f$ is small, then the dynamics as we vary the $D$ value would be similar to the original case, i.e., when all edges share the varying $D$ value. We computed sentinel node sets when $D$ is globally tuned (i.e., $f=0$) and tested the performance of the obtained sentinel node sets against simulations with $f>0$.
We considered five methods to select a fraction $f$ of edges for which the coupling strength is fixed over the course of simulations. We select the fraction $f$ of edges with (i) the lowest edge betweenness centrality, (ii) highest edge betweenness centrality, (iii) lowest degree sum, (iv) highest degree sum, or (iv) uniformly at random. The degree sum of the edge is defined as the sum of the degree of two nodes forming the edge.

\begin{figure}
  \centering
  \includegraphics[width = \textwidth]{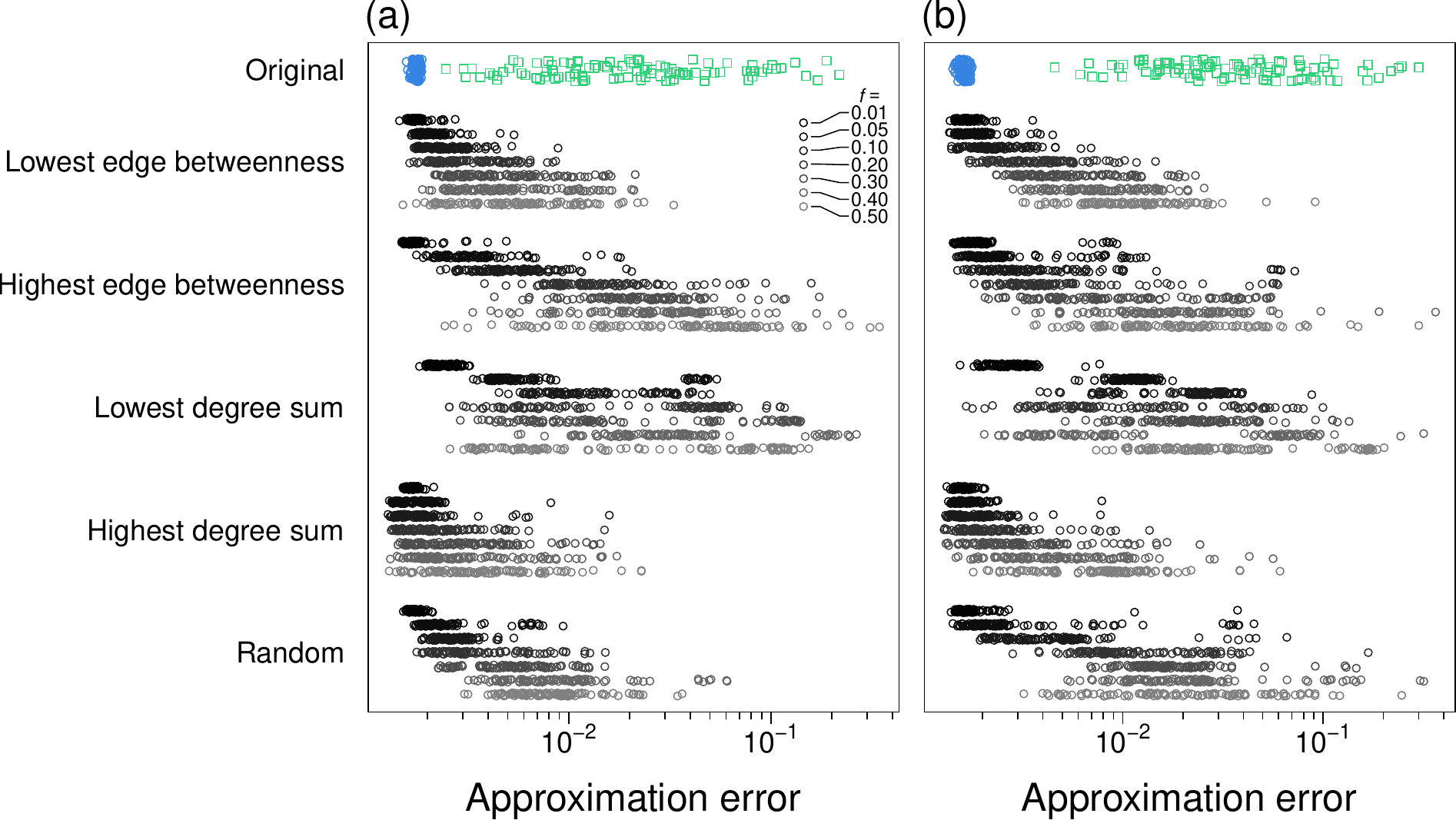}
  \caption{Sentinel node approximation when the coupling strength gradually varies only for a fraction $1-f$ of edges. (a) Proximity network. (b) Email network. Each gray and blue circle represents a sentinel node set optimized for the original network. The darker gray circles correspond to a larger value of $f$. The green squares represent completely random node sets. The blue and green symbols show the approximation error, $\varepsilon$, assessed for the original case, i.e., when the coupling strength is shared by all edges. The gray symbols show $\varepsilon$ assessed for the case in which a fraction $f$ of edges have a fixed coupling strength, and the remaining edges have a shared varying coupling strength. We use the coupled double-well dynamics and set $n = \lfloor \ln N \rfloor$.}  \label{fig:some-D}
\end{figure}

We show the approximation error for combinations of $f = \{0.01, 0.05, 0.1, 0.2, 0.3, 0.4, 0.5 \}$ and one of the five edge selection methods for the Proximity network in Fig.~\ref{fig:some-D}a. We used the coupled double-well dynamics. For $100$ optimized node sets determined under the globally tuned coupling strength, which is used in all other parts of this study, the approximation error under the same globally tuned coupling strength is shown by the blue circles. Each row of the $100$ circles of the same darkness of gray circles corresponds to a value of $f$. A darker gray corresponds to a smaller value of $f$. For example, the black circles in the first of the seven rows of circles for each edge selection method correspond to $f=0.01$. The lightest gray circles in the seventh row correspond to $f=0.5$. The $100$ gray circles in each row correspond to the same $100$ optimized node sets shown in blue, but evaluated for the scenario in which a fraction $f$ of edges has a fixed coupling strength while the other fraction $1-f$ of edges has a varying coupling strength. The green squares show the approximation error for $100$ completely random node sets in the case of the globally shared $D$ as reference.

Figure~\ref{fig:some-D}a suggests that the approximation error when some edges stick to its initial coupling strength remains smaller than the completely random case when $f \le 0.1$ overall for four of the five edge selection methods (i.e., except for ``lowest degree sum''). Note that the horizontal axis is on the logarithmic scale. For the ``lowest degree sum'', the approximation error is smaller than the completely random node sets only for $f\le 0.05$. The results are similar for a larger network, as shown in Fig.~\ref{fig:some-D}b. We consider that optimized node sets determined with $f=0$ is transferrable only up to a relatively small value of $f$ because selectively changing edges with the highest/lowest betweenness or degree sum distorts the effective weighted network considerably even if $f$ is small. In fact, when a fraction $f$ of edges is uniformly randomly selected, a reasonable quality of transferability persists up to $f = 0.4$ for both networks (see ``Random'' in Fig.~\ref{fig:some-D}). Therefore, we conclude that our sentinel node selection method bears some robustness (i.e., up to $f=0.05$ to $0.1$ even when particular edges are chosen to have a fixed coupling strength and up to $f=0.4$ when edge selection is uniformly at random) under this scenario.

\newpage
\clearpage

\section{Sentinel node approximation for brain data}

To demonstrate transfer learning with multivariate time series data for which the dynamics generating the data is unknown, we used the resting-state functional magnetic resonance imaging (fMRI) data recorded from healthy human individuals collected in the Human Connectome Project (HCP)~\cite{Van_Neuroimage2013wu}. The participants are $1,200$ adults between $22$–$35$ years old who underwent four sessions of 15-min EPI sequence with a 3T Siemens Connectome-Skyra (TR $= 0.72$ s, TE $= 33.1$ ms, $72$ slices, $2.0$ mm isotropic, field of view (FOV) $= 208 \times 180$ mm) and a T1-weighted sequence (TR $= 2.4$ s, TE $= 2.14$ ms, $0.7$ mm isotropic, FOV $= 224 \times 224$ mm). The resting-state fMRI data of each participant are composed of two sessions, and each of the two sessions is broken down into a Left-Right (LR) and Right-Left (RL) phases. Each phase provides a time series with 1,200 time points. 
We used the preprocessed time series data provided by the HCP, called the ``node timeseries (individual subjects)'' data. These data are from 1,003 participants having complete resting-state fMRI data over four phases. The HCP provides the data from the four phases as a concatenated time series of 4,800 time points. Each node is an ICA (independent component analysis) component as a result of parcellation of the brain. We used the time series with the largest number of nodes, i.e., $N=300$. 

Some participants have nodes showing outlier behavior. To remove the data from such participants, for each participant, we first computed $\overline{x}_i$, $\forall i$ for the test set, i.e., the last 100 time points. We then computed the mean and standard deviation of the 100 values of $\overline{x}_i$, denoted by $\mu$ and $\sigma$, respectively. If any $\left| x_i \right|$ at any time point is larger than $\mu + 5\sigma$, we removed that participant's data from further consideration. Using this rule, we removed $103$ out of $1,003$ participants, leaving $900$ participants.

For each of the $900$ participants, we estimated a network by the Pearson correlation coefficient between each pair of nodes. We used the first 4,700 time points from each node to calculate the correlation. Although the Pearson correlation is a naive method, various methods for estimating networks from correlation data have pros and cons \cite{masuda2025}. Because our goal of the data analysis is to demonstrate transfer learning in the setting where the dynamics producing time series data is unavailable, we resorted to a simplest network estimation method, i.e., Pearson correlation. We discarded negative correlation coefficient values to obtain a (positively) weighted network. Although the  ``node timeseries (individual subjects)'' data set contains two networks per participant, they are correlation-based. We computed the correlation network from the time series data on our own because, in this manner, we can separate the training data (i.e., the first 4,700 time points) to estimate the network and the test data (i.e., the last 100 time points) for running our optimization algorithm and assessing the performance of the obtained sentinel node sets.

We ran the coupled double-well dynamics on the network estimated for each participant for the range of $D$ values used in the main text. Then, we ran the optimization algorithm to identify a set of $n = \lfloor \ln N \rfloor = 5$ sentinel nodes. We removed $\sum_{\ell=1}^L \overline{x}_{\ell}$ in the definition of $\varepsilon$, i.e., Eq.~(3) in the main text, because the present $x_i$ can take both positive and negative values and this sum only contributes a normalization factor without affecting the optimization. For each participant, we repeated this procedure $100$ times, and computed $\varepsilon$ for each optimized node set. We also generated the same number of degree-preserving and completely random node sets and calculated $\varepsilon$ for each node set. We visually compare the average $\varepsilon$ between the node set types across the $900$ participants in the main text. 
We find that, although the tendency is weak, optimal node sets tend to yield a smaller $\varepsilon$ than completely random node sets. We used a mixed effects model in which the dependent variable was approximation error, the fixed independent variable was node set type (reference: completely random node sets), and there was a random intercept for each individual. Because $\varepsilon$ is not widely distributed, we used $\varepsilon$, not $\ln \varepsilon$, as the dependent variable. We found that $\varepsilon$ for the optimal node sets was 5.81\% smaller than that for the completely random node sets (see Table \ref{tab:fmri-models}).

To assess the generality of these results with respect to the choice of the dynamics, we ran the same analysis for the mutualistic species, SIS, and gene-regulatory dynamics. In addition, we ran the same analysis for the coupled Wilson-Cowan model for neuronal dynamics given by
  \begin{equation}
    \frac{\text{d}x_i}{\text{d}t} = -x_i + \sum_{j=1}^N \frac{A_{ij}}{1 + e^{\mu - \tau x_j}},
  \end{equation} 
where $x_i$ represents the activation level of a mass of neurons, $\mu$ controls the activation threshold, and $\tau$ controls the slope of the activation function \cite{laurence2019}. We set $\mu=3$ and $\tau=1$ following \cite{laurence2019}. We show the results in Table~\ref{tab:fmri-models}. The table indicates that the result that optimal and degree-preserving node sets, in particular the former, provide a better approximation to $\overline{x}$ than completely random node sets remains the same for the different dynamics.

In Table~\ref{tab:fmri-models}, we analyzed each dynamics separately. We further confirmed these results by evaluating the same statistical model as the one used in Table~\ref{tab:fmri-models} (i.e., mixed effects with random intercepts for participants, etc.) for all five dynamics combined, except that we added the model of dynamics used as the training dynamics as another categorical independent variable. The results of this analysis, shown in Table~\ref{tab:fmri-all}, are similar to those of analyzing the dynamics separately, shown in Table~\ref{tab:fmri-models}. In particular, the training dynamics used to generate node sets does not have a strong effect on the approximation error when compared to the type of node set. Over all dynamics, optimized node sets on average achieve 6.40\% smaller approximation error than completely random node sets; degree-preserving node sets achieve 3.13\% smaller approximation error.

\begin{table}
  \centering
  \caption{
    Improvement, with respect to completely random node sets, of optimized and degree-preserving node sets on the fMRI data for each of five training dynamics. The coefficient estimates $b$, as well as the associated $t$ and $p$ values, come from separate mixed-effects models for each training dynamics.
  }
  \label{tab:fmri-models}
  \begin{tabular}{llSSSc}
    \toprule
    Training dynamics & Node set type & {\% improvement} & {$b$} & {$t$} & $p$\\
    \midrule
    Coupled double-well & Optimized & 5.81 & -103.7 & -7.727 & $< 10^{-7}$\\
    Coupled double-well & Degree-preserving & 3.33 & -59.48 & -4.431 & $9.96 \times 10^{-6}$\\
    Mutualistic species & Optimized & 5.95 & -106.7 & -8.429 & $< 10^{-7}$\\
    Mutualistic species & Degree-preserving & 1.72 & -30.87 & -2.440 & 0.0148\\
    SIS & Optimized & 3.94 & -70.43 & -5.656 & $< 10^{-7}$\\
    SIS & Degree-preserving & 3.96 & -70.74 & -5.680 & $< 10^{-7}$\\
    Gene-regulatory & Optimized & 7.79 & -139.1 & -12.23 & $< 10^{-7}$\\
    Gene-regulatory & Degree-preserving & 3.45 & -61.58 & -5.413 & $< 10^{-7}$\\
    Wilson-Cowan & Optimized & 8.5  & -152.2 & -11.95 & $< 10^{-7}$\\
    Wilson-Cowan & Degree-preserving & 3.18 & -56.81 & -4.461 & $8.68 \times 10^{-6}$\\
    \bottomrule
  \end{tabular}
\end{table}

\begin{table}
  \centering
  \caption{Statistical results for a single mixed effects model, including all five dynamics, for the fMRI data. The dependent variable is the approximation error. There were two categorical fixed effects: training dynamics (reference: coupled double-well dynamics) and node set type (reference: completely random node sets). There was a random effect for participant.}
  \label{tab:fmri-all}
  \begin{tabular}{lSSc}
    \toprule
    & $b$ & $t$ & $p$\\
    \midrule
    Intercept & 1787 & 146.5 & $< 10^{-7}$\\
    Training dynamics & & & \\
    \hspace{1 em} Mutualistic species & 15.14 & 2.108 & 0.0350\\
    \hspace{1 em} SIS & 8.727 & 1.215 & 0.2244\\
    \hspace{1 em} Gene-regulatory & -10.94 & -1.523 & 0.1277\\
    \hspace{1 em} Wilson-Cowan & -10.84 & -1.509 & 0.1313\\
    Node set type & & &\\
    \hspace{1 em} Optimized & -114.4 & -20.57 & $< 10^{-7}$\\
    \hspace{1 em} Degree-preserving & -55.90 & -10.05 & $< 10^{-7}$\\
    \bottomrule
  \end{tabular}
\end{table}

\newpage
\clearpage

\section{Optimizing node weights\label{sec:SIweights}}

In this section, we analyze the effect of optimizing node weights in additional to optimizing node selection.

\subsection{An example\label{sub:weighted-example}}

\begin{figure}[h]
  \centering
  \includegraphics[width=0.6\textwidth]{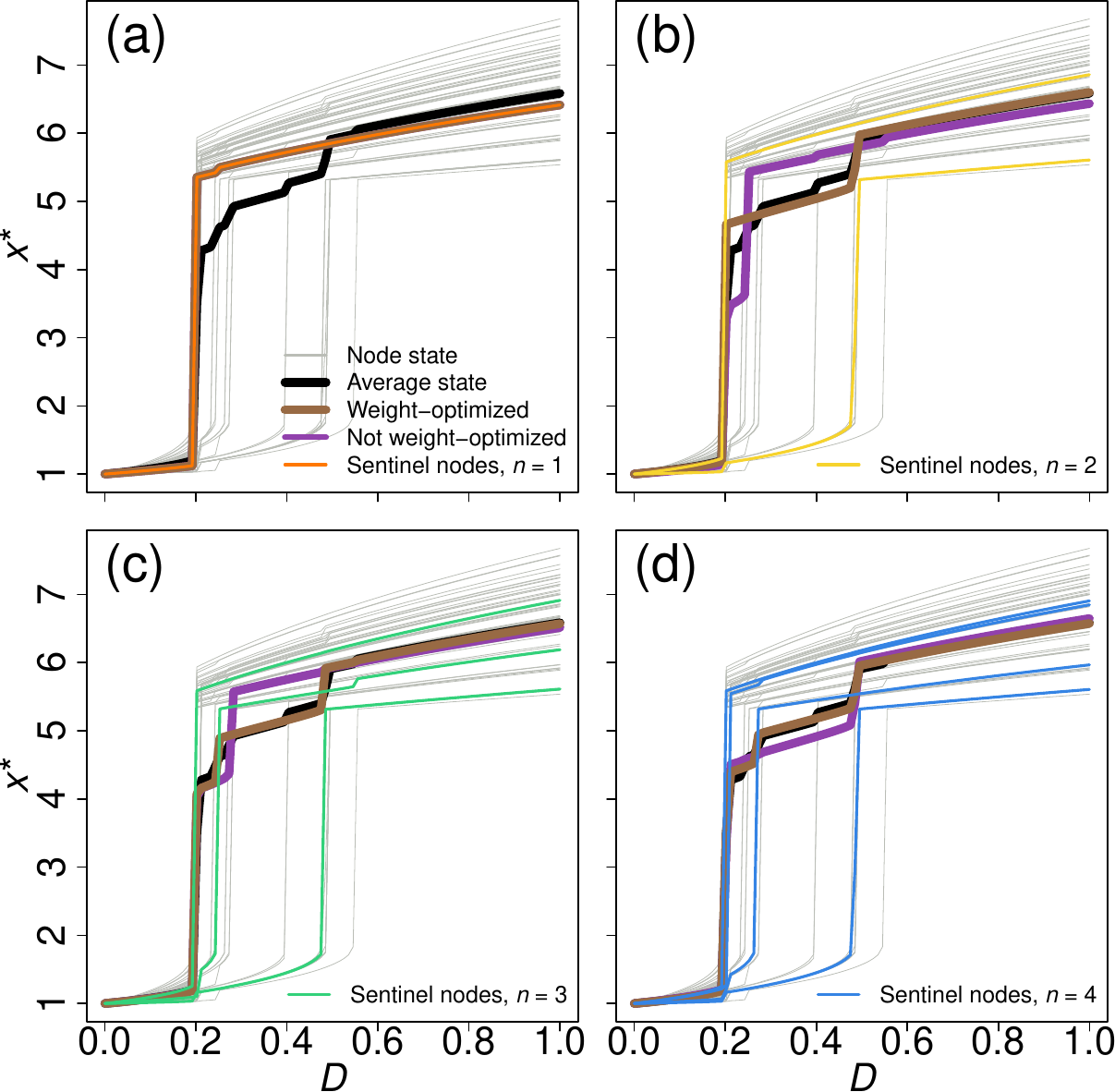}
  \caption{Approximation of the average state, $\overline{x}$, of the coupled double-well dynamics on the dolphin network when we also optimize node weights. (a) $n = 1$. (b) $n = 2$. (c) $n = 3$. (d) $n = 4$. The tan lines represent the weight-optimized approximation. The purple lines represent the corresponding approximation when we do not optimize node weights. In (a), the tan line completely covers the purple line.
}
  \label{fig:weighted-dolphin-demo}
\end{figure}

In Fig.~1 
in the main text, we demonstrate our algorithm on sentinel node sets of size $n \in \{1, 2, 3, 4\}$ for the coupled double-well dynamics on the dolphin network. In Fig.~\ref{fig:weighted-dolphin-demo}, we show a similar example when we also optimize node weights. When we use $n \in \{2, 3, 4\}$, the additional optimization step---that is, optimizing node weights in addition to the combinatorial optimization---allows us to even more closely approximate the averaged network activity, $\overline{x}$. 

\newpage
\clearpage

\subsection{Effect of node set type on approximation error}
\label{sub:home-dynamics-w}

To quantify performances of the optimized node sets in which we also optimize the node weights, we built an ANOVA model using $\ln \varepsilon$ as the dependent variable, as we did in section~\ref{sec:structural-features}\ref{sub:SI-heuristic}. We generated 100 weight-optimized node sets for each pair of dynamics and network. For comparison, we use the same completely random and combinatorially optimized node sets (i.e., without optimizing node weights) from section \ref{sec:structural-features}\ref{sub:SI-heuristic}. The three independent variables were dynamics, network, and node set type (random, optimized, and weight-optimized). All three independent variables were significant (dynamics: $df = 3$, $F = 5,436.06$, $p < 10^{-7}$; network: $df = 8$, $F = 155.76$, $p < 10^{-7}$; node set type: $df = 2$, $F = 12,362.80$, $p < 10^{-7}$). 

We computed the differences between the three node set types with Tukey's honestly significant difference test. We show the results in Table \ref{tab:HSD-home-w}. As described in the main text, the average $\ln \varepsilon$ for the weight-optimized node sets is small, i.e., $1/e^{-3.832} = 46.16$ times smaller than the optimized node sets without the additional weight optimization step.

\begin{table}[b]
  \centering
  \caption{Differences between node set types in terms of average $\ln \varepsilon$ when we optimize node weights and know the test dynamics. CI: confidence interval.}
  \label{tab:HSD-home-w}
  \begin{tabular}{lScc}
    \toprule
    & {Difference} & CI & $p$\\
    \midrule
    Optimized $-$ Random & -5.625 & $[-5.767, -5.483]$ & $< 10^{-7}$\\
    Weight-optimized $-$ Random & -9.456 & $[-9.598, -9.315]$ & $< 10^{-7}$\\
    Weight-optimized $-$ Optimized & -3.832 & $[-3.973, -3.690]$ & $< 10^{-7}$\\
    \bottomrule
  \end{tabular}
\end{table}

\clearpage

\subsection{Transfer learning\label{sub:alternate-dynamics-w}}

In this section, we investigate performances of weight-optimized node sets when one does not know the test dynamics. As in section \ref{sec:alternate-dynamics}, we generated 100 weight-optimized node sets optimized with a training dynamics and evaluated the approximation error of those node sets on a test dynamics. We did this for each combination of training dynamics, test dynamics, and network. For comparison, we include the completely random and optimized node sets generated in the analysis in section \ref{sec:alternate-dynamics}.

All of the independent variables of the constructed ANOVA model are significant (training dynamics: $df = 3$, $F = 1,353.61$, $p < 10^{-7}$; test dynamics: $df = 3$, $F = 11,278.18$, $p < 10^{-7}$; network: $df = 8$, $F = 285.22$, $p < 10^{-7}$; node set type: $df = 2$, $F = 5,171.33$, $p < 10^{-7}$), and the ANOVA model fits the data well ($R^2 = 0.61$). As reported in the main text, there is no significant difference between weight-optimized node sets and our standard optimized node sets (i.e., without node-weight optimization) when we optimize on a dynamics that is not the test dynamics ($p = 0.591$; see Table~\ref{tab:HSD-other-w}).

\begin{table}[b]
  \centering
  \caption{Differences between node set types in terms of average $\ln \varepsilon$ when we optimize node weights and do not know the test dynamics. CI: confidence interval.}
  \label{tab:HSD-other-w}
  \begin{tabular}{lScc}
    \toprule
    & {Difference} & CI & $p$\\
    \midrule
    Optimized $-$ Random & -1.748 & $[-1.794, -1.701]$ & $< 10^{-7}$\\
    Weight-optimized $-$ Random & -1.728 & $[-1.775, -1.682]$ & $< 10^{-7}$\\
    Weight-optimized $-$ Optimized & 0.019 & $[-0.027, 0.066]$ & $0.591$\\
    \bottomrule
  \end{tabular}
\end{table}

\newpage
\clearpage

\section{Time complexity\label{sec:time-complexity}}

\subsection{Node set optimization algorithm}

In the simulated annealing algorithm to determine the sentinel node set, $S$, we iterate the steps of the algorithm $h_{\max} = 50N$ times. Each step of the algorithm consists in selecting a node to be kicked out from $S$ and a different node to be recruited to $S$, both of which need $O(\ln N)$ time by a standard binary search algorithm. Each step also involves the computation of $\varepsilon$ for the tentative new sentinel node set, which requires $O(N)$ time. Therefore, we expect that the time complexity of the entire optimization algorithm is $O(N^2)$.

\begin{figure}[b]
\centering
  \includegraphics[width=0.6\textwidth]{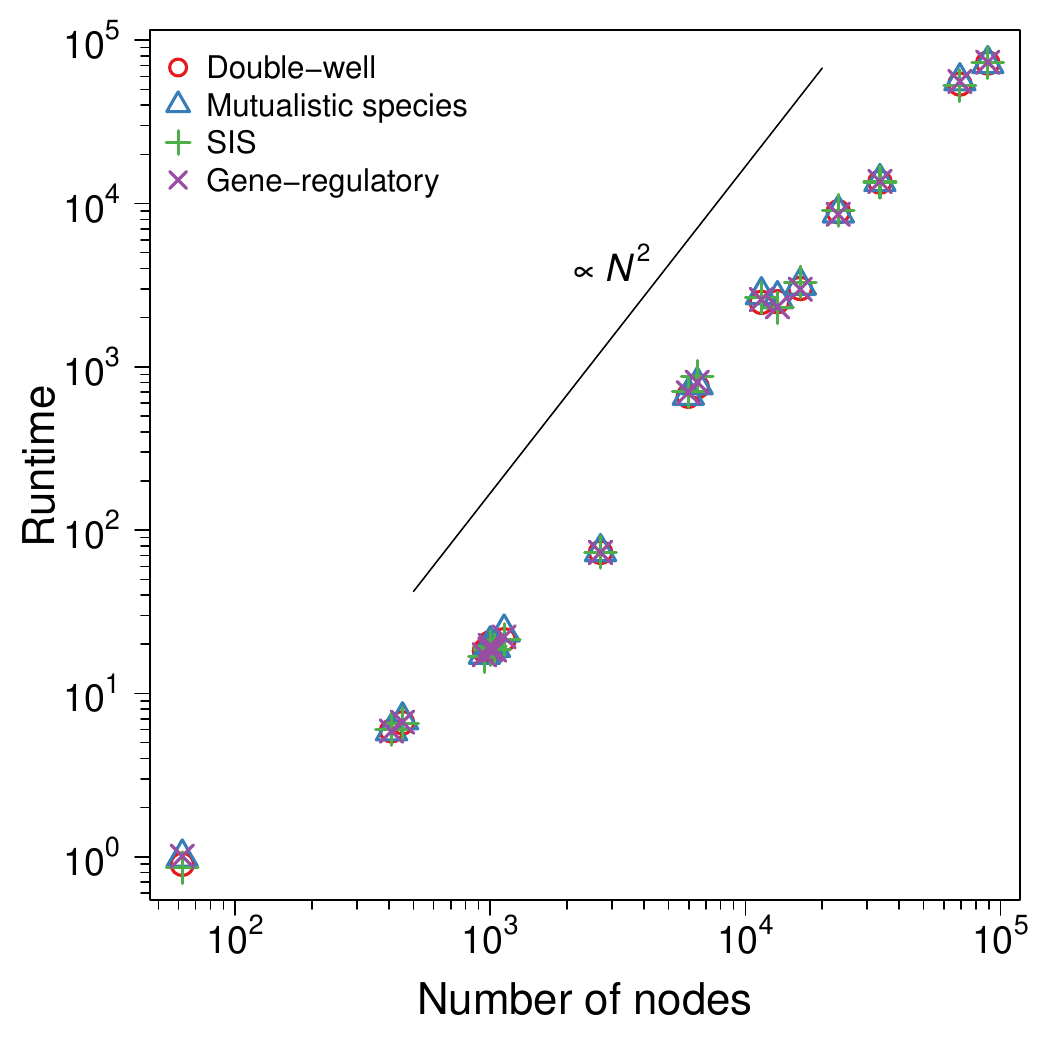}
  \caption{
    Runtime of the simulated annealing algorithm to optimize the sentinel node set for each of the four dynamics models and each of the 20 networks. A symbol represents the largest runtime among 100 runs for a pair of the dynamics model and network. We used Intel Xeon Gold 6448Y processors with 32 central processing units, 64 threads, and 512 GB of memory.
  }
  \label{fig:runtime-SA}
\end{figure}

We investigated the actual runtime of the optimization algorithm for all the combinations of dynamics and network.
We show the results in Fig.~\ref{fig:runtime-SA}. Each symbol in the figure represents the largest runtime among 100 independent runs of the algorithm. The figure indicates that the runtime is $O(N^2)$ or somewhat smaller.

In contrast, the GBB reduction runs in $O(N)$ time, which is devoted to the calculation of the first and second moment of the node's degree, yielding $\beta_{\text{eff}}$. The DART requires the leading eigenvalue and eigenvector, which one can obtain by the power method. One iteration of the power method consists of the multiplication of the adjacency matrix and a vector of size $N$, which costs $O(N^2)$ and $O(N)$ time for dense and sparse networks, respectively, followed by normalization. This multiplication sets the time complexity of DART.

\begin{figure}
\centering
  \includegraphics[width=0.7\textwidth]{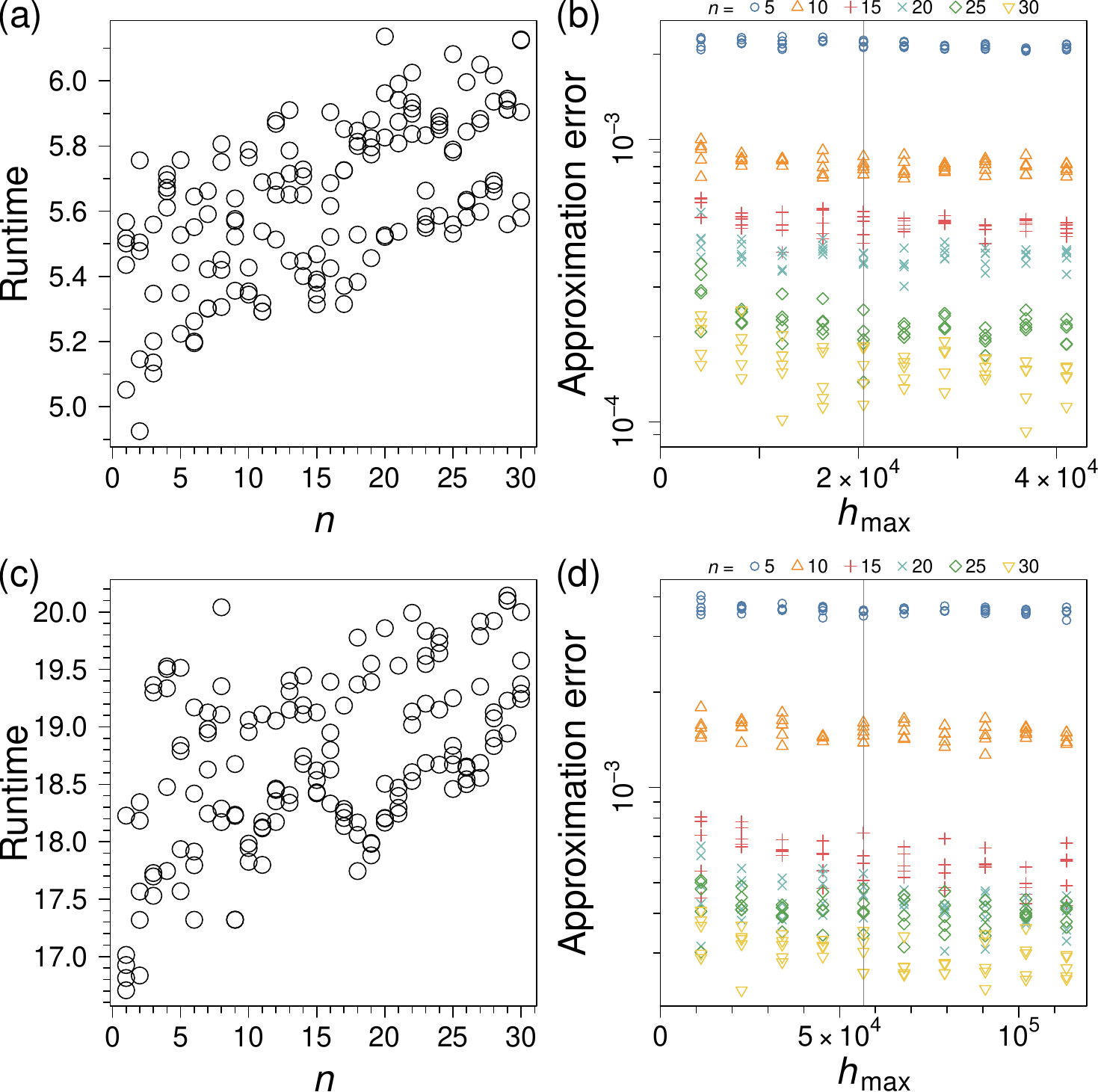}
  \caption{Runtime of the simulated annealing algorithm as a function of the number of nodes in the sentinel node set, $n$. (a) Runtime for the Proximity network. A symbol represents one run of the simulated annealing algorithm. We carried out five runs for each $n$ value used in the figure. (b) Approximation error, $\varepsilon$, as a function of $h_{\max}$ for the Proximity network. (c) Runtime for the Email network. (d) Approximation error for the Email network. We used the coupled double-well dynamics. We set $h_{\max} = 50N$ in (a) and (c). See the caption of Fig.~\ref{fig:runtime-SA} for the hardware used.}
  \label{fig:runtime-SA-n}
\end{figure}

For two arbitrarily selected networks, we also investigated the runtime as a function of $n$, the number of nodes in the sentinel node set. We show in Fig.~\ref{fig:runtime-SA-n}a the runtime of the simulated annealing algorithm for the Proximity network. Each circle represents the runtime for a single run of the simulated annealing. We used the coupled double-well dynamics. We find that the runtime only modestly increases as $n$ increases. The number of iterations of the algorithm, $h_{\max}$, necessary for sufficient convergence of the approximation error, $\varepsilon$, may depend on $n$. Therefore, we also examined $\varepsilon$ as a function of $h_{\max}$ in Fig.~\ref{fig:runtime-SA-n}b for some representative values of $n$. We recall that we set $h_{\max} = 50N$ in generating Fig.~\ref{fig:runtime-SA-n}a as well as all the other results in the main text and the SI. The vertical solid line in Fig.~\ref{fig:runtime-SA-n}b shows $h_{\max} = 50N$. The figure suggests that, for all the values of $n$ shown in the figure, $h_{\max} = 50N$ is large enough for the convergence of the simulated annealing algorithm. Based on Fig.~\ref{fig:runtime-SA-n}a and b, we conclude that our optimization algorithm is not excessively slow as $n$ increases. We have verified that these results remain qualitatively the same for another larger network (see Fig.~\ref{fig:runtime-SA-n}c and d).

\clearpage

\subsection{Dynamics simulations}

\begin{figure}
\centering
  \includegraphics[width=0.6\textwidth]{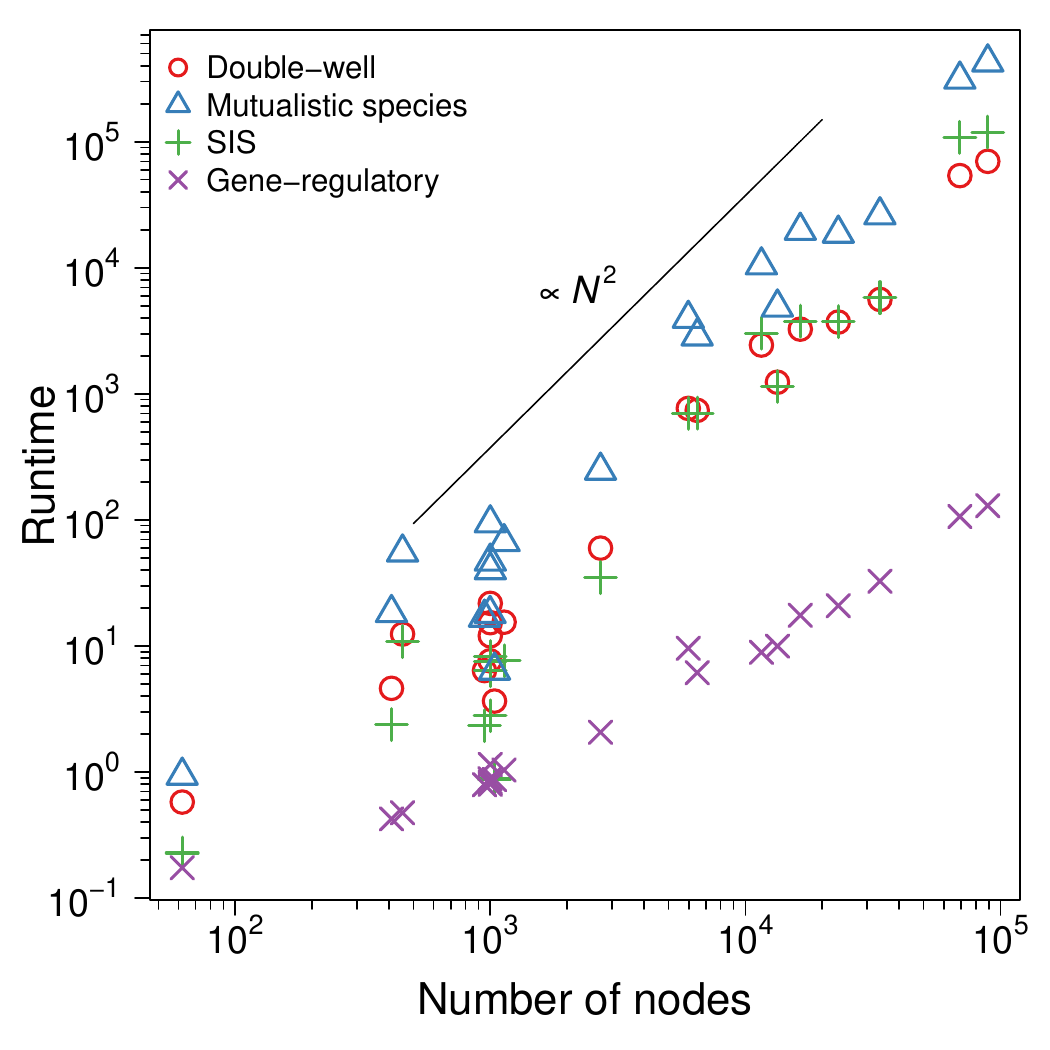}
  \caption{Runtime of the dynamics on networks. For each pair of dynamics and network, we measured the runtime summed over all the control parameter values. A symbol represents the runtime summed over all 100 control parameter values for a pair of dynamics and network. See the caption of Fig.~\ref{fig:runtime-SA} for the hardware used.}
  \label{fig:runtime-ground-truth}
\end{figure}

For practical application of transfer learning, one needs to select a dynamics and numerically simulate it for each control parameter value to obtain $x_1^*$, $\ldots$, $x_N^*$ before running the simulated annealing algorithm to determine a sentinel node set. Therefore, we assessed the runtime of the dynamics as a function of $N$. We remind that we used the implicit Adams method and a fixed total simulation time of $T=15$ (see section IV-B in the main text). We show the runtime for the 20 networks in Fig.~\ref{fig:runtime-ground-truth}. The figure indicates that the runtime scales as somewhat more slowly than $O(N^2)$ as $N$ increases. This result in combination with Fig.~\ref{fig:runtime-SA} supports that the dynamics simulations are not a computational bottleneck as $N$ becomes large when we use our framework for transfer learning assuming no knowledge about the real dynamics.

\clearpage



\end{document}